\documentclass[notitlepage, superscriptaddress, 11pt, single-spaced,floatfix]{revtex4-1}
\usepackage[utf8]{inputenc}
\usepackage{microtype}
\usepackage{amsmath}
\usepackage{amsfonts}

\usepackage{nameref}
\usepackage{enumitem}
\usepackage{graphicx} 
\usepackage{booktabs} 
\usepackage{tabularx}
\usepackage{xcolor} 
\usepackage{multirow} 
\usepackage{placeins}[section]
\usepackage[ruled]{algorithm2e} 
\DeclareMathOperator{\aleft}{\mathit{left}}
\DeclareMathOperator{\aright}{\mathit{right}}

\usepackage{xr-hyper} 
\usepackage{xr}
 
\usepackage{hyperref}
\usepackage{cleveref}

\hypersetup{colorlinks = true, linkcolor = NavyBlue,
            urlcolor  = gray,
            citecolor = NavyBlue,
            anchorcolor = NavyBlue}

\definecolor{Orange}{rgb}{0.9,0.5,0}
\definecolor{NavyBlue}{rgb}{0.1, 0.4, 0.8}
\definecolor{Magenta}{rgb}{0.8, 0.1, 0.6}
\definecolor{Cyan}{rgb}{0.2, 0.85, 0.85}

\makeatletter
\AtBeginDocument{%
	\expandafter\renewcommand\expandafter\subsection\expandafter
	{\expandafter\@fb@secFB\subsection}%
	\newcommand\@fb@secFB{\FloatBarrier
		\gdef\@fb@afterHHook{\@fb@topbarrier \gdef\@fb@afterHHook{}}}%
	\g@addto@macro\@afterheading{\@fb@afterHHook}%
	\gdef\@fb@afterHHook{}%
}
\makeatother

\usepackage[caption=false]{subfig}

\begin{document}

\title{Living in a pandemic: adaptation of individual mobility and social activity in the US}

\author{Lorenzo Lucchini}\thanks{These two authors contributed equally.\\  Correspondence to: lorenzo.f.lucchini.work@gmail.com and work@marcodena.it.}
\affiliation{Fondazione Bruno Kessler (FBK), Trento, Italy}

\author{Simone Centellegher}\thanks{These two authors contributed equally.\\  Correspondence to: lorenzo.f.lucchini.work@gmail.com and work@marcodena.it.}
\affiliation{Fondazione Bruno Kessler (FBK), Trento, Italy}

\author{Luca Pappalardo}
\affiliation{Institute of Information Science and Technologies, National Research Council (ISTI-CNR), Pisa, Italy}

\author{Riccardo Gallotti}
\affiliation{Fondazione Bruno Kessler (FBK), Trento, Italy}

\author{Filippo Privitera}
\affiliation{Cuebiq Inc., New York, NY, USA}

\author{Bruno Lepri}
\affiliation{Fondazione Bruno Kessler (FBK), Trento, Italy}

\author{Marco De Nadai}
\affiliation{Fondazione Bruno Kessler (FBK), Trento, Italy}


\maketitle

\textbf{The non-pharmaceutical interventions (NPIs), aimed at reducing the diffusion of the COVID-19 pandemic, has dramatically influenced our behaviour in everyday life. 
In this work, we study how individuals adapted their daily movements and person-to-person contact patterns over time in response to the NPIs.
We leverage longitudinal GPS mobility data of hundreds of thousands of anonymous individuals in four US states and empirically show the dramatic disruption in people's life.
We find that local interventions did not just impact the number of visits to different venues but also how people experience them. 
Individuals spend less time in venues, preferring simpler and more predictable routines and reducing person-to-person contact activities. 
Moreover, we show that the stringency of interventions alone does explain the number and duration of visits to venues: individual patterns of visits seem to be influenced by the local severity of the pandemic and a risk adaptation factor, which increases the people's mobility regardless of the stringency of interventions.}

\section*{Introduction}

The COVID-19 pandemic has prompted many countries to implement numerous Non-Pharmaceutical Interventions (NPIs) such as international travel restrictions, physical distancing mandates, closures of business venues, and stay-at-home orders to prevent the spread of the virus~\cite{brauner2020inferring,chinazzi2020effect,dehning2020,flaxman2020nature,haug2020ranking,haushofer2020interventions,lai2020effect}. 
These policies have had a profound impact on numerous aspects of human life including employment~\cite{baek2020unemployment,forsythe2020labor}, economy~\cite{dunn2020measuring, chetty2020economic, bonaccorsi2020economic,brodeur2020literature} and people's social behaviour~\cite{feehan2021quantifying,zhang2021impact,Wellenius2021}. 

Previous literature has exploited mobile phone data to simulate the evolution of the epidemic~\cite{aleta2020modelling,chang2020mobility,grantz2020use,huang2021integrated, oliver2020mobile, xiong2020pnas} and the effectiveness of physical distancing interventions, which reduced people's mobility between and within cities~\cite{googlereports,fraiberger2020uncovering,schlosser2020covid,Wellenius2021,woskie2021}, physical activity~\cite{hunter2020effect, tison2020worldwide} and person-to-person contact patterns~\cite{klein2020reshaping, salje2020estimating,pepe2020covid,benzell2020rationing}. 
Thus, there is no doubt that NPIs proved effective in reducing people's mobility and increasing physical distancing. 
However, much less is known about an individual's behaviour.
Most studies have limited their focus on how much people were moving and how many visits each Point of Interest (POI) experienced (e.g.~\cite{googlereports, klein2020reshaping, cacciapaglia2020mining, hunter2020effect}), with little evidence on how individual habitual leisure activities and social interactions changed, and adapted, over time during the pandemic. 

This paper studies the changes in the daily routine activity of people, focusing on the number and the chronological sequences of visits to Points Of Interest (POIs), the time spent in them and the places where person-to-person contacts happen.
We combine geographical open data from OpenStreetMap (OSM) with privacy-enhanced longitudinal GPS mobility traces of more than 837,000 anonymous opted-in individuals, measured for nine months from 3 January 2020 to 1 September 2020.
Our dataset has an average accuracy of 22 meters and covers 16 hours of activity per day, allowing us to describe human mobility at a fine spatial and temporal granularity while ensuring users' privacy (see SI \Cref{SI:sec:datasets} for additional details).
We analyse and compare the individual's mobility in four US states with the highest and lowest values of daily COVID-19 death rate and NPIs stringency, specifically on Arizona (many deaths and low stringency), Oklahoma (few deaths and low stringency), Kentucky (few deaths and high stringency), and New York (many deaths and high stringency) (see SI \Cref{SI:sec:states} for details).

Our results describe the disruption effect of the COVID-19 pandemic on human behaviour over time at an unprecedented level of detail. 
Overall, we find that individuals profoundly changed their daily routine activities preferring shorter and more predictable movements between places.
People's reduced the number of visits to POIs but also changed their duration, which struggles to recover even during the re-opening phases.
Since visits and social interactions at POIs are intertwined, we model the expected number of person-to-person contact activity through co-location events between two individuals. We find that individuals reduce the number and the duration of co-location events more than expected. Notably, this reduction is strongly correlates with the daily number of deaths. 
Finally, we find that the number of visits increases over time even when NPIs policies do not change.
To explain such contradictory behaviour, we model the visits to POIs and find that people might adapt to the COVID-19 risk and increase their time spent outside an individual's residential area over time regardless of the NPIs stringency.

\section*{Results}

\begin{figure}[!ht]
    \centering
    \includegraphics[width=\textwidth]{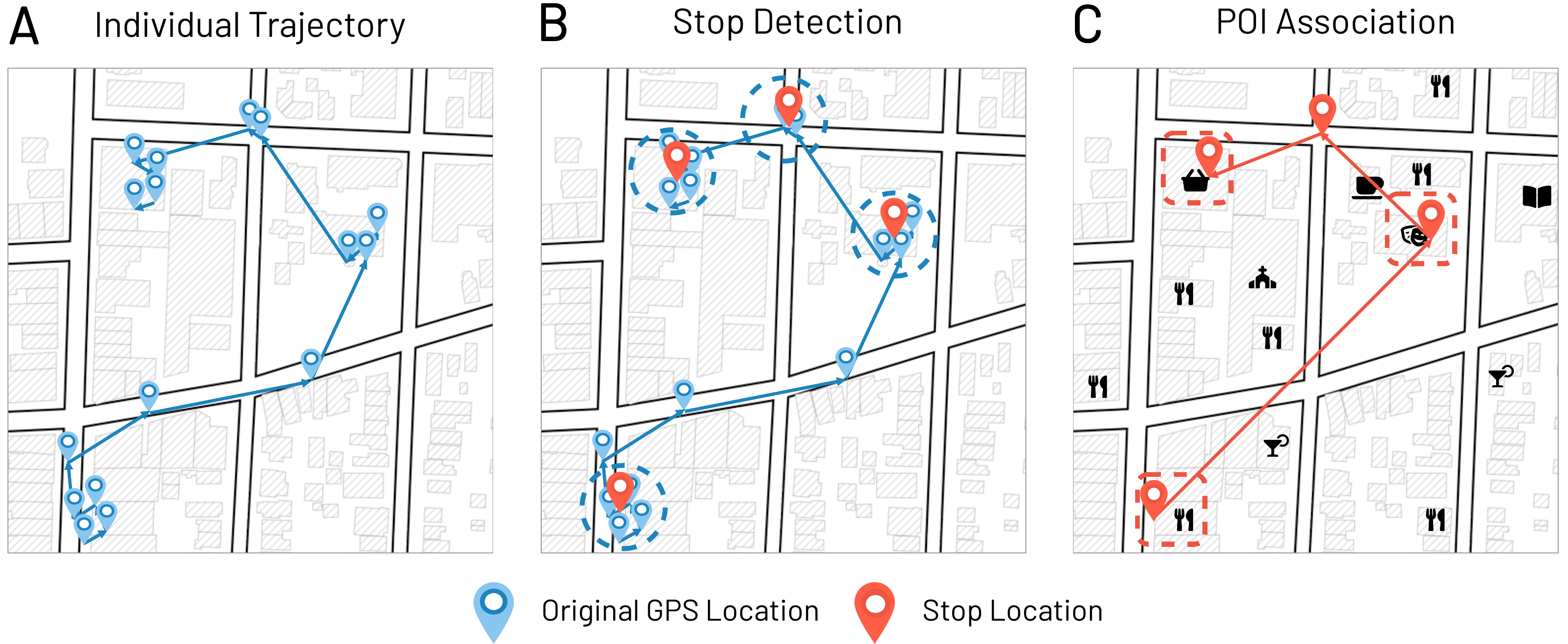}
    \caption{(A) From an individual's original GPS trajectory, we detect a stop whenever the individual spends at least 5 minutes within a distance of 65 meters from a given trajectory point. (B) We detect stop locations through a combination of the Lachesis~\cite{hariharan2004project} and DBSCAN algorithms~\cite{ester1996density}.  
    (C) If present, we associate each stop location to the nearest Point Of Interest (POI) within a distance of 65 meters.}
    \label{fig:schema_data}
\end{figure}

We explore and characterize human mobility from an individual's set of stop locations, defined as places where a person stays for at least 5 minutes within a distance of 65 meters. 
From the original GPS data (see \Cref{fig:schema_data}A), we detect the stop locations of each individual through a combination of the Lachesis~\cite{hariharan2004project} and DBSCAN algorithms~\cite{ester1996density} (see Methods and \Cref{fig:schema_data}B). 
As a result, stop locations are described as tuples \textit{(lat, lon, start-time, end-time)}, where an individual stays in a particular location with latitude (\emph{lat}) and longitude (\emph{lon}) from \textit{start-time} to \textit{end-time}. \emph{lat} and \emph{lon} are the mean latitude and longitude values of the GPS points found within the specified distance of 65 meters (we refer to the Methods and SI \Cref{SI:sec:stop-detection} for the  details).
Then, we focus on the home and work locations of people. To preserve privacy, the data provider obfuscates users' precise home and work locations by transforming it to the centroid of the corresponding Census Block Group. 
Thus, we identify the home census block group of users, from now on called \emph{Residential} area, by looking at the most visited locations during the nights (from 8 pm to 4 am) with a moving time window of 28 days. Similarly, we also identify the \emph{Workplace}, defined as the most visited census block group during the week (from 9 am to 5 pm), which is not marked \emph{Residential}. We refer to SI \Cref{SI:sec:home} for additional details.

Finally, we add semantic meaning to individuals' mobility trajectories associating each stop location to the nearest Point of Interest (POI), extracted from OpenStreetMap (OSM) \cite{OpenStreetMap}, whenever a POI lies within 65 meters from the stop location  (see \Cref{fig:schema_data}C). 
POIs are commonly described as public locations that people may find interesting, for example, for business or recreational activities~\cite{poidef}. 
Since the OSM POIs taxonomy is not hierarchically organized, we create a human-curated mapping from the OSM tagging system\cite{osm_tagging} to the Foursquare venue category hierarchy~\cite{4sq_hierarchy} (see Methods and SI \Cref{SI:sec:stop-detection} for details).
SI \Cref{SI:sec:popularity} shows the popularity of POIs in different states, highlighting the variety of the visiting behaviour in the US.

To validate the data provided by Cuebiq and our pre-processing, we compute the correlation of the time-series of visits to POIs, residential areas and workplace areas between our data,  Google data~\cite{googlereports}, and Foursquare data~\cite{visitdata4sq}. We report an average Pearson correlation in New York state of 0.91 and an average Pearson correlation of 0.84 in all four selected states (see SI \Cref{SI:sec:correlationGoogle}). 

Starting from the visits to POIs of each individual in the selected US states and periods, we address the following questions: How did the pandemic change how individuals visit POIs? How did their mobility routines change? How did the physical social contacts adjust with the pandemic? What are the main factors influencing the visits to POIs throughout the pandemic? 

\subsection*{Changes in POIs visit patterns}\label{poi_visits_and_time}

\begin{figure}[!ht]
    \centering
    \includegraphics[width=\textwidth]{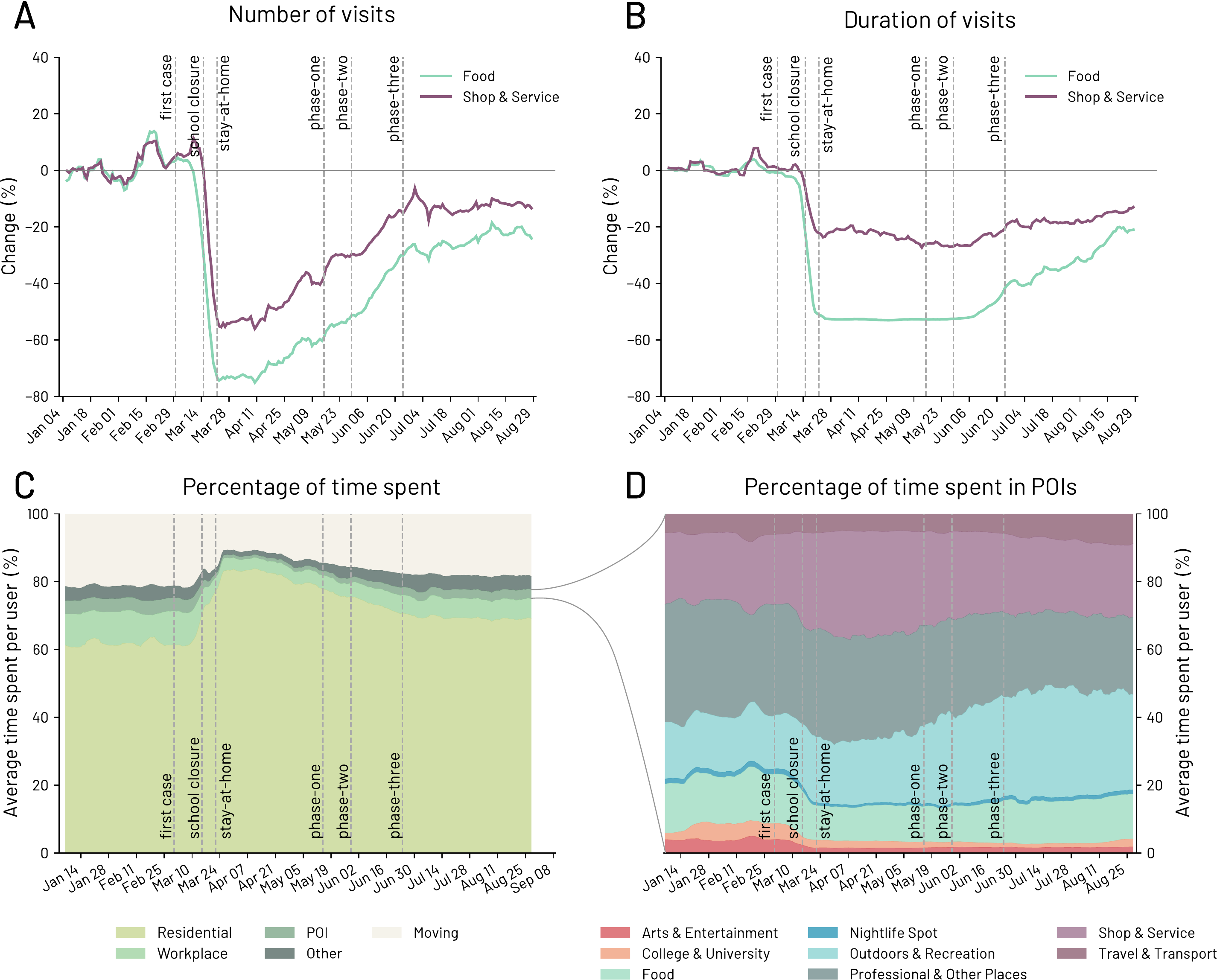}
    \caption{Changes in number and duration of visits to POIs in the state of New York. A) Percent change over time in the number of visits with respect to the baseline period (3 January 2020 - 28 February 2020) for venues in the \emph{Food} and \emph{Shop \& Service} categories. 
    B) Change in percentage of the duration of visits over time with respect to the baseline period of the median duration of visits to \emph{Food} and \emph{Shop \& Service} POIs. 
    For visualization purposes, the original curves are smoothed using a rolling average of seven days.
    Vertical dashed lines indicate the date of restrictions and orders imposed by the state of New York government.
    C) Percentage of time spent by people at \emph{Residential}, \emph{Work}, \emph{POIs}, \emph{Other}, and \emph{Moving} (i.e. people in movement). 
    D) Percentage of time spent by people in venues under the eight first-level categories of POIs.}
    \label{fig:decreases_ny}
\end{figure}

We begin our analysis by describing the impact of the COVID-19 pandemic on the visits to POIs over time.
Overall, we observe that NPIs are associated with fewer visits to POIs, confirming previous results obtained with similar datasets~\cite{chang2020mobility, hunter2020effect, googlereports}.
For example, in the state of New York, POIs belonging to the \emph{Food}  (e.g., restaurants) and \emph{Shop \& Service} (e.g., book stores and supermarkets) categories experienced low points of $-75\%$ and $-56\%$ in the number of visits, respectively. 
These reductions in the number of visits are very heterogeneous. For example, essential shops such as supermarkets faced a lower reduction in the number of visits than the non-essential shops (with low points of $-38.3\%$ vs $-67\%$ shortly after the stay-at-home order) and even increased the number of visits before the stay-at-home order (see SI \Cref{SI:sec:essential_nonessential}).
However, the physical distancing policies not only changed how much people visit POIs but also how people experienced them. \Cref{fig:decreases_ny}B shows that the duration of visits to POIs drastically dropped and reached low points of $-53\%$ and $-27.3\%$ for \emph{Food} and \emph{Shop \& Service}, respectively.

Since two weeks after the stay-at-home order, we observe an increase in the number of visits, which is correlated with the expansion of the list of the essential businesses categories, and the progressive relaxation of government restrictions (see \Cref{fig:decreases_ny}A-B) throughout the various re-opening phases, conveniently named phase-one, phase-two, and phase-three (see the Methods Section).
Interestingly, while the number of visits gradually recover in the period between 25 March 2020 and 20 May 2020 (see \Cref{fig:decreases_ny}A), the duration of visits remains stable and recover much slowly.
At the end of the period of study, the duration of visits reaches $-21\%$ (\emph{Food}) and $-13\%$ (\emph{Shop \& Service}) with respect to the pre-pandemic values (see Figure \ref{fig:decreases_ny}A-B). We find similar results for other relevant POIs (see SI \Cref{SI:sec:pois_visits_NY}) and for the other states (see SI \Cref{SI:sec:fourstates}).
We note that the number of visits to \emph{Food} and \emph{Shop \& Service} starts increasing between the stay-at-home order and phase-one, despite the absence of significant changes in the restrictions.

Figure \ref{fig:decreases_ny}C shows the distribution of time spent by individuals over time in the state of New York. We group the time spent into five categories: \textit{Residential}, \textit{Workplace}, \textit{POI}, \textit{Other} (i.e. stop locations not matched with a POI and not detected neither as \textit{Residential} or \textit{Workplace}), and \textit{Moving} (i.e. time spent moving from place to place).
After the stay-at-home order, we find that the percentage of time spent in residential areas increases significantly to the expense of the time spent at \textit{Workplace}, \textit{POI}, \textit{Other}, and \textit{Moving}.
After phase three, we find that the time spent in residential areas gradually reduces while other categories increase. The time spent in each category, however, does not return to the pre-pandemic period. 

Figure \ref{fig:decreases_ny}D shows the percentage of time spent by people in the venues of the eight first-level categories of POIs.
The time spent at \emph{Shop \& Service} increases at the expense of the other POIs. 
Then, after phase three, people significantly increase the time spent in some categories such as \emph{Outdoors \& Recreation} and \emph{Travel \& Transport}, probably also due to seasonal effects, but other categories such as \emph{Arts \& Entertainment} do not recover to pre-pandemic levels.
We refer the readers to SI \Cref{SI:sec:newyork} for a concrete example of the reduction of visits in New York City.

We also observe significant differences between states. 
For example, New York experienced the most significant drop in the number of visits (e.g. -75\% in \emph{Shop \& Services}), while all other states experienced different reductions (e.g. around -40\% in \emph{Shop \& Services} for the other states), see SI \Cref{SI:sec:fourstates}) for details.
Noteworthy differences are present also in the recovery phases due to the different re-opening measures.
As an example, the state of Arizona experienced a smaller drop in the number of visits during the safe-at-home phase ($-60.7\%$ for \textit{Food} and $-37.6\%$ for \textit{Shop \& Service}), but the number of visits recovers better and much faster than the state of New York (due to early re-opening phases and the absence of restrictions after the end of the stay-at-home order on 15 May 2020). However, following a record high in the number of hospitalizations, on 29 June 2020, the government issued a partial reversal of the stay-at-home that halted the recovery and resulted in $-22.5\%$ for \textit{Food} and $-11.7\%$ for \textit{Shop \& Service} at the end of the period of study.
Interestingly, we observe that individuals allocate the everyday time spent similarly across states, regardless of the NPIs implemented, but again we found differences due to different re-opening strategies. 
We refer to SI \Cref{SI:sec:fourstates} for the results in the four states, and to SI \Cref{sec:behaviour_and_change_metrics} for a set of additional metrics (i.e. individuals' unique stop locations, diversity of visits, and radius of gyration) supporting our findings.

\subsection*{Individual mobility routines}

By looking at the aggregated mobility, we have only a partial view of the disruption in people's life due to COVID-19.
Thus, we here focus on individual's sequences of visits.
We transform the individual's chronological sequence of visits to places into a sequence of symbols (e.g., \textit{Food}, \textit{Residential}, \textit{Workplace}). Then, we apply the Sequitur algorithm~\cite{nevill1997identifying, di2018sequences} to generate a hierarchical representation of the original sequence, compressing repeated occurrences into new symbols called \textit{words}.  Here, each \textit{word} represents a routine, namely a chronological sequence of two or more places that frequently appears into the patterns of visits of an individual (see \Cref{SI:sec:sequitur_supplementary} for an example).

Due to computational constraints, we focus our attention on two 4-week periods, before (from 1 February 2020 to 28 February 2020) and during pandemic (from 21 March 2020 to 17 April 2020), and compute the significant routines. First, we generate 1000 randomized POI sequences with length equals to the original sequence for each individual. Then, we define an individual's sequence as significant if it has a z-score $<2$ between the occurrence of the real routines with respect to the randomized ones. Finally, we filter out all the non-significant routines. We refer to the Methods for additional details.

\begin{figure}[!ht]
	\centering
	\includegraphics[width=0.93\textwidth]{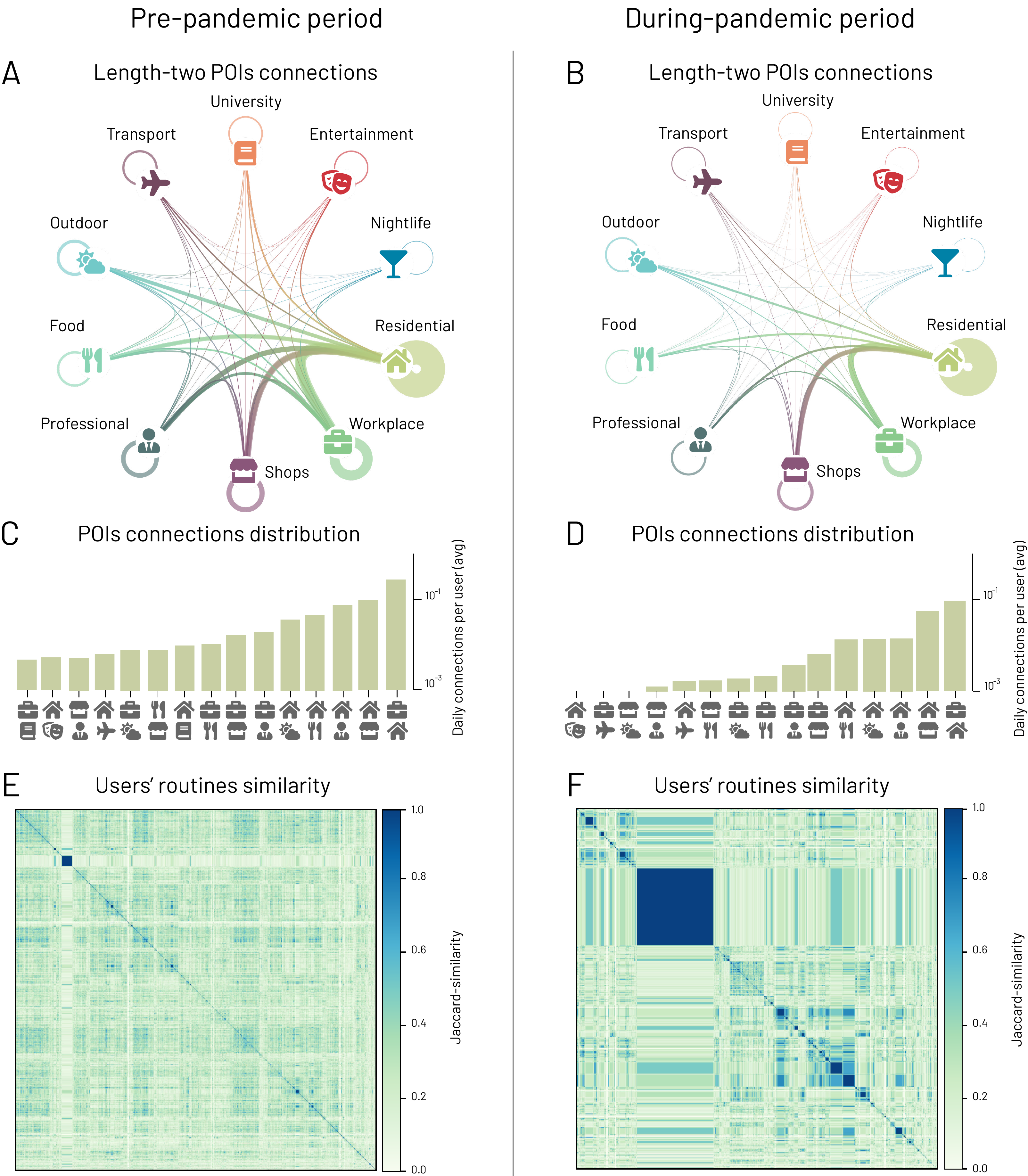}
	\caption{Changes in the routine behaviour between the before and during the pandemic periods. (A-B) Network of the subsequent movements between POI categories for all users in the state of New York before and during the pandemic. The thickness of the links is proportional to the number of movements between the two POI categories. Pre-pandemic (C) and during-pandemic (D) distributions of these intensities excluding self-loops (e.g., \textit{Residential $\leftrightarrow$ Residential}, \textit{Food $\leftrightarrow$ Food} connections). Jaccard similarity matrix between individuals' routines before the pandemic (E) and during the pandemic (F). 
	}
	\label{fig:routines_clustering}
\end{figure}

The selected routines represent meaningful sub-sequences of an individual's mobility and allow to better understand, at a micro-level, how human mobility preferences changed with COVID-19.
To that end, we model the ordered sequences of visits to POIs as a weighted undirected network in which the nodes represent the POI categories, and a link between category $c_1$ and category $c_2$ exists if there is at least a sequence where $c_1$ immediately follows $c_2$  or vice-versa. The weight of the link represents the daily average proportion of sequences containing $[c_1, c_2]$ or $[c_2, c_1]$.
\Cref{fig:routines_clustering}A-B shows the weighted network of the POI categories in the state of New York, where the size of the links is proportional to the intensity of the relationship between the two POI categories. We observe that all links reduce their intensity (on average -79\%) with the exception of the \textit{Residential $\leftrightarrow$ Residential}, which increased by 5\%. 
\Cref{fig:routines_clustering}C-D shows the distributions of the network weights in the pre-pandemic (from 1 February 2020 to 28 February 2020) and during pandemic (from 21 March 2020 to 17 April 2020) periods in New York state, from which we excluded all self-loops (e.g., \textit{Residential $\leftrightarrow$ Residential}, \textit{Food $\leftrightarrow$ Food} connections).

We observe the mildest reduction for \textit{Shop \& Service $\leftrightarrow$ Residential} (-40\% visits), \textit{Shop \& Service $\leftrightarrow$ Workplace} (-61\%), and \textit{Outdoors \& Recreation $\leftrightarrow$ Residential} (-63\%), which might be interpreted as essential venues.
Higher levels of reduction are found in those POIs connections which include sectors strongly afflicted by NPIs: \textit{Arts \& Entertainment $\leftrightarrow$ College \& University} (-95\%), \textit{College \& University $\leftrightarrow$ Nightlife Spot} (-95\%), \textit{College \& University $\leftrightarrow$ Residential} (-93\%), and \textit{Arts \& Entertainment $\leftrightarrow$ Food} (-92\%) (for further details refer to \Cref{fig:SI_routine_reduction_analysis}).

The dramatic change of people's behaviour also emerges from the analysis of the similarity between the characteristic routine of different individuals. 
We represent each individual's significant routine behaviour regarding the presence or absence of two-elements sequences between POI categories.
Then, we compute the Jaccard similarity between a randomly selected sample of 10,000 individuals (due to computational constraints). 
Finally, we apply agglomerative hierarchical clustering \cite{tan2005introduction} to find relevant groups of individuals with similar routine behaviour.
By comparing \Cref{fig:routines_clustering}E and \Cref{fig:routines_clustering}F, we observe that \Cref{fig:routines_clustering}F contains larger clusters, which means that mobility routines simplify and people's behaviour gets more homogeneous. This result is also quantitatively confirmed by the larger silhouette score~\cite{rousseeuw1987silhouettes} of all individuals during-pandemic period compared with the pre-pandemic period (see SI \Cref{SI:clustering_distance_silhouette} for details).
By inspecting the everyday routines in the two biggest clusters, which include almost 34\% of users, we observe that individuals limit their mobility to \textit{Residential $\leftrightarrow$ Residential}, \textit{Residential $\leftrightarrow$Shops \& Services} and 
\textit{Shops \& Services $\leftrightarrow$ Shops \& Services} routines.
We refer to SI \Cref{SI:sec:topclusters} for a comparison between the six biggest clusters.

During the pandemic period, we also find that individuals tend to favour simpler and more redundant sequences.
We measure the compression ratio~\cite{gallotti2013entropic} defined as the length of the original sequence divided by the length of the Sequitur compressed sequence. 
On average, in the state of New York, the individual's sequences before the pandemic have a compression ratio of $2.75$, while during the COVID-19 pandemic, the compression ratio increases by 40\%, reaching $3.79$ (see SI~\Cref{SI:fig:sequitur_ratios_NY}). 
Similar results apply in all the other states (see SI~\Cref{SI:sequitur_compression}).

\subsection*{Co-location events}
 
\begin{figure}[!ht]
	\centering
	\includegraphics[width=\textwidth]{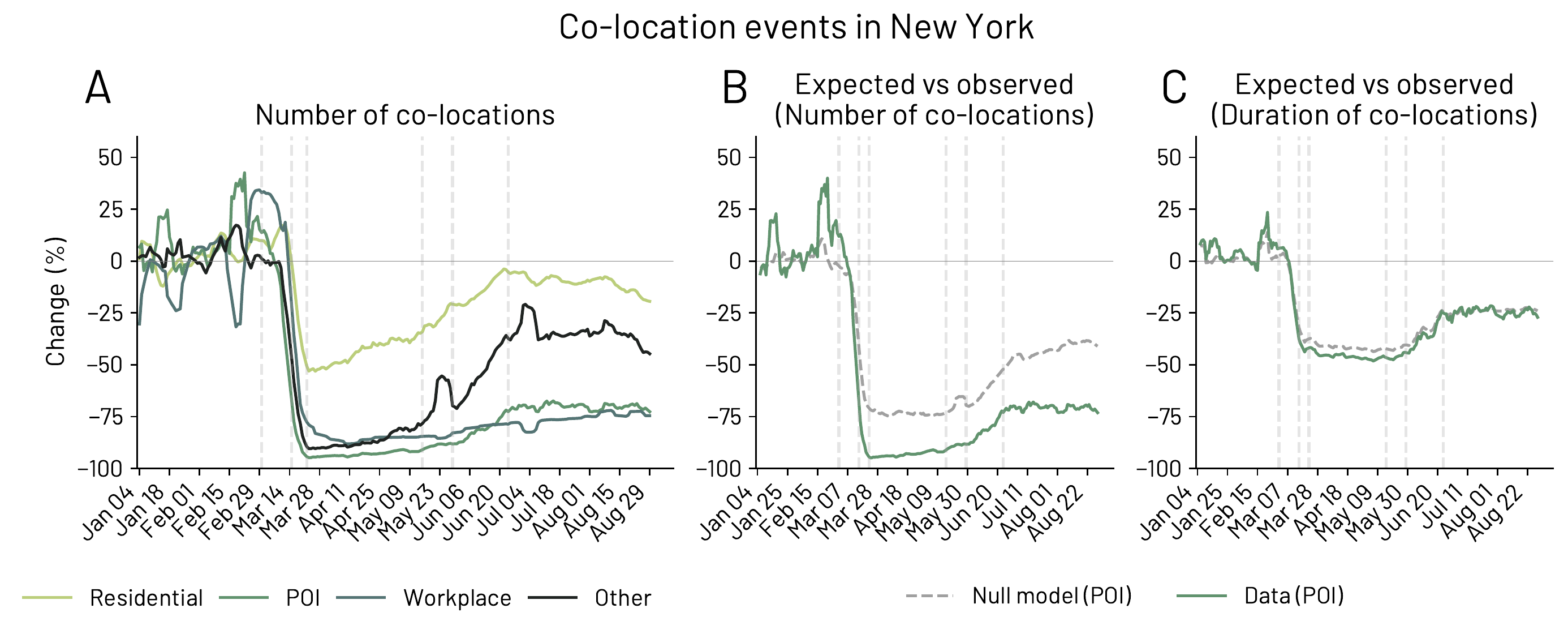}
	\caption{A) Percentage change of co-location events from the baseline (until 29 February 2020) in the New York state. We measure the change for Residential areas (one of the two individuals in proximity is at their residential area), POI (both individuals are at the same POI), Workplace (both of having the same work location) and Other co-location events. B-C) We compare the difference between the expected and observed number and duration of co-location events. During the pandemic, individuals tend to have fewer and shorter co-locations than expected. }
	\label{fig:social_contacts}
\end{figure}

As a proxy to understand how much people engage in physical and social activities, we define a co-location event as when two individuals stop for at least fifteen minutes and are at most 50 meters apart from each other. 
These co-location events are aggregated 
in four different categories depending on the place where the possible social contact took place: (i) \textit{Residential}, a co-location event where only one of the two individuals have the stop marked as \emph{Residential} location; (ii) \textit{Workplace}, a co-location event that happened in a venue labeled as a workplace for both the individuals; (iii) \textit{POI}, a co-location event where both the individuals are in the same POI; and (iv) \textit{Other}, a co-location event in which the two individuals meet in a place that it is neither a \textit{Residential} nor a \textit{Workplace} nor a \textit{POI} (see SI~\Cref{SI:sec:co-locations} for additional details). We here note that \emph{Residential} and \emph{Workplace} co-locations are defined at a coarse level, due to the anonymization process done by Cuebiq.

\Cref{fig:social_contacts}A shows the abrupt change of the co-location events in the New York state, starting at the school closure day on 15 March and reaching low points of $-92\%$, $-67\%$ and $-83\%$ for \textit{POI}, \textit{Workplace} and \textit{Other} co-location events respectively. 
Interestingly, we observe that \textit{Residential} co-location events, namely between people who do not live together, experience the smallest reduction, decreasing at most by $32\%$ from the pre-pandemic levels. 
During the strictest measures put in place in New York state, we notice that people maintain their co-location events inside other's people residential areas and in places which are not marked as POIs (see \textit{Residential} and \textit{Other} in \Cref{fig:social_contacts}A). 
This presumably happens because of the impossibility of having co-location events in venues such as pubs and restaurants (see \textit{Food} and \textit{Nightlife Spot} categories in SI \Cref{SI:sec:co-location-number}) due to the NPI interventions including physical distancing measures and closures of POIs such as \textit{Arts \& Entertainment} and \textit{College \& University}.

Again, we find some differences between the number and the duration of co-location events. First, the duration decrease less than the number of co-location events, reaching low-points of $-7\%$, $-46\%$, $-5\%$, $-3\%$ for \emph{Residential}, \emph{POI}, \emph{Workplace} and \emph{Other} co-location events respectively (see SI \Cref{SI:sec:co-location-duration}). 
We obtain similar results in the other states (see SI \Cref{SI:sec:co-locations}), although with some differences. 
For example, in Oklahoma, Kentucky and Arizona, the \emph{Residential} co-locations events experience a lower reduction, with a low point around $-20\%$.
Notably, we observe a decrease in co-locations events from the partial reversal of the reopening from 29 June 2020 in Arizona.

We correlate the daily number of co-location events with the NPIs stringency in each state and find a strong negative Spearman correlation ($-0.83$ with p-value $p < 0.001$). 
Similarly, we find that the daily number of co-location events is also negatively correlated with the daily number of new cases and deceased ($-0.58$ and $-0.67$ respectively, with p-values $p < 0.001$). 

The number of co-location events and the visits to places are inherently connected. Therefore, as soon as the number and the duration of visits decrease, it becomes less probable to have co-location events. 
Thus, we estimate the daily expected number of co-location events through a null model and compare it with the observed co-locations.
The number of co-location events $e_{i,d}$ occurring at a POI $i$ on a specific day $d$ can be estimated from both the number of individuals visiting the POI $n_{i,d}$, and the median duration of their stops there $\bar{d}_{i,d}$.
We can then have an estimate of the co-location events following $e_{i,d} = \binom{n_{i,d}}{2} p_{i,d}$, where $p_{i,d}$ is the probability of having a co-location event given two individuals visiting POI $i$ on day $d$ and $p_{i,d}$ is computed assuming a Uniform distribution for the time-interval of visit of two individuals potentially having a co-location event (see SI \Cref{SI:sec:co-location-null} for additional details). We follow a similar reasoning to model the expected duration of the individual's co-location events.

\Cref{fig:social_contacts}B shows that, during the pandemic, the observed number of co-location events at POIs are lower than expected. Similarly, \Cref{fig:social_contacts}C shows that the duration of co-location events is slightly lower than expected. 
Thus, we compare the deviation from the expected number of co-location events and find an higher Spearman correlation with the daily new cases and deaths (0.66 $p < 0.001$ and 0.48 $p < 0.001$, respectively) than with the NPIs stringency (0.28 $p < 0.001$).
While the intensity of the local restrictions is the main driver for reducing the number of co-location events, the discrepancy between the theoretical and observed number of co-location events seems largely driven by the epidemic burden (i.e. daily new cases and deaths).

\subsection*{Fitting the curve: the factors influencing the visits to POIs}

We have previously shown that the recovery in the number of visits to POIs seems to have just a loose connection with the NPIs stringency.
To explain this unexpected behaviour, we now shift our attention to the different factors that might influence the number of visits to POIs. 
Using a multivariate Bayesian linear mixed model, we investigate the combined effect of the NPIs, the daily death ratio, the weather (i.e., daily max temperature and precipitation) on the daily number of visits to POIs in each state.
Our model also accounts for the different mobility behaviour of people across states and day of the week, by including a random effect for the state and a random effect for each day of the week.

We select, as a baseline, a model that includes as fixed effects only the NPIs stringency and the death ratio over the state population. 
We evaluate the model through the well-established Bayesian $R^2$~\cite{gelman2019r} and the PSIS-LOO information criterion~\cite{vehtari2017practical}. 
\Cref{tab:ablation} shows that this simple model achieves $R^2 = 0.67$ and PSIS-LOO $= 610.60$ and, as expected, shows that the NPIs stringency correlates negatively with the mobility of people. Interestingly, we also find that the death ratio influences the number of visits to POIs.

Then, we account for seasonal effects that might influence the visits to POIs. The Weather model adds the daily precipitations and maximum temperature to the baseline model. \Cref{tab:ablation} shows that these two variables significantly increase the model's performance ($R^2 = 0.72$, PSIS-LOO $= 747.44$) that grow by 7.46\% and 22.41\%, respectively. 

As mentioned in previous sections, despite the absence of significant changes in state restrictions in some phases of the pandemic, we find an increase in the number of visits to POIs. To explain this behaviour, we hypothesise the presence of progressive behavioural relaxation and adaptation to the epidemic risk, also observed by previous literature~\cite{klein2020reshaping, petherick2021worldwide}.

We model the effects of this risk adaptation as a function of time with a sigmoid function, fitted by our model. We refer to the Methods and SI \Cref{SI:sec:model_selection_all} for additional details.
\Cref{tab:ablation} shows that the Full model provides the highest performance ($R^2 = 0.78$, PSIS-LOO $= 846.76$) and that the second-most important factor in understanding the daily visits to POIs is the risk adaptation factor. 
This result also holds even when we predict the time spent outside the home and when we hypothesise that the risk adaptation depends on the cumulative NPIs stringency, which varies from state to state (see SI \cref{SI:sec:model_selection_all}).

\begin{table}[!h]
    \centering
    \setlength{\tabcolsep}{6pt}
    \begin{tabular}{@{}l rrr@{}}
	\toprule
		 & \textbf{Baseline} & \textbf{Weather} & \textbf{Full}\\
		\midrule
		NPIs stringency & $-.148 \pm .004$ & $-.204 \pm .004$ & $-.242 \pm .009$\\
		Death ratio & $-.072 \pm .004$ & $-.053 \pm .003$ & $-.040 \pm .007$\\
		Max temperature & - & $.009 \pm .004$ & $-.001 \pm .099$\\
		Precipitations & - & $-.002 \pm .003$ & $.001 \pm .100$\\
		Risk adaptation & - & - & $.118 \pm .004$\\
		\hline 
		$R^2$ & $.67 \pm .02$ & $.72 \pm .02$ & $\mathbf{.78 \pm .02}$\\
		PSIS-LOO & $610.60 \pm 31.94$ & $747.44 \pm 28.78$ & $\mathbf{846.76 \pm 29.59}$\\
    \bottomrule      
    \end{tabular}
    \caption{Quantitative results of the Bayesian multivariate linear mixed model to explain the daily number of visits to POIs. We report the mean and 95\% confidence intervals of all the $\beta$ coefficients. We report the mean and standard deviation for $R^2$ and PSIS-LOO.
    }
    \label{tab:ablation}
\end{table}

\section*{Discussion}

This paper digs into the COVID-19 induced changes in human behaviour at an unprecedented scale and detail. 
We exploit a privacy-enhanced longitudinal GPS dataset of more than 837,000 anonymous opted-in individuals to show how individuals changed the patterns of visits to places and the person-to-person contact activity over time.

We show that, as previously found~\cite{googlereports,perra2021non}, the COVID-19 pandemic dramatically reduced the number of visits to POIs, which only partially recovered the pre-pandemic levels reaching, at the end of the period under study, a $-28\%$ in the state of New York. 
However, our analysis reveals that while the time spent in POIs decreases less than the number of visits, it only shows a modest recovery after the reopening phases ($-23\%$ at the end of the period).
This result suggests people's behavioural change goes beyond the number of visits as people seem to be less willing to spend time in POIs, perhaps to minimise proximity contacts in public venues.
This result is confirmed by our findings in the co-location events, which are fewer and shorter than what expected from the null-model.

We also find that changes in co-location events, used as a proxy of social contacts, are often place-dependent, and they are not well aligned with the risk level identified by the literature. 
For example, on average, co-locations at other's people residential areas, namely between people who do not live together, reduce by 46\% while co-locations in POIs and at work are reduced on average by 73\% and 81\% respectively after the stay-at-home order, despite being considered less risky than contacts at home~\cite{fisher2020community, galmiche2021exposures}. 
Seminal literature such as Chang \emph{et al.} ~\cite{chang2020mobility} have focused on simulating the epidemic diffusion of COVID-19 in POIs.
However, considering the epidemiological risk in private gatherings~\cite{galmiche2021exposures, fisher2020community}, our results highlight the importance of considering \emph{all} co-location places when modeling or studying optimal intervention strategies.

Overall, our data show that human routines during the COVID-19 pandemic get shorter, more predictable and change in structure. 
These changes may have widespread consequences on previous empirical research describing the regularities of mobility~\cite{ gonzalez2008understanding, song2010limits,  pappalardo2015returners, alessandretti2018evidence, alessandretti2020scales, schlapfer2021human}, social interactions~\cite{deville2016scaling, pan2013urban} and human predictability~\cite{cuttone2018understanding, song2010limits}, which are all assumed to be almost universal. Thus, an open question is wherever these mobility regularities are ``resilient'' to altered mobility, such as during a pandemic.

Finally, we find people adapt to the pandemic risk over time, revealing a two-fold behaviour in the visits to POIs.
On the one hand, individuals
visits and time spent outside the home are influenced by the NPIs stringency as well as the number of deaths in the state. On the other hand, people increase the time spent outside the residential area, the number and duration of visits to POIs despite no significant changes in the NPIs even when we account for the deaths and the weather.
Multiple reasons may explain this risk adaptation, so we can only speculate about its causes.
One hypothesis is that the risk adaptation in the patterns of visits results from a change in the risk assessment. 
As the pandemic lasts for months, people might get more used to the number of deaths, reduce their self-protection and act less prudently. However, there is no evidence of a reduction of mask-use over time in the US~\cite{protectionmeasures}, and our results show that the number of co-locations is lower than what is expected from the null-model. Thus, people protect themselves against person-to-person contacts.
Another hypothesis is motivated by the sustained economic burden, which may get people to decrease policy adherence to get back at work. 
Finally, the risk adaptation might also be a consequence of the psychological burden, which reduces the ability or motivation to perform self-protective behaviour~\cite{harvey2020behavioral, michie2020concept}.
We cannot dismiss any of these hypotheses. However, our evidence well aligns with previous self-reported results~\cite{petherick2021worldwide} and suggests that epidemiological models and public health communication campaigns should consider people's relaxation to governmental orders.

Analysing everyday activities from GPS data does not come without limitations. 
First, the analysis from smartphones data might be biased towards younger adults and fail to capture the mobility of those people who do not carry their phones while visiting places.
Second, our \emph{Residential} and \emph{Workplace} are just an estimate of the real home and work locations, which are obfuscated for privacy reasons by the data provider.
Third, we acknowledge that our co-location events are a loose proxy of social interactions, and people might share the same location even without knowing each other (i.e., familiar strangers~\cite{milgram1977familiar}).

It is well-known that COVID-19 disrupted the lives of millions of people. However, the medium and long term effects on people's everyday life are not well understood.
Our work shows that the pandemic-induced changes are not just about how much people stay at home and visit POIs. Instead, the COVID-19 risk and policy interventions seem to have reshaped people's conventions and habits, changing the way individuals experience POIs, places and social interactions. On the one hand, individual behaviour constantly adapted to follow the NPIs and reduce the risk of person-to-person contacts, while on the other hand, it increased the time outside the home regardless of the NPIs stringency.
Future work is needed to clarify how people adapt to the epidemic risk and what are the long-term effects of the COVID-19 pandemic and restriction measures on human behaviour.

\section*{Methods}
\subsection*{Selection of the important dates}
We identify several important dates that help readers with the interpretation of our results. In the state of New York, on March 1, the first positive cases were broadly discussed by the public opinion. On March 15, the Governor Andrew Cuomo  announced that New York City schools would have closed from the following day~\cite{NYSCHOOLS}. On March 22, Andrew Cuomo announced the statewide stay-at-home order, also known as the NYS on Pause Program, with a mandate that all non-essential workers work from home beginning at 8 p.m. and that only businesses declared as essential were allowed to remain open~\cite{NYPAUSE, NYPAUSE2}.

On May 15, Andrew Cuomo announced a gradual plan of reopening also called Phase I. From May 28, the New York State Department of Health released the guidelines for the reopenings of Phase II, which included the reopening of professional services including finance and insurance, retail, administrative support, and real estate/rental leasing~\cite{NYPHASE2}. We note that New York City met the criteria on June 22.
On June 15, Andrew Cuomo announced that regions upon entry of Phase 3 will be allowed non-essential gatherings of up to 25 people and that on-location restaurants would open~\cite{NYPHASE3}. New York City entered this phase on July 6.

We highlight some of these dates in the plots and we describe the important days for all the states in \Cref{si:sec:important dates}.

\subsection*{Stop location detection}
We use GPS location data provided by Cuebiq, a location intelligence company that shared a dataset consisting of anonymized GPS locations from users that opted-in to share the data anonymously for research purposes through a CCPA (California Consumer Privacy Act) compliant framework.
To further preserve privacy, the data provider obfuscates the precise home and work locations of users by transforming it to the centroid of the corresponding Census Block Group. 

The dataset span a period of 9 months, from January 2020 to September 2020 (details in SI \Cref{SI:sec:stop-detection}).
The data is provided through the Cuebiq Data for Good COVID-19 Collaborative program, which provides access to de-identified and privacy-enhanced mobility data for academic research and humanitarian initiatives only.

To ensure the data well describes the mobility of people throughout the pandemic, we filter out all users with less than one month of data before the declaration of national emergency (March 13, 2020) and less than four months after it. We also require users have 5 hours per day covered by at least one GPS location.
The resulting dataset includes more than 837,000 anonymous, opted-in individuals. 

For all users, we extract their stop events with an algorithm based on Hariharan and Toyama~\cite{hariharan2004project}. A stop event is defined as a temporal sequence of GPS coordinates in a radius of $\Delta_s$ meters where a user stayed for at least $\Delta_t$ minutes. 
The algorithm, its optimization, and its computational complexity are explained in detail in SI  \Cref{SI:sec:stop-detection}. To define a stop event, we used $\Delta_s=65$ meters and $\Delta_t = 5$ minutes due to the distribution of accuracy of the underlying data (see SI \Cref{SI:sec:stop-detection}). 

For each user, we then define their stop locations as the sequences of stop events that can be considered as part of the same place. To determine a stop location from a sequence of stop events we use the DBSCAN algorithm~\cite{ester1996density}. With DBSCAN, we group points within a distance of $\epsilon = \Delta_s - 5$ meters to form a cluster with at least $\text{minPoints}=1$ stop event (see SI \Cref{SI:sec:stop-detection} for more details). 
    
\subsection*{Point Of Interest (POIs) association}
We extract all POIs from OpenStreetMap (OSM)~\footnote{\url{https://www.openstreetmap.org/}} and then, due to the lack of structure in OSM POIs, we map each OSM POI to the corresponding  Foursquare Venue Category Hierarchy~\footnote{\url{https://developer.foursquare.com/docs/build-with-foursquare/categories/}}. After the association, each OSM POI  is mapped to the Foursquare categorization with 8 first-level categories and 178 second-level categories.
Further details are described in the SI \Cref{sec:poi_extraction_and_match}.

Then, for all users, we associate each stop location to its nearest POI whenever the Haversine distance is less or equal than 65 meters. Since in OpenStreetMap POIs can be represented as Points or Polygons, sometimes nested, we assign POIs to stop locations with the heuristic described in SI Section~\ref{sec:poi_extraction_and_match}.

\subsection*{Percent change}
Throughout the analysis we compute the change with the same methodology of the Google mobility reports~\cite{googlereports}.
Specifically, we compute the percent change as: 
$$P_{i} = \frac{v_{i} - b_{w(i)}}{b_{w(i)}} * 100$$ where $v_i$ is the original value at day $i$ and  $b_{w}$ is the median value at day of the week $w(i)$, going from 0 to 6, computed during the baseline period (i.e., before the pandemic).

\subsection*{Sequitur, significant routine patterns, and user-user similarity}\label{sec:sequitur}
Sequitur is a compression algorithm that reduces a sequence size by introducing new symbols/words when in the original sequence appear repetitions of short sub-sequences and motifs~\cite{nevill1997identifying}.
We represent an individual's mobility through a sequence of symbols that maps a stop location into a category (e.g., \textit{Residential}, \textit{Workplace}, \textit{Food}). 
From these sequences of symbols we extracted recurrent patterns of visits consisting of sub-sequences of length $l\ge2$ following Di Clemente \emph{et al.}~\cite{di2018sequences}. 

We focus our attention on two 4-week periods. The first one starts on February 1, 2020 and excludes January, which might display some unusual patterns due to seasonal effects (e.g., the end of the holiday period). The second one starts at the beginning of each state emergency response orders, i.e. the ``stay-at-home'' order (in Kentucky, we selected the ``healthy-at-home'' order since no ``stay-at-home'' was ever issued). This period terminates before the first re-openings to include the most stringent early regulatory phase for each state (for more details on relevant dates see SI~\Cref{tab:important_dates}). 

For each individual sequence of visits, we randomly shuffle the sequence $1000$ times and apply the sequitur algorithm on each one of them \cite{nevill1997identifying}. 
For each symbol $s$ (a visit in our case), we compute the mean number of occurrences, $\mu_s$ and its standard deviation, $\sigma_s$, across all random synthetic sequences. We use this quantity to compute the standard score, $z_s=\frac{o_s-\mu_s}{\sigma_s}$, where $o_s$ is the number of occurrences of each symbol $s$ within the original sequences. Significant routines are selected from all the sub-sequences if $z_s$ is greater than $2$. We refer to significantly recurrent sub-sequences as ``routines''.



\subsection*{Modeling expected number and duration of co-location events}
The expected number of co-location event is assumed to depend on both the number of individuals visiting a POI, $i$ on day $d$ and the average duration of their stop there, $d_{i,d}$. The first quantity is used to compute the combinations of possible co-location between different individuals. The second is used to estimate the probability $p_{i,d}$ of having two temporal interval of duration $\bar{d}_{i,d}$ overlapping by at least 15 minutes over the entire time span of a day. Combined together, our estimate of the number of co-location events, $e_{i,d}$, can be written as: $e_{i,d} = \binom{n_{i,d}}{2} p_{i,d}$.

Following a similar reasoning, the average duration of co-location events is estimated conditioning on the occurrence of the event and assuming that the temporal overlap of two individuals, $o_{i,d}$, depends on their average permanence at POI $i$, $\hat{d}_{i,d}$. Thus, the average overlap area can be straightforwardly computed as: $o_{i,d} = \frac{\hat{d}_{i,d}-15}{2}$. More information about the formulation and computations at support of these models can be found in SI~\Cref{SI:sec:co-location-null}.

\subsection*{Modeling visits to Points Of Interest}
We model the daily average number of visits to POIs $Y$ with a Bayesian linear regression formulated for each day of year $d$, state $s$ and weekday $w$ as:
\begin{equation}\label{eq:full_model}
Y_{d, s, w} \simeq \alpha_s + \beta_{\text{policy}} S_d + \beta_{\text{deaths}} D_{d-1} + \beta_{\text{temp}} T_d + \beta_{\text{prec}} P_d + \frac{\beta_{\text{adapt}}}{1+e^{-\gamma_s (d-\phi_s)}}  + \rho_w + \epsilon,
\end{equation}
where $\epsilon$ is the residual error, $\alpha_s$ is the state-specific intercept, $\beta_{\text{policy}}$, $\beta_{\text{deaths}}$, $\beta_{\text{temp}}$, $\beta_{\text{prec}}$ and $\beta_{\text{adapt}}$ are the $\beta$ coefficients for each independent variable in the regression, while $\rho_w$ is the week-day random effect controlled by the variable $w$, the day of the week of (from 0 to 6 where 5 is Saturday and 6 is Sunday).
$S_d$ is the value of the Stringency Index, measuring local enacted regulations and preventive and informative campaigns (see \cite{hale2021global} for more details), $D_{d-1}$ is the death ratio at the previous day (measured over a population of $100$k people), $T_d$ is the maximum temperature in Celsius degrees and $P_d$ represents the millimeters of precipitations. $T_d$ and $P_d$ account for the seasonal effects of POI visits.
Then, $\frac{1}{1+e^{-\gamma_s (d-\phi_s)}}$ is a standard sigmoid function that models the collective behavioural adaptation of people to the perceived epidemic risk. 
The sigmoid function depends on time where $\phi_s$ and $\gamma_s$ model the location and the sharpness of the sigmoid, respectively.
All the independent variables are z-score standardized.

We extract the daily temperature and precipitation from the PRISM Climate Group~\cite{di2008constructing}, which provides the maximum temperature and precipitations with a 4km grid. We compute the average maximum temperature and precipitations for each state. The average is weighted with the population of each county to account for the number of people that were exposed to the measured temperature and precipitations.

We assess the out of sample predictive accuracy through the Pareto-smoothed importance sampling Leave-One-Out cross-validation (PSIS-LOO)~\cite{vehtari2017practical}. This metric overcome the issues of the Deviance Information Criterion (DIC)~\cite{spiegelhalter2002bayesian} such as its lack of consistency and the fact that is not a proper predictive criterion~\cite{spiegelhalter2014deviance, vehtari2017practical}, and it has rapidly become the state of the art for evaluating Bayesian models.
The PSIS-LOO is defined in the log score as:
    \begin{equation}
        \label{LOO}
    \text{PSIS-LOO} = \sum_{i=1}^n \ln \left(\frac{\sum_{s=1}^S w_i^s p(y_i|\theta^s)}{\sum_{s=1}^S w_i^s}\right).
    \end{equation}
    where $n$ is the number of data points, $\theta^s$ are draws from the full posterior $p(\theta|y)$, $s=1,\dots,S$ represent the $S$ draws, and $w_i^s$ is a vector of weights that are the Pareto Smoothed importance ratios built through an algorithm described in the PSIS-LOO original paper~\cite{vehtari2017practical}. The best model is associated with the highest PSIS-LOO value.
We also report Bayesian $R^2$~\cite{gelman2019r} as an additional and easy-to-interpret measure of goodness of fit.

\subsection*{Data and source code Availability}
Replication code is available on GitHub at \url{https://github.com/denadai2/living-the-pandemic}. All the data sources are freely available on the Internet while the mobility data from Cuebiq can be accessed only through the Data for Good initiative of the company~\footnote{\url{https://www.cuebiq.com/about/data-for-good/}}. Limitations apply to the availability of this data, due to the rigorous anonymity constraints.

\section*{Acknowledgements}
We would like to thank Cuebiq for allowing the use of anonymized data through their COVID19 Collaborative program.
L.L., S.C. and M.D.N. thank Matteo Saloni for the help in scaling the algorithms seamlessly.
L.L., S.C. and M.D.N. also thank Brennan Lake for the insightful comments on the article.

\section*{Author contributions}
S.C.,  M.D.N., L.L., and L.P. conceived the original idea and planned the experiments.
F.P. provided the mobility data.
S.C.,  M.D.N., and L.L. pre-processed the mobility data, carried out the experiments and made the Figures.
S.C.,  M.D.N., L.L., and L.P. contributed to the interpretation of the results and wrote the manuscript. 
All authors provided critical feedback and helped shape the manuscript.

\section*{Competing interests}
All authors declare no competing interests.

\section*{Materials \& Correspondence}
Correspondence and material requests can be addressed to: \url{lorenzo.f.lucchini.work@gmail.com}  and \url{work@marcodena.it}.

\bibliographystyle{plain}
\bibliography{POI_bib}

\clearpage
\section*{Supplementary Notes}
\appendix

\section{New York City before and during the pandemic}
\label{SI:sec:newyork}
\begin{figure}[!ht]
	\centering        \includegraphics[width=\textwidth]{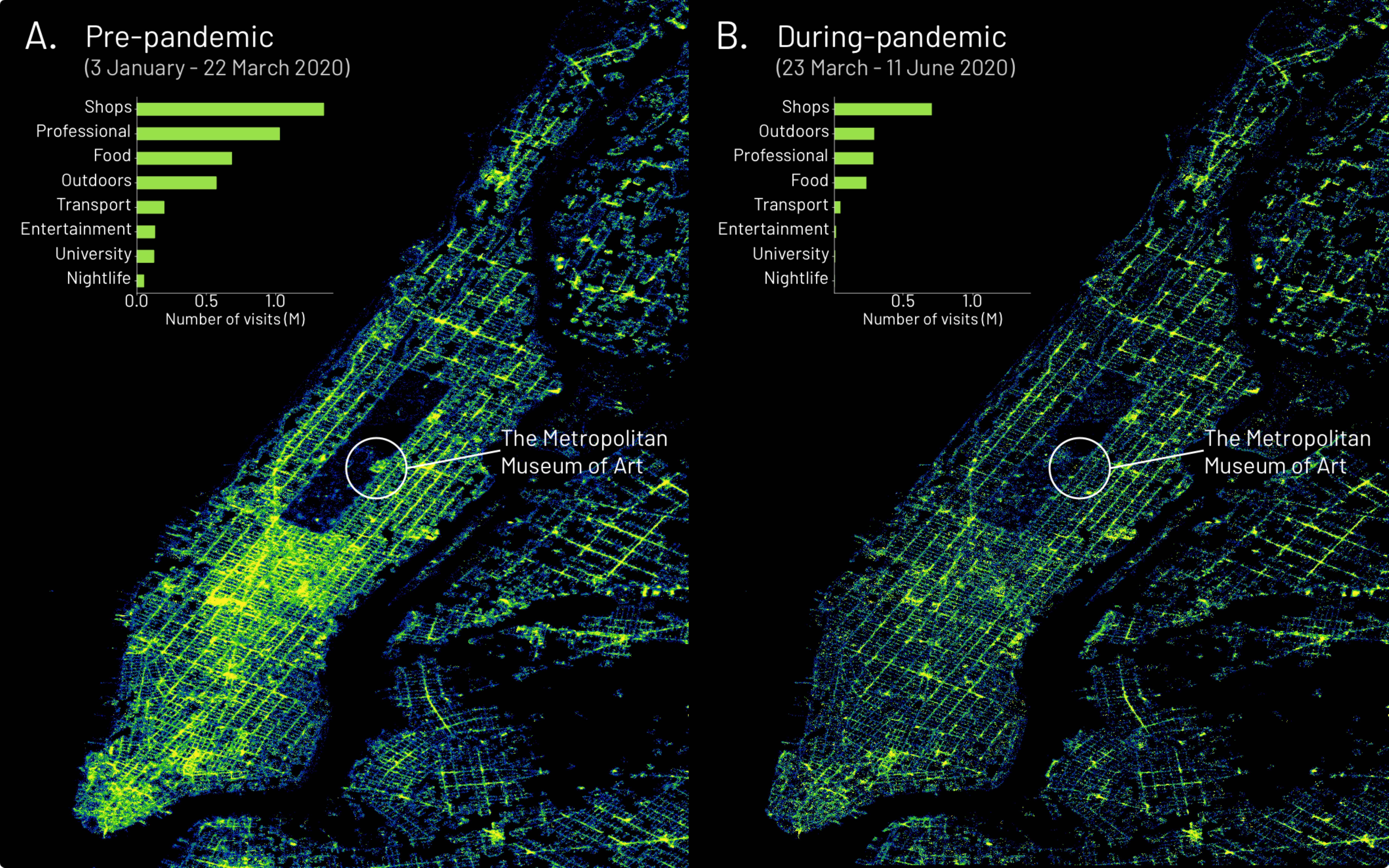}
	\caption{(A) Number of visits to POIs in New York City in the 82 days before the stay-at-home order (22 March 2020)
		and (B) the 82 days after the stay-at-home order. 
		In the state of New York, we find a decrease of 60.6\% of the visits to the POIs, revealing the dramatic change in visiting habits brought about by the pandemic in 2020. In the insets, we show the distribution of the visits to the POI categories in the state of New York.
		As a concrete example, we highlight how activity at the Metropolitan Museum of Art in New York City experiences a drastic reduction of visits during the pandemic.}
	\label{fig:example_pois_newyork}
\end{figure}

\Cref{fig:example_pois_newyork} shows more than 9 million stop locations to POIs in New York City before and after the stay-at-home order. 
On average, we observe a $60.6\%$ reduction of visits to POIs in the state of New York, with reduction of $-87.7\%$, $-83.7\%$, and $-77.7\%$ for \emph{College \& University}, \emph{Nightlife Spot}, and \emph{Arts \& Entertainment}, respectively. 
As a concrete example, \Cref{fig:example_pois_newyork} highlights the number of stop locations at the Metropolitan Museum of Art, one of the most popular \emph{Arts \& Entertainment} venues in New York City; during the pandemic, it almost disappears from the map.

\section{Datasets}
\subsection{Location data}
\label{SI:sec:datasets}
The location data is provided by Cuebiq Inc., a location intelligence and measurement company. The dataset was shared within the Cuebiq Data for Good program, which provides access to de-identified and anonymized mobility data for academic and research purposes. 

The location data used consists of users in the US over nine months, from January 2020 to August 2020, and includes only users who have opted-in to share their data anonymously. The data is General Data Protection Regulation (GDPR) and California Consumer Privacy Act (CCPA) compliant. Furthermore, to increase and preserve users' privacy, Cuebiq obfuscates home and work locations to the census block group level. 

The data is collected through the Cuebiq Software Development Kit (SDK) that collects user locations through GPS and Wi-Fi signals in Android and iOS devices. 

\begin{figure}[!ht]
	\centering
	\subfloat{\includegraphics[width=0.45\textwidth]{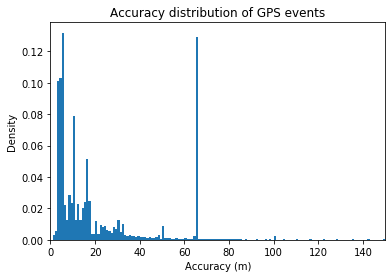}}
	\subfloat{\includegraphics[width=0.45\textwidth]{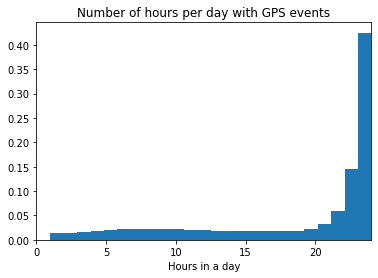}}
	\caption{(left) Accuracy distribution of Cuebiq GPS events. (right) Distribution of the average number of hours (per user) covered by at least one GPS event.}
	\label{fig:accuracy}
\end{figure}

The device determines the location accuracy, which varies from 0 to more than 100 meters. 
\Cref{fig:accuracy} (left) shows the accuracy in meters of the original GPS events in the dataset. We can see that the accuracy distribution is bimodal, with one peak around 5 meters and the other peak at 65 meters. We speculate that the latter peak is caused by the home obfuscation mechanism of Cuebiq to preserve users' privacy. 
\Cref{fig:accuracy} (right)  shows the average number of hours per day with at least one GPS location per user. We can see that most users have almost all the hours covered, enabling us to describe human behaviour accurately.

\subsection{COVID-19 government response measures}

During the COVID-19 pandemic, governments have put in place Non-pharmaceutical Interventions (NPIs) to contain the spread of the virus. These measures include, among others, school closures, travel and movement restrictions, bans on public gatherings. 
We used the dataset provided by the Oxford COVID-19 Government Response Tracker (OxCGRT), which provides a systematic set of measures of governments' policies and interventions from 1 January 2020, providing a standardized series of indicators. The dataset covers over 180 countries, including sub-national measures enacted in the United States. The state-specific information included in the dataset describes the overall policy stringency under which residents of the state are subjected. The US states policies are quantified using indicators inherited from higher-level governments and their policies. State-level policies are built from regulations made at the state level and below (e.g., indexes also include decisions of county or city governments)~\cite{covidpolicytrackercodebook}. As an aggregate measure of the strength of the enacted regulations, we use the "Stringency Index "~\cite{hale2021global}. The index is composed of nine different indicators, each quantifying the strength of the enacted policies on a specific regulatory category both concerning the containment and closure policies and the health care system policies. Here we report the nine indicators (and their respective range of values) from which the "Stringency Index" is constructed (for a more detailed description on how these indexes are aggregated, see ~\cite{hale2021global} or ~\cite{covidpolicytrackercodebook}):
\begin{itemize}[leftmargin=*,noitemsep,topsep=0pt]
	\item school closure (from 0 to 3);
	\item workplace closure (from 0 to 3);
	\item cancel public events (from 0 to 2);
	\item restrictions on gatherings (from 0 to 4);
	\item public transport closure (from 0 to 2);
	\item stays at home requirements (from 0 to 3);
	\item restriction of internal movements (from 0 to 2);
	\item restriction on international travels (from 0 to 4);
	\item intensity of public information campaigns (from 0 to 3).
\end{itemize}

\section{Stops detection}
\label{SI:sec:stop-detection}
A single original GPS location does not contain any information about people's movements, and it is thus impossible to know, from a single point, whenever a user is stationary in a place or not. Moreover, the coordinates of GPS points can fluctuate over time even when the user is not moving. 
Therefore, we apply a stop location detection algorithm to transform a sequence of original GPS data into a sequence of locations in which a user stops. 

We detect the users' stops in a two-step algorithm composed of detecting (i) stop events and (ii) stop locations.\\

\noindent\textbf{Stop events detection.} From a sequence of ordered time events $T= \left [t_0, t_1, \ldots, t_n \right] \mid t_j \leq t_i$, a corresponding set of GPS locations $R=\left [r_0, r_1, \ldots, r_n\right]$, and a geographical distance function $d(i,j)$, we define a \emph{stop event} as a maximal set of locations $S=\left [r_i, r_{i+1}, \ldots, r_j\right] \mid d(r_i, r_j) < \Delta s \land t_j - t_i \geq \Delta t \ \forall r_i, r_j \in S$. 

Then $\mathcal{S} = \{S_i \mid S_i \text{ is a stop event} \land r_i \in S_i \land r_j \in S_j \land i<j \}$ represents the set of \emph{stop events}. 
To form a \emph{stop event} we heuristically choose to group locations in a time-ordered fashion.
In other words, in this step, we aim at finding all those places at most $\Delta s$ meters large where people stopped for at least $\Delta t$ minutes. 
Each \emph{stop event} is composed of at least two locations, and the locations can belong only to at most one \emph{stop event}.

To extract \emph{stop events} we base our method on Hariharan and Toyama's work~\cite{hariharan2004project}. The algorithm is depicted in \Cref{alg:one} and can be summarised as follows: for each user, we first order their GPS locations by timestamp, and then we select groups of GPS sequences with the desired spatial ($\Delta s$) and temporal ($\Delta t$) thresholds to form \emph{stop events}. 

The \texttt{Diameter} function computes the greatest distance between points, while the \texttt{Medoid} function selects the GPS location with the minimum distance to all other points in the set.

\begin{algorithm}[t]
	\SetAlgoNoLine
	\KwIn{Time-ordered list of a user's original GPS positions $R=[r_0, r_1, \ldots, r_n]$, their timestamp $T=[t_0, t_1, \ldots, t_n]$, a spatial threshold $\Delta s$ and a temporal threshold $\Delta t$.}
	\KwOut{The set $S$ of a user's stop events.}
	$\aleft$ = 0; $S \leftarrow \emptyset$ \;
	\While{$\aleft < n$}{
		$\aright \leftarrow$ minimum $j$ such that $t_j \geq t_{\aleft} + \Delta t$\;
		\If{Diameter($R$, $\aleft$, $j$) $> \Delta s$\;
		}{
			$\aleft\leftarrow \aleft + 1$\;
		}\Else{
			$\aright \leftarrow$ maximum $j$ such that $j \leq n$ and Diameter($R$, $\aleft$, $j$) $ < \Delta s$\;
			$S \leftarrow S \cup$ (Medoid($R$, $\aleft$, $\aright$), $t_{\aleft}$, $t_{\aright}$) \;
			$\aleft \leftarrow \aright + 1$\;
		}
	}
	\caption{Algorithm for extracting the \emph{stop events} from original GPS sequences.}
	\label{alg:one}
\end{algorithm}

The computational complexity of the \emph{stop event} algorithm~\cite{hariharan2004project} is $\mathcal{O}(n^3)$, because of the repeated \texttt{Diameter} function that computes a distance matrix, whose complexity is $\mathcal{O}(n^2)$.

To reduce the computational burden of the algorithm, we divide the sequence of points of each user into buckets/chunks.
To do so, we use the algorithm we depict in \Cref{alg:two}.
Then, we horizontally parallelize the \texttt{Diameter} algorithm across users and chunks, as each chunk can compute the stop locations independently from the others.

\begin{algorithm}[t]
	\SetAlgoNoLine
	\KwIn{Time-ordered list of a user's original GPS positions $R=[r_0, r_1, \ldots, r_n]$ and the spatial threshold $\Delta s$.}
	\KwOut{The set $B$ buckets of an user's stop events.}
	$i$ = 0; $j$ = -1;
	\While{$i < n$}{
		\If{$i = 0$ or $d(r_{i-1}, r_i) > \Delta s$\;
		}{
			$j \leftarrow j + 1$\;
			$B_j \leftarrow \emptyset$\;
		}
		$B_j \leftarrow B_j \cup \left\{r_i \right\}$ \;
		$i \leftarrow i + 1$\;
	}
	\caption{Algorithm for dividing an user's \emph{stop events} into buckets.}
	\label{alg:two}
\end{algorithm}

The \texttt{Diameter} $d(i,j)$ is computed through the Haversine great-circle distance between $r_i$ and $r_j$.
Given the average radius of the Earth $r$ and two points with latitude and longitude $\varphi_1$, $\varphi_2$ and $\lambda_1$, $\lambda_2$ respectively, the Haversine distance $d$ between them is:
\begin{equation*}
	d(r_1, r_2) = 2 r \arcsin\left(\sqrt{\sin^2\left(\frac{\varphi_2 - \varphi_1}{2}\right) + \cos(\varphi_1) \cos(\varphi_2)\sin^2\left(\frac{\lambda_2 - \lambda_1}{2}\right)}\right).
\end{equation*}

\noindent\textbf{Stop location detection.} For each user, we define \emph{stop locations} as the sequences of \emph{stop events} that can be considered part of the same place. 
For example: if user $U$ goes to the Colosseum in Rome multiple times, $U$ could have multiple \emph{stop events} (e.g., northern entrance, southern entrance) that can be grouped in a unique \emph{stop location} (i.e., the Colosseum).

To detect a \emph{stop location} from \emph{stop events} we use the DBSCAN~\cite{ester1996density} algorithm that groups points within $\epsilon = \Delta s -5$ meters of distance to form a cluster with at least $minPoints = 1$ \emph{event}.
The complexity of DBSCAN is $\mathcal{O}(n)$. We horizontally scale the computation through different cloud machines thanks to Apache Spark.

Following the distribution of accuracy in \Cref{fig:accuracy}, we choose $\Delta s = 65$ meters. We also choose $\Delta t = 5$ minutes to detect all stops from the shortest to the longest ones.

We select $\epsilon = \Delta s -5$ meters to avoid the creation of an extremely --and incorrect-- long chain of sequential \emph{stop events}.  

\section{Residential/Workplace detection}
\label{SI:sec:home}
To analyze human mobility in detail, we need to classify each user's stop location semantically. For this reason, we computed the most probable residential and workplace areas for each user. Furthermore, to capture possible changes in residential and workplace areas due to the pandemic, we computed these areas multiple times for a moving window of 28 days.
We proceed as follow, for each user $i$ and stop $s_i(t)$ in a day $t$, we aggregate the time spent in $[t-14, t+14]$ days into two buckets:
\begin{enumerate}
	\item \textbf{Potential residential time.} We sum the time between 8 pm and 4 am spent in stop $s_i$;
	differently from previous literature (e.g., TimeGeo~\cite{jiang2016timegeo}), we do not assume that the entire weekend is potential residential time as the US Bureau of Labor Statistics recently estimated that between 34\% of employed people work in the weekend~\cite{bureau}.
	\item \textbf{Potential work time.} We sum the weekday time between 9 am and 5 pm spent in stop $s_i$. Moreover, we assume that a potential work stay should last at least 30 minutes and have a frequency of 5 visits per week. These assumptions are similar to previous literature~\cite{jiang2016timegeo}. We choose the aforementioned working hours since they represent the most popular working time in the US~\cite{officeHoursUS}.
\end{enumerate}

For each day $t$, we label a stop $s_i(t)$ as \emph{Residential stop} if this stop has the largest \emph{potential Residential time}.
Then, we label a stop $s_i(t)$ as \emph{work stop} if this stop is not \emph{Residential stop} and it has the largest \emph{potential work time}. 

\section{States selection}
\label{SI:sec:states}
We aim at understanding whether there were differences in the mobility behaviour patterns of individuals in states that put in place different strategies in response to the COVID-19 pandemic. 

\begin{figure}[!ht]
	\centering
	\includegraphics[width=0.64\textwidth]{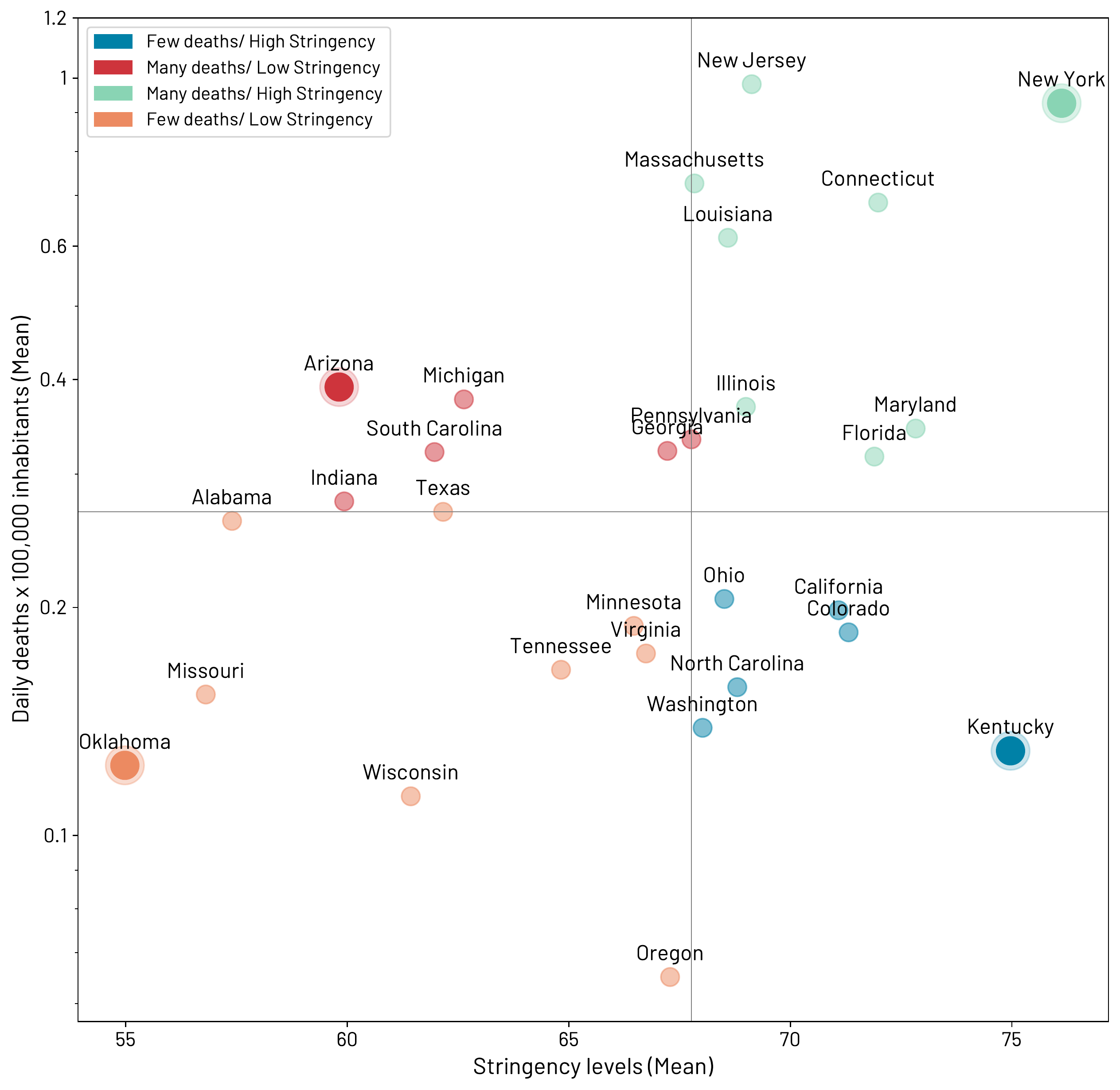}
	\caption{State selection (AZ, KY, NY, OK) based on daily deaths per 100K inhabitants and mean stringency level for states that have a population $>$ 1\% of US total population. }
	\label{fig:sup_states_selection}
\end{figure}

To select these four US states, we took into account (i) the daily COVID death rate and (ii) the stringency measures the states adopted to slow down the spread of the virus. Both information was retrieved from the OxCGRT dataset~\cite{hale2021global}. As shown in Fig~\ref{fig:sup_states_selection} we selected the states of Arizona (many deaths/low stringency), Oklahoma (few deaths/low stringency), Kentucky (few deaths/high stringency) and New York (many deaths/high stringency).

\subsection{Important dates}
\label{si:sec:important dates}
We retrieved individual state dates for relevant events concerning the pandemic and the local regulatory response from state-dedicated web pages on the Wikipedia website~\cite{ImportantDatesWiki}. We here summarize the main events, focusing on the first case reported within the state borders and the main regulatory responses. 
The dates are conveniently reported also in \Cref{tab:important_dates}.

\noindent\textbf{Arizona}
\begin{itemize}[noitemsep,topsep=0pt]
	\item First case reported: 26 January 2020;
	\item Universities start to move to online courses: 12 March 2020;
	\item Governor issues a statewide ``stay-at-home'' order, barring Arizonans from leaving their houses except for essential activities: 31 March 2020;
	\item A partial reopening for a selected set of non-essential activities is announced: 4 May 2020;
	\item The ``stay-at-home'' order expires: 15 May 2020;
	\item Following a record high in the number of hospitalizations a partial reversal of the ``stay-at-home'' order is announced: 29 June 2020.
\end{itemize}

\noindent\textbf{Kentucky}
\begin{itemize}[noitemsep,topsep=0pt]
	\item First case reported: 6 March 2020;
	\item School closure: 16 March 2020;
	\item Governor issues a ban for mass-gatherings: 20 March 2020;
	\item Non-essential business closure is enacted and a ``healthy-at-home'' order (similar to a ``shelter in place'' policy): 26 March 020;
	\item Reopening starts requiring business to follow public health guidelines (``healthy-at-work'' initiative): 9 May 2020;
	\item Restaurants reopening: 22 May 2020;
	\item Bars reopening operating (similarly to restaurants) at a 50\% indoor capacity: 29 June 2020.
\end{itemize}

\noindent\textbf{New York}
\begin{itemize}[noitemsep,topsep=0pt]
	\item First case reported: 1 March 2020;
	\item School closure in New York city: 15 March 2020;
	\item Statewide ``stay-at-home'' order is declared, all non-essential business are ordered to close: 22 March 2020;
	\item ``Phase one'', a county-level partial reopening upon meeting qualifications is announced: 15 May 2020;
	\item ``Phase two'', if meeting qualifications counties can reopen restaurants outdoor activity as in-store activities for specific shop categories: 29 May 2020;
	\item ``Phase three'', if meeting qualifications counties can reopen indoor restaurant activity and bars at 50\% capacity: 24 June 2020.
\end{itemize}

\noindent\textbf{Oklahoma}
\begin{itemize}[noitemsep,topsep=0pt]
	\item First case reported: 7 March 2020;
	\item Announcement of a "safer-at-home" order requiring vulnerable people to remain at their residences except for essential activities, "school closure" order and ban of mass gatherings are enacted on the following day: 24 March 2020 (and 25 March 2020);
	\item "Phase one" of business reopening allows for outdoor activities, personal care facilities, restaurants, cinemas, gyms, and places of worship to restart activities subject to physical distancing and sanitation protocols (specific guidelines are issued for the industry): 24 April 2020;
	\item "Phase two" allows religious ceremonies, bars to reopen, and organized sports to restart activities subject to physical distancing: 15 May 2020;
	\item "Phase three" reopens business that has been restricted to appointments only, summer caps and (although limited) also visitation at hospitals: 1 June 2020.
\end{itemize}

\begin{table}[!ht]
	\centering
	\begin{tabularx}{\linewidth}{Xccccc}
		\toprule
		& \textbf{AZ} & \textbf{NY} & \textbf{OK} & \textbf{KY}\\
		\midrule
		\textbf{first case} &  2020-01-26 &  2020-03-01 &  2020-03-07 &  2020-03-06 \\
		\textbf{stay-at-home} &  2020-03-31 &  2020-03-20 &  2020-03-24 & -\\
		\textbf{healthy-at-home} & - & - & - &  2020-03-26\\
		\textbf{school closure} & - &  2020-03-15 & - &  2020-03-16\\
		\textbf{university online} &  2020-03-12 & - & - & - \\
		\textbf{stay-at-home expired} &  2020-05-15 & - & - & -\\
		\textbf{partial reversal stay-at-home} & 2020-06-29 & - & - & -\\
		\textbf{partial reopening} &  2020-05-04 & - & - & - \\
		\textbf{phase-one} & - &  2020-05-15 &  2020-04-24 & -\\
		\textbf{phase-two} & - &  2020-05-29 &  2020-05-15 & - \\
		\textbf{phase-three} & - &  2020-06-24 &  2020-06-01 & -\\
		\textbf{start reopening} & - & - & - &  2020-05-09 \\
		\textbf{restaurants reopening} & - & - & - &  2020-05-22\\
		\textbf{bars reopening} & - & - & - &  2020-06-29\\
		\bottomrule
	\end{tabularx}
	\caption{Dates of the non-pharmaceutical intervention phases for each state.}
	\label{tab:important_dates}
\end{table}

\section{Users selection}
\label{SI:sec:users}
For each US state we studied, we performed all our analyses on panels of individuals that:
\begin{enumerate}
	\item have been seen for at least 7 days before the stay-at-home declaration and that have more than 5 hours of activity on average;
	\item have been seen for at least 7 days after the stay-at-home declaration and that have more than 5 hours of activity on average.
\end{enumerate}
This pre-processing step leaves us with individuals for which we have information before and after the stay-at-home declaration, enabling a more sound analysis of how mobility behaviour changed during the pandemic.

\section{Points of interest}
\label{sec:poi_extraction_and_match}

\subsection{Extraction of Points of interest (POIs)}

We downloaded the POIs from OpenStreetMap (OSM) for the entire US. More precisely, we extracted the POIs with the Python package Pyrosm\footnote{\url{https://pyrosm.readthedocs.io}}, which reads the data from an OSM public dump.

OSM is a collaborative project with maps created by everyone. The nature of the project and its tagging system - which has a quite free form - create some inconsistencies since it allows a user to insert a POI in different ways. As an example, if users want to add a \emph{hospital} in OSM, they can do it setting \emph{healthcare=hospital} or setting \emph{amenity=hospital}. 

Another disadvantage of the OSM tagging system is that the POI taxonomy is not organized hierarchically and thus it does not exists a unique way to organize the POIs.
To overcome these problems, we decided to create a manual mapping between the OSM tags \footnote{\url{https://wiki.openstreetmap.org/wiki/Map_features}} and the Foursquare Venue Category Hierarchy \footnote{\url{https://developer.foursquare.com/docs/build-with-foursquare/categories/}}. In all the analyses in the manuscript, we will therefore use Foursquare categories. 

Finally, we retained only POIs that represent actual places where people can meet or go, such as for leisure activities (e.g. parks, sports facilities) or for business reasons (e.g. offices). We decided to remove POIs such as highways, sheds, landuse, fountains, as they do not represent interesting venues. Moreover, we kept only POIs that had a mapping with at least a Foursquare second-level categorization.

\subsection{Popularity of POIs in the US}
\label{SI:sec:popularity}

\begin{figure}[!h]
	\centering
	\includegraphics[width=0.8\textwidth]{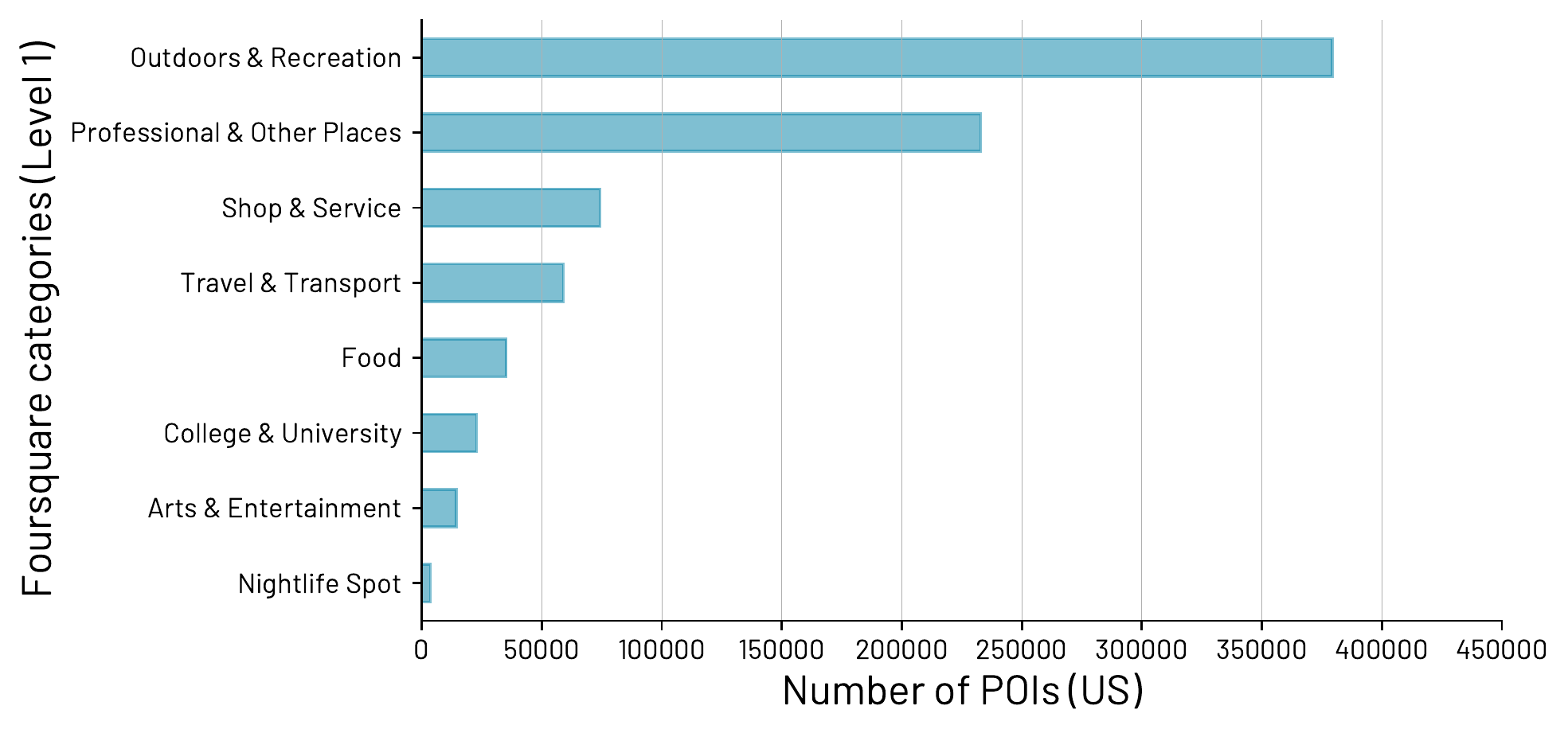}
	\caption{Level 1 Foursquare categories mapped from Open Street Map POIs downloaded  for the four states analysed.}
	\label{fig:poi_us_lvl1}
\end{figure}

\begin{figure}[!h]
	\centering
	\includegraphics[width=0.8\textwidth]{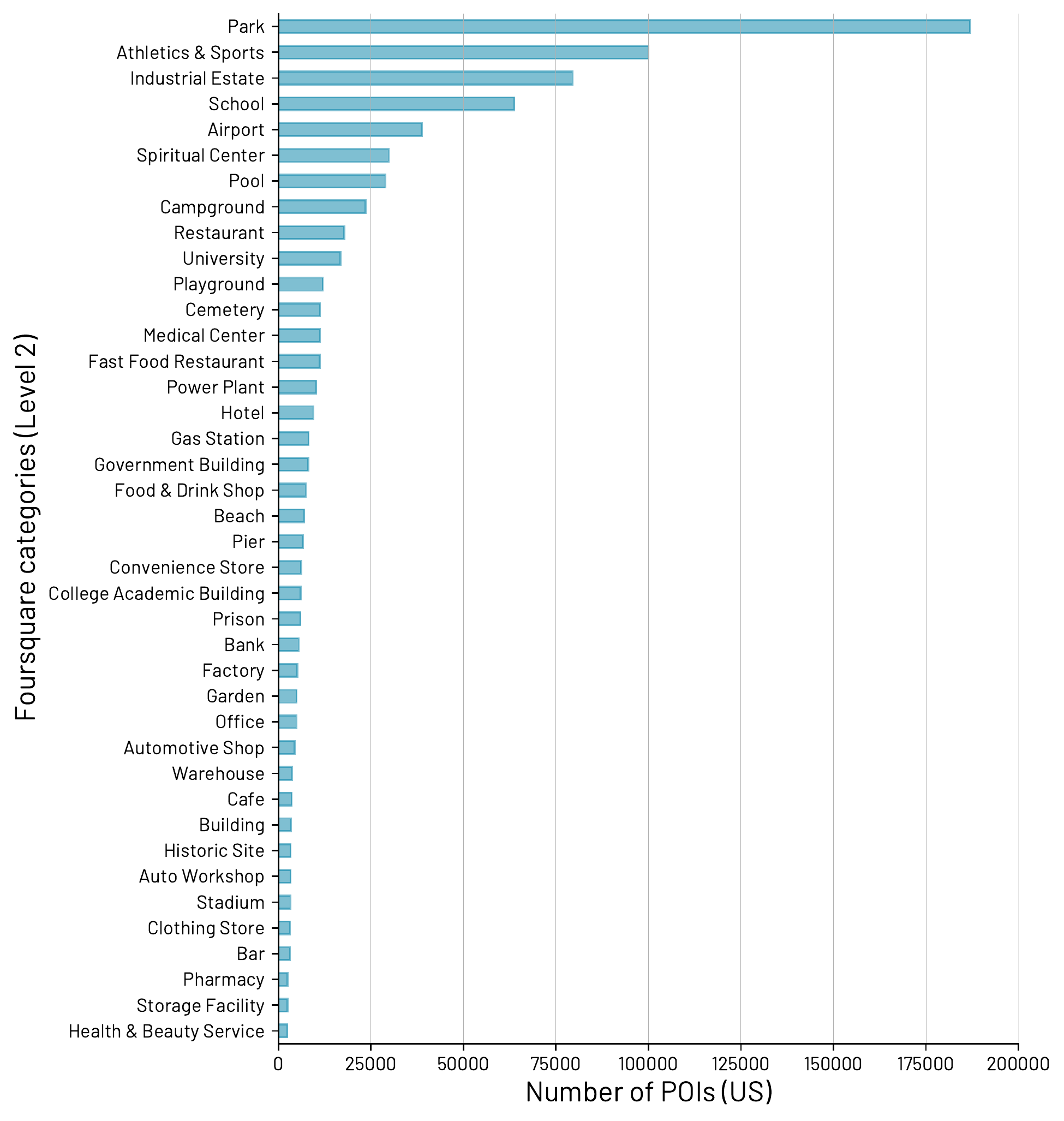}
	\caption{Top-40 Level 2 Foursquare categories mapped from Open Street Map POIs for the four states analysed.}
	\label{fig:poi_us_lvl2}
\end{figure}

After the POIs extraction phase, we remain with a total of 3,353,502 POIs for the entire US.
In Fig~\ref{fig:poi_us_lvl1} we show the number of OSM POIs mapped to Level 1 Foursquare categories that we extracted for the four states analyzed. In the same way, Fig~\ref{fig:poi_us_lvl2} shows the number of OSM POIs mapped to Level 2 Foursquare categories. 

As shown in Fig~\ref{fig:poi_us_lvl1}, the majority of POIs extracted from OSM are related to \textit{Outdoors \& Recreation} activities as also reflected in the Level 2 Foursquare categories \textit{Park} and \textit{Athletics \& Sports} (Fig~\ref{fig:poi_us_lvl2}).

The second most common category is \textit{Professional \& Other Places} which contains venues such as \textit{Office}, \textit{Factory} and \textit{School} (e.g., Elementary School, High School, Music School, etc.)  which appears at the top of Fig~\ref{fig:poi_us_lvl2}.

\subsection{Matching stop locations with points of interest}
To further enrich our analysis, we added semantic meaning to individuals' trajectories associating stop locations to their nearest POI when looking at the distance between the stop location and the POI.

When we associate a POI to a stop location, we prioritize a Point POI over a Polygon POI, checking whether a POI is inside another POI. This situation happens, for example, when an individual is inside a shopping mall. 

In this case, we will have a POI representing the entire shopping mall (Polygon POI), and the distance between the stop location of the individual and the POI will be 0. Now, suppose that we have a cafeteria (Point POI) inside the shopping mall, which is at 3 meters distance. In this case, we will assign the stop location to the cafeteria since it gives us a fine-grained piece of information on the visit patterns of the individual.

\begin{figure}[!h]
	\centering
	\includegraphics[width=0.9\textwidth]{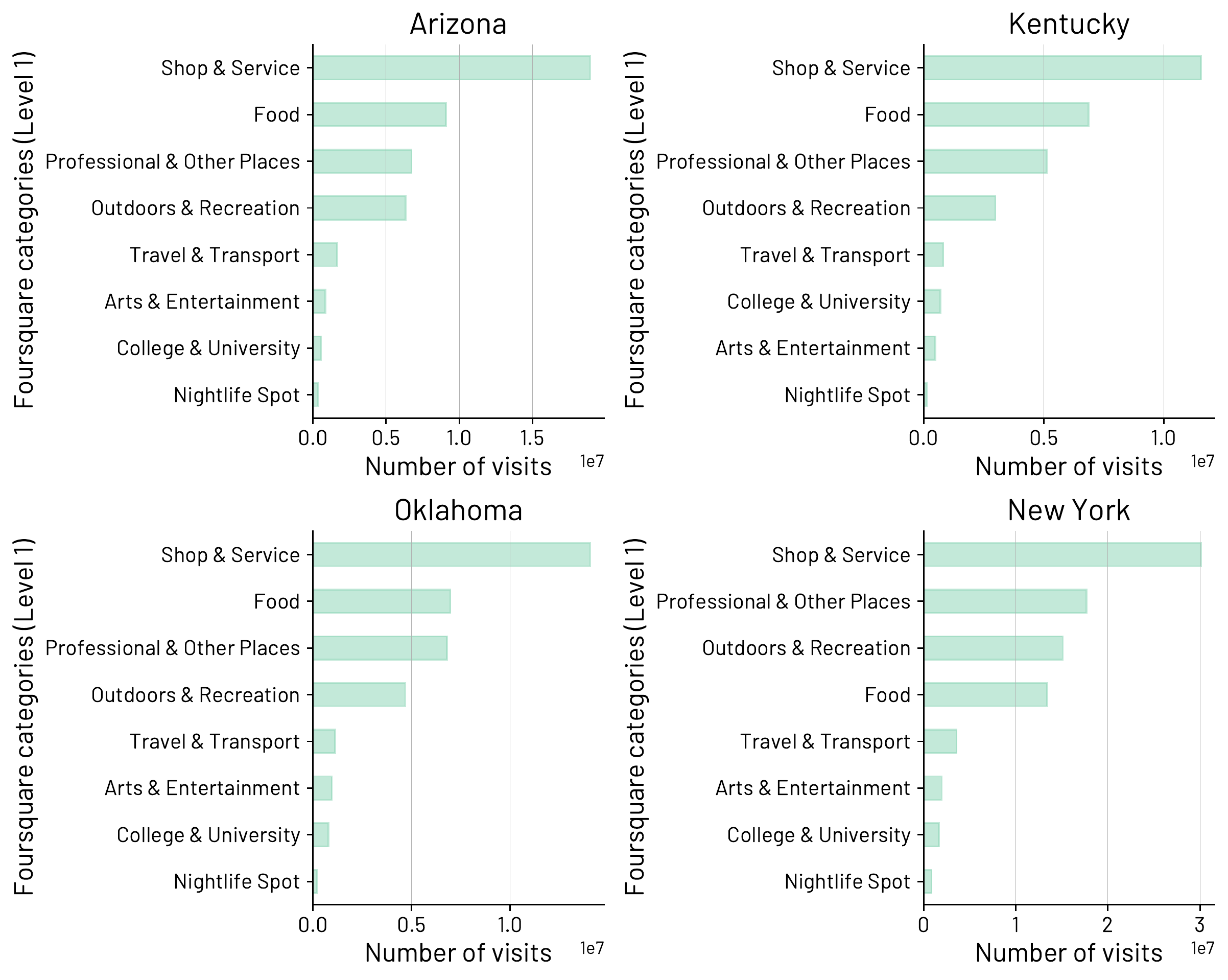}
	\caption{Number of visits made by individuals in the four selected US states (AZ, KY, NY, OK) to POIs mapped at the Level 1 Foursquare categories}
	\label{fig:poi_visits_lvl1}
\end{figure}

\begin{figure}[!ht]
	\centering
	\includegraphics[width=0.9\textwidth]{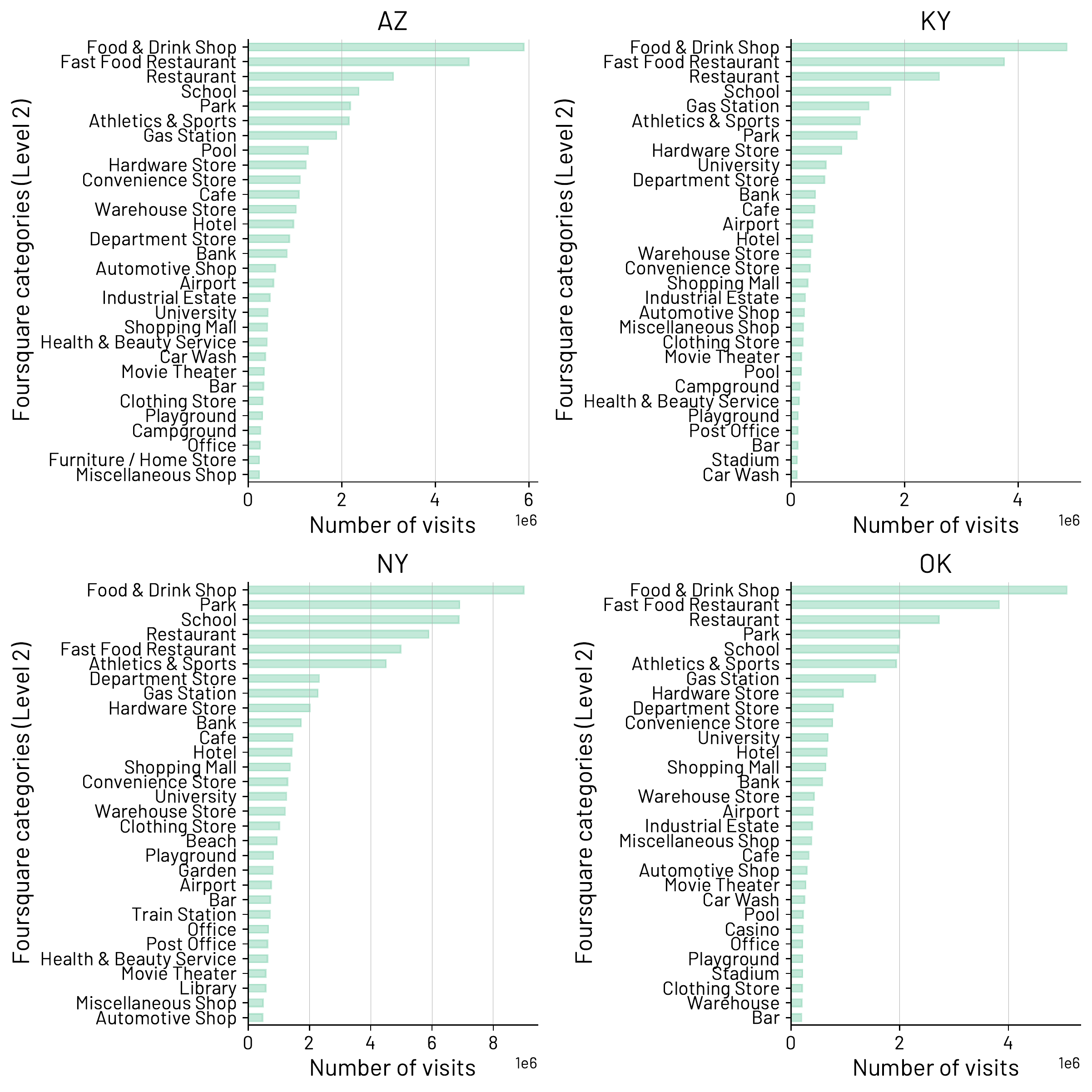}
	\caption{Number of visits made by individuals in the four selected US states (AZ, KY, NY, OK) to POIs mapped at the Level 2 Foursquare categories. From level 2 Foursquare categories, \textit{health-related} categories, ``Spiritual Centers'', ``Governmental Buildings'', and ``Prisons'' were removed due to the sensitive nature of those locations.}
	\label{fig:poi_visits_lvl2}
\end{figure}

In Fig~\ref{fig:poi_visits_lvl1} we show the number of visits made by individuals in the four selected US states (AZ, KY, NY, OK) to POIs mapped at Level 1 Foursquare categories, and in Fig~\ref{fig:poi_visits_lvl2} at Level 2 Foursquare categories. Following Cuebiq's policy we removed  \textit{health-related} categories, ``Spiritual Centers'', ``Governmental Buildings'', and ``Prisons''~\cite{cuebiqPrivacyPolicy}.

\subsection{New York state - Number and duration of visits to POIs}
\label{SI:sec:pois_visits_NY}
\begin{figure}[!ht]
	\centering
	\includegraphics[width=0.9\textwidth]{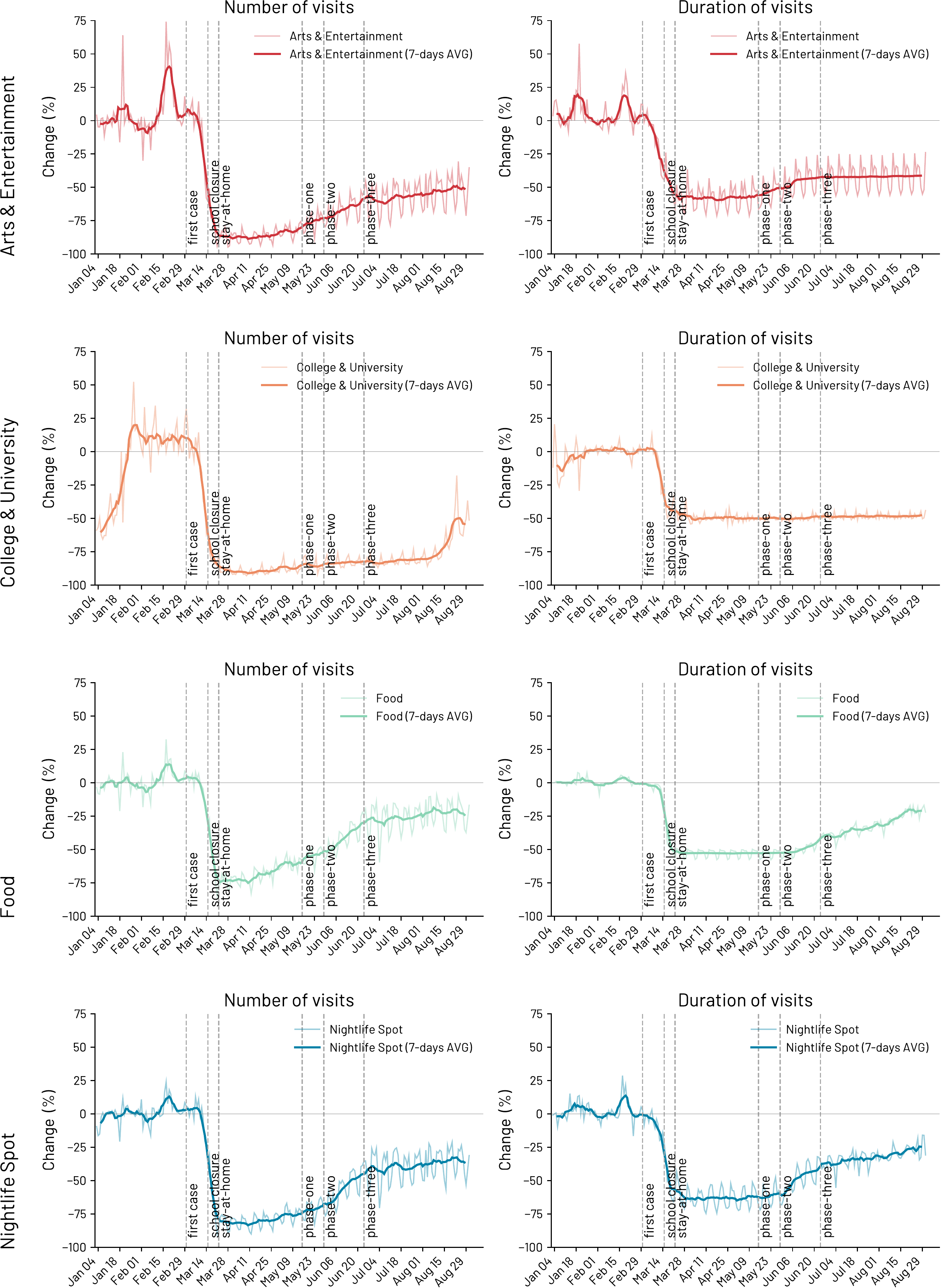}
	\caption{Number of visits and duration of visits to POIs categories in the state of New York}
	\label{fig:number_duration_visits_NY_1}
\end{figure}

\begin{figure}[!ht]
	\centering
	\includegraphics[width=0.9\textwidth]{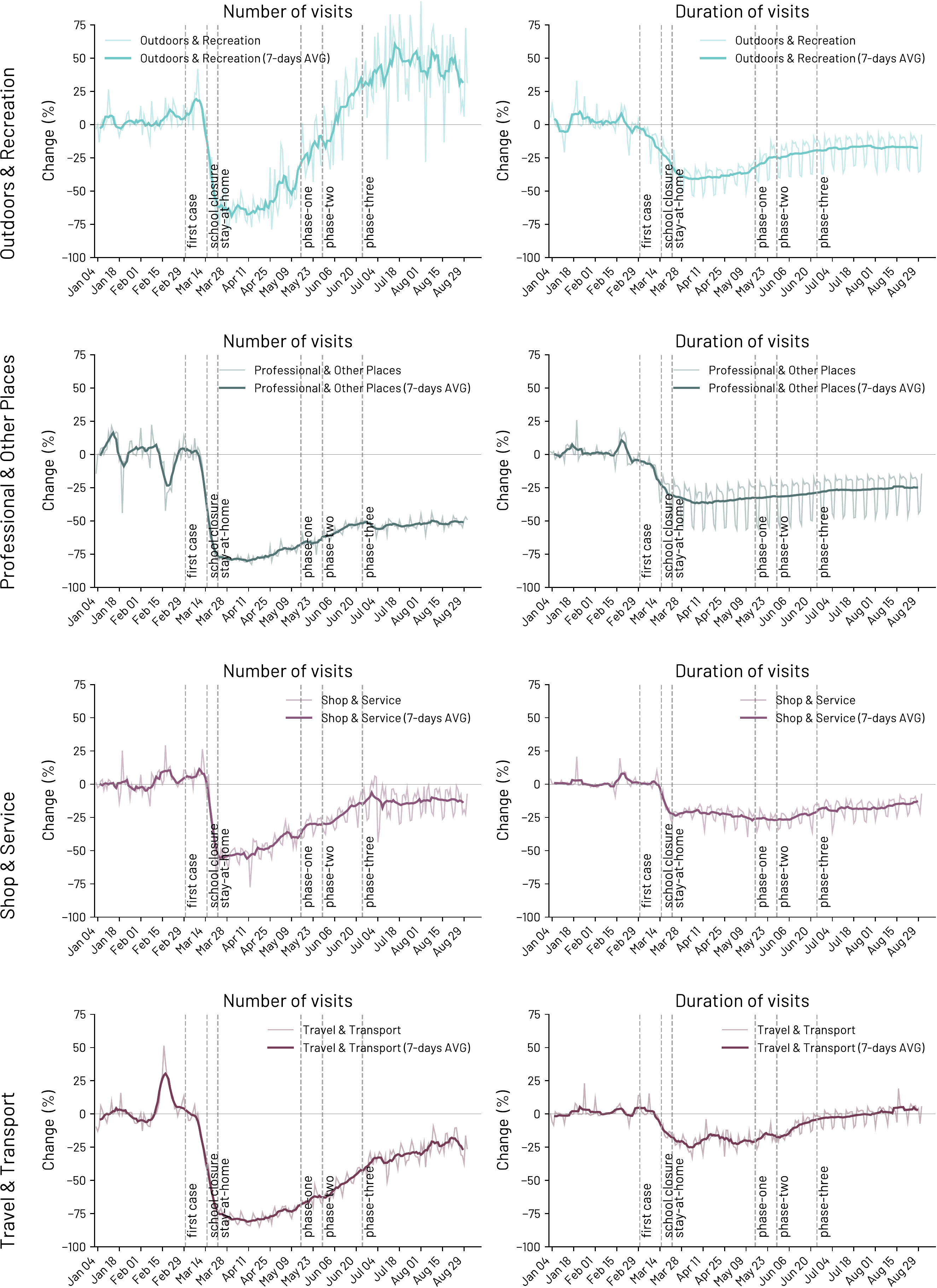}
	\caption{Number of visits and duration of visits to POIs categories in the state of New York}
	\label{fig:number_duration_visits_NY_2}
\end{figure}

Here we report the number of visits and the duration of visits to POIs over time for individuals in the state of New York. Figure~\ref{fig:number_duration_visits_NY_1} shows these quantities for the \textit{Arts \& Entertainment}, \textit{College \& University}, \textit{Food}, \textit{Nightlife Spot} business venues, while Fig.~\ref{fig:number_duration_visits_NY_2} shows the same quantities for the \textit{Outdoors \& Recreation}, \textit{Professional \& Other Places}, \textit{Shop \& Service}, and \textit{Travel \& Transport} venue categories.


\subsection{Change in POIs' visits in Arizona, Kentucky and Oklahoma}
\label{SI:sec:fourstates}

\begin{figure}[!ht]
	\centering
	\includegraphics[width=0.9\textwidth]{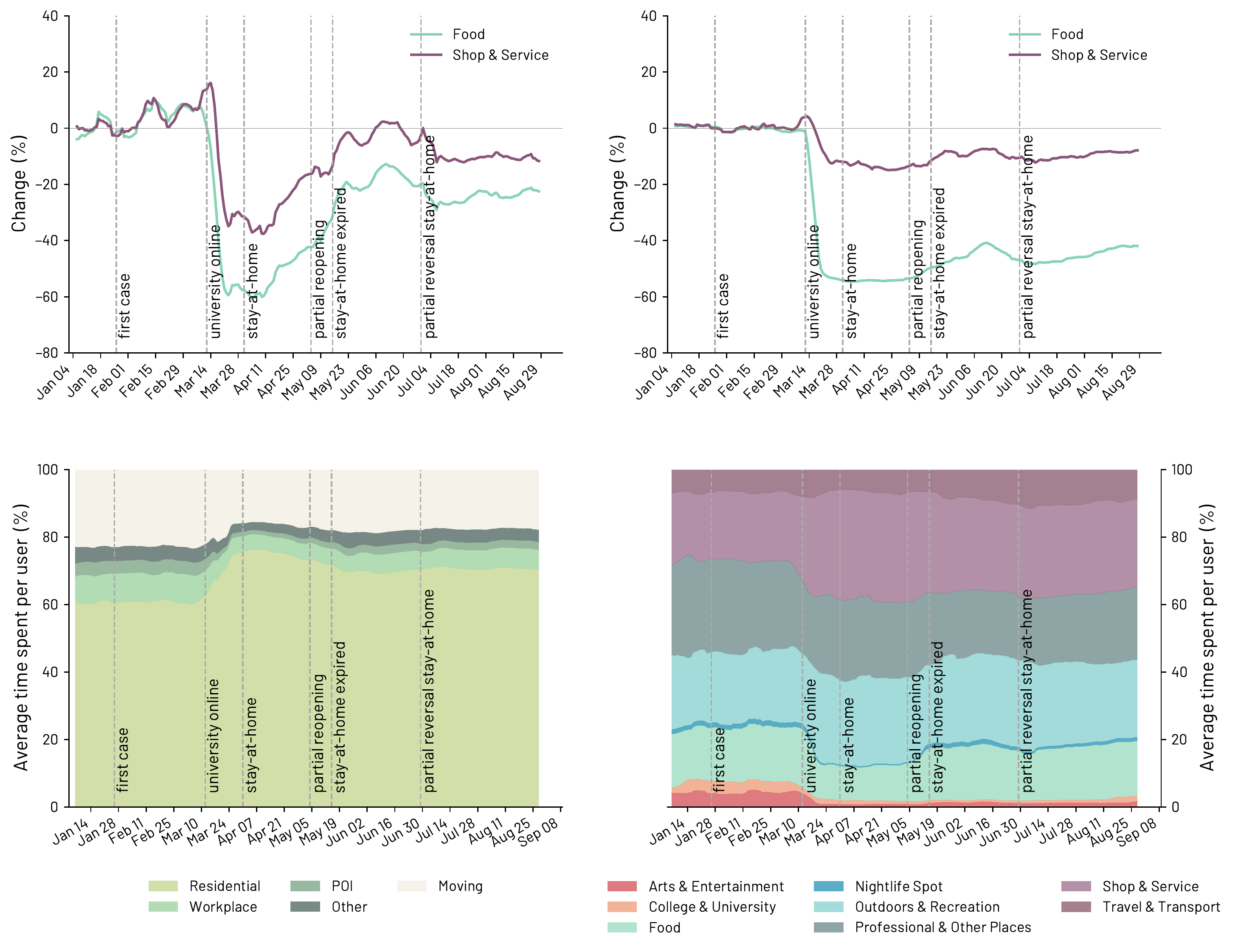}
	\caption{Changes in number and duration of visits to POIs in the state of Arizona. A) Percent change over time in the number of visits with respect to the baseline period (January 3, 2020 - February 28, 2020) for venues in the \emph{Food} and \emph{Shop \& Service} categories. 
		B) Change in percentage of the duration of visits over time with respect to the baseline period of the median duration of visits to \emph{Food} and \emph{Shop \& Service} POIs. 
		For visualization purposes, the original curves are smoothed using a rolling average of seven days.
		Vertical dashed lines indicate the date of restrictions and orders imposed by the state of Arizona government.
		C) Percentage of time spent by people at \emph{Residential}, \emph{Work}, \emph{POIs}, \emph{Other}, and \emph{Moving} (i.e. people in movement). 
		D) Percentage of time spent by people in venues under the eight first-level categories of POIs.}
	\label{fig:fig3_AZ}
\end{figure}

\begin{figure}[!ht]
	\centering
	\includegraphics[width=0.9\textwidth]{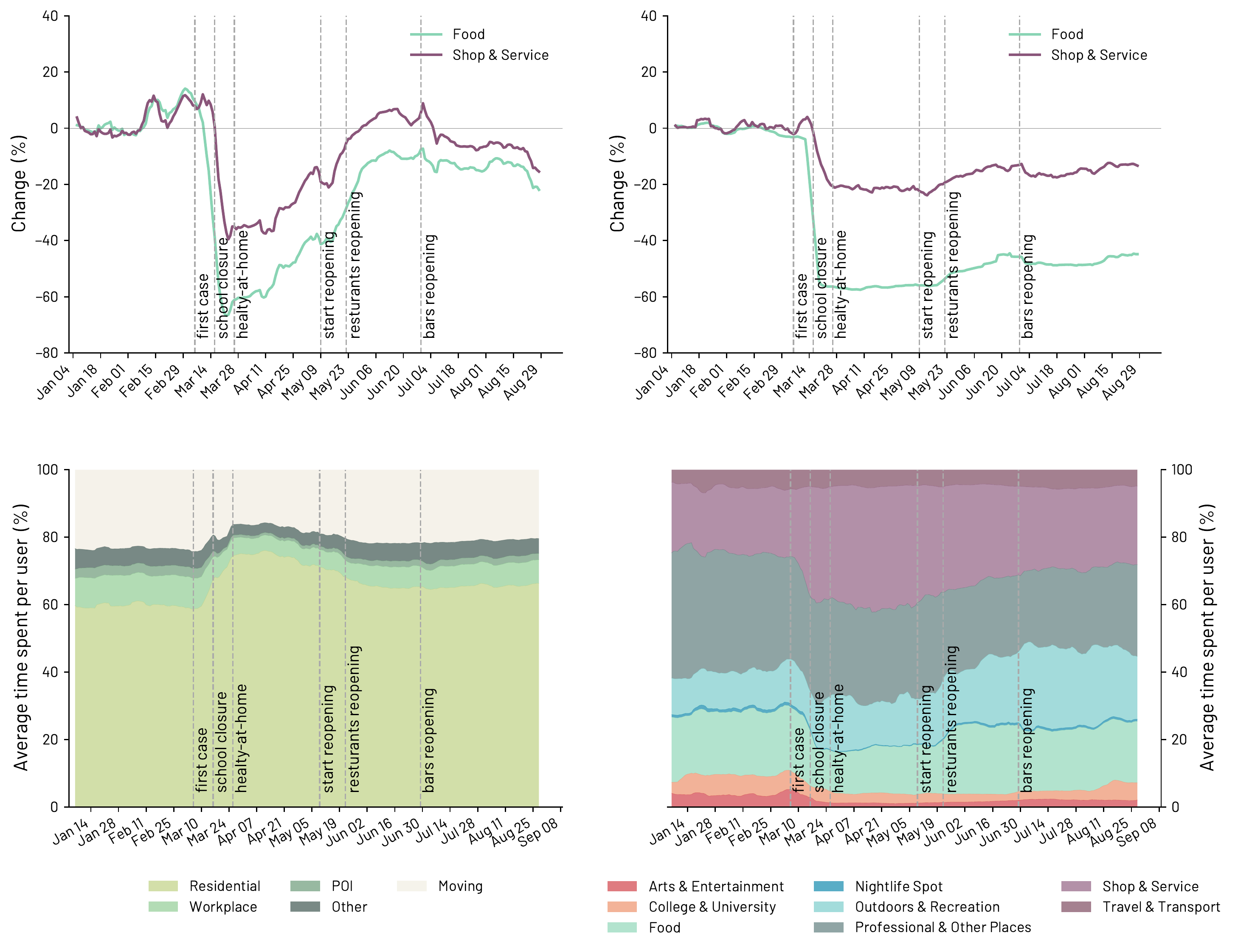}
	\caption{Changes in number and duration of visits to POIs in the state of Kentucky. A) Percent change over time in the number of visits with respect to the baseline period (January 3, 2020 - February 28, 2020) for venues in the \emph{Food} and \emph{Shop \& Service} categories. 
		B) Change in percentage of the duration of visits over time with respect to the baseline period of the median duration of visits to \emph{Food} and \emph{Shop \& Service} POIs. 
		For visualization purposes, the original curves are smoothed using a rolling average of seven days.
		Vertical dashed lines indicate the date of restrictions and orders imposed by the state of Kentucky government.
		C) Percentage of time spent by people at \emph{Residential}, \emph{Work}, \emph{POIs}, \emph{Other}, and \emph{Moving} (i.e. people in movement). 
		D) Percentage of time spent by people in venues under the eight first-level categories of POIs.}
	\label{fig:fig3_KY}
\end{figure}

\begin{figure}[!ht]
	\centering
	\includegraphics[width=0.9\textwidth]{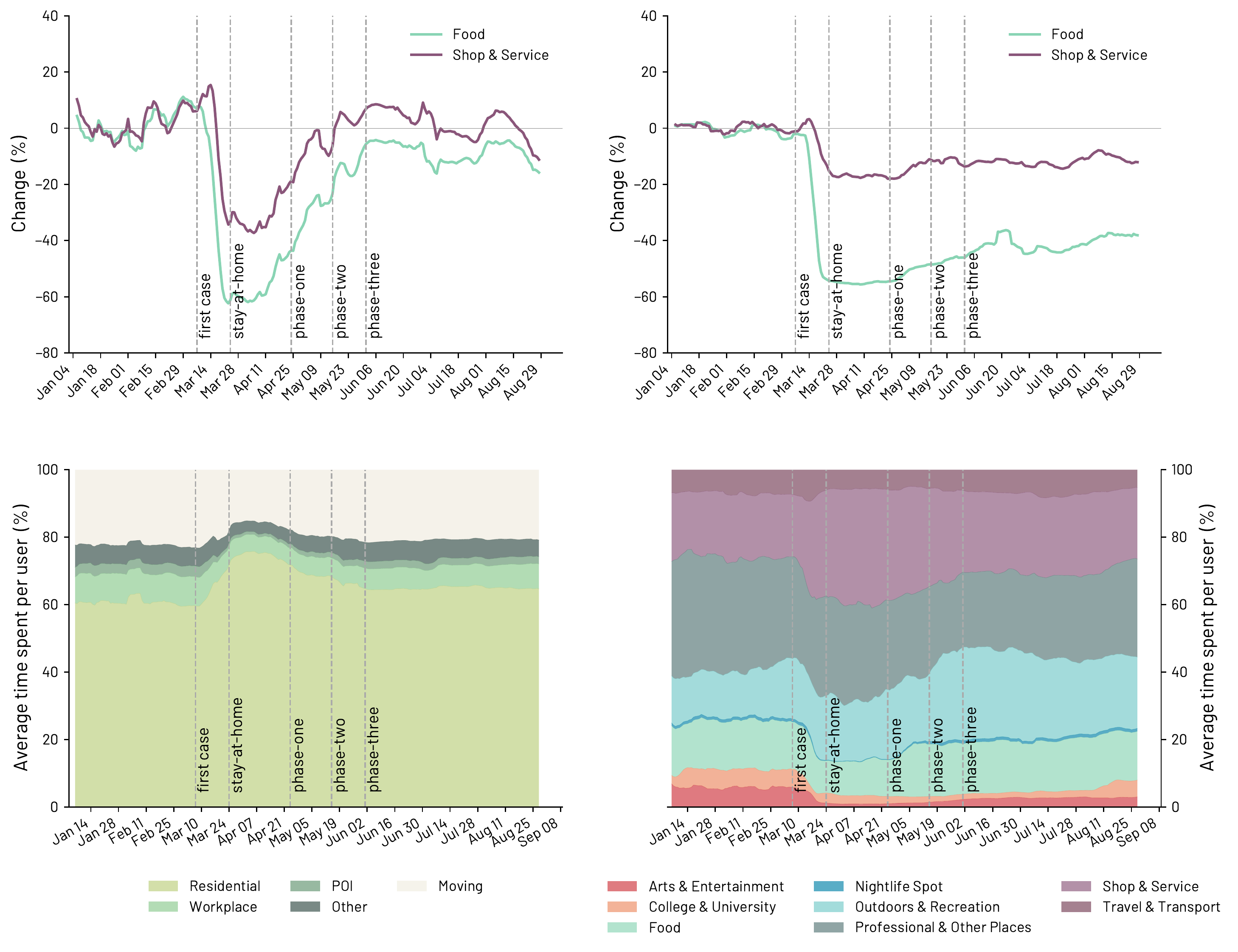}
	\caption{Changes in number and duration of visits to POIs in the state of Oklahoma. A) Percent change over time in the number of visits with respect to the baseline period (January 3, 2020 - February 28, 2020) for venues in the \emph{Food} and \emph{Shop \& Service} categories. 
		B) Change in percentage of the duration of visits over time with respect to the baseline period of the median duration of visits to \emph{Food} and \emph{Shop \& Service} POIs. 
		For visualization purposes, the original curves are smoothed using a rolling average of seven days.
		Vertical dashed lines indicate the date of restrictions and orders imposed by the state of Oklahoma government.
		C) Percentage of time spent by people at \emph{Residential}, \emph{Work}, \emph{POIs}, \emph{Other}, and \emph{Moving} (i.e. people in movement). 
		D) Percentage of time spent by people in venues under the eight first-level categories of POIs.}
	\label{fig:fig3_OK}
\end{figure}

\Cref{fig:fig3_AZ}, \Cref{fig:fig3_KY} and \Cref{fig:fig3_OK} show the changes in the number and duration of visits to POIs in Arizona, Kentucky and Oklahoma respectively.

\subsection{Essential vs non-essential Shop \& Service}
\label{SI:sec:essential_nonessential}

We compare the time series of the number and duration of the visits to POIs between non-essential and essential \textit{Shop \& Services}.
We define as essential \textit{Shop \& Services} all those POIs belonging to the \textit{Shop \& Services} category and with a second-level classification belonging to one of the following categories: \textit{Food \& Drink Shop}, \textit{Miscellaneous Shop}, \textit{Fruit \& Vegetable Store}, \textit{Market}, \textit{Pharmacy}, \textit{Miscellaneous Shop}, \textit{Warehouse store}, \textit{Supplement shop}, \textit{Drugstore}, \textit{Convenience Store}, or \textit{Big Box Store}.

\begin{figure}[!ht]
	\centering
	\includegraphics[width=\textwidth]{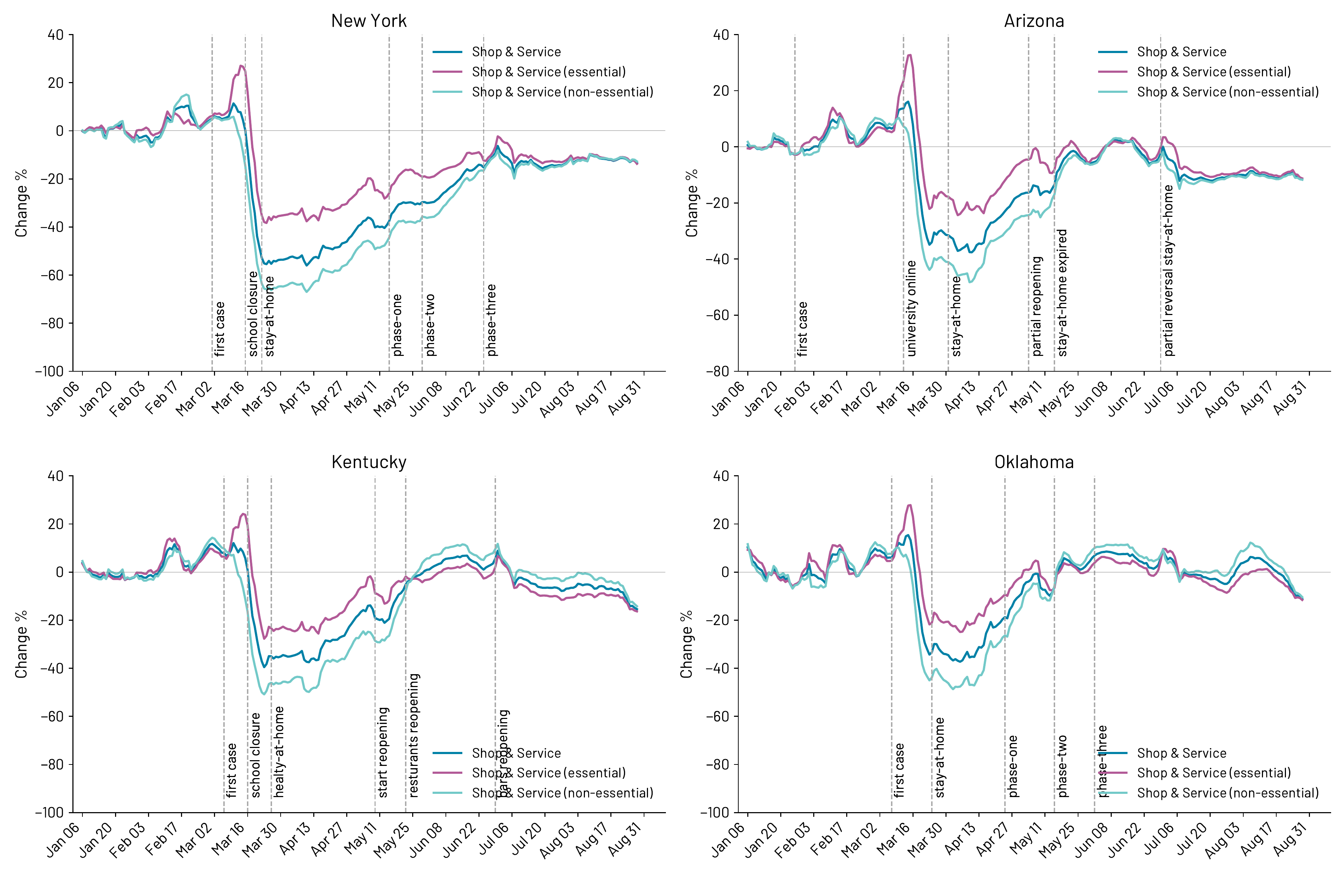}
	\caption{Comparison of the number of visits for POIs that are essential and non-essential in the \textit{Shop \& Service} category for the selected four states, namely AZ, KY, NY, OK.}
	\label{fig:shop_essential_non_ess}
\end{figure}

\Cref{fig:shop_essential_non_ess} shows that just before the pandemic, stay-at-home orders sparked a $27\%$ increase of visits to essential \textit{Shop \& Service} POIs, while non-essential POIs had a lower increase. 
During the pandemic, essential \textit{Shop \& Service} POIs faced a lower reduction than other POIs.

\section{Comparison with Google and Foursquare mobility reports}
\label{SI:sec:correlationGoogle}

To assess the reliability of the Cuebiq data, we compared the mobility data of Cuebiq to the mobility reports provided by Google and Foursquare. 

Google provides the relative change in the number of visits to a specific category by comparing the actual number of visits to a baseline computed for each day of the week, over 5 weeks from 3 January to 6 February 2020 (e.g., 5 values for Monday, 5 values for Tuesday). The only exception is for residential places, for which Google computes the change metric in terms of the fraction of time spent at home on a specific day.

We apply the same procedure to our mobility data, and we compare the results with the data provided by Google (comparing residential and workplaces) and Foursquare (comparing the different visits to POIs categories). 

We computed the Pearson correlation between the relative change in the number of visits for each US state and category according to Google/Foursquare and Cuebiq.

\begin{figure}[!h]
	\centering
	\includegraphics[width=0.65\textwidth]{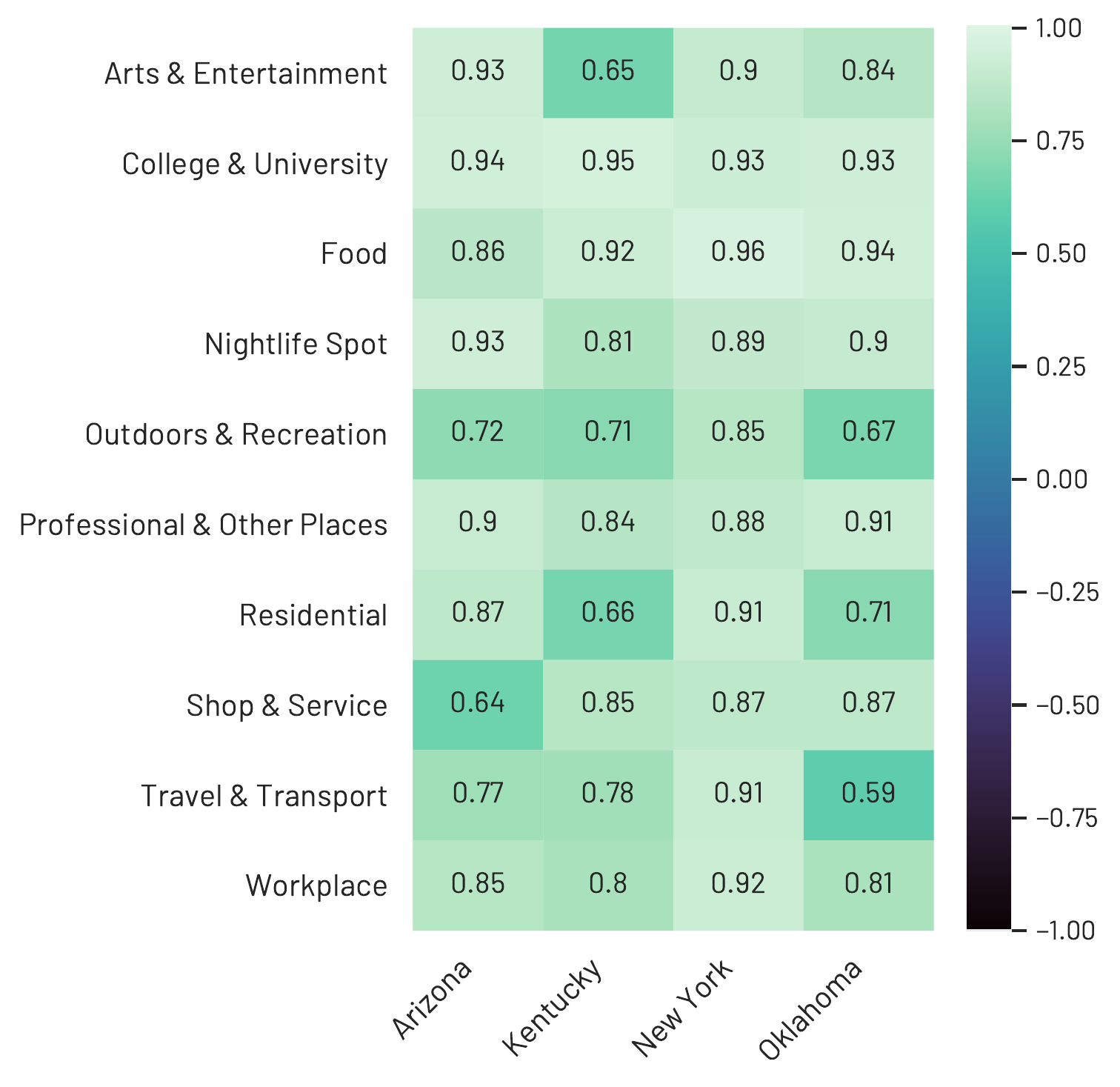}
	\caption{Pearson correlation of the daily number of visits to POIs, number of visits to workplace, and time spent at Residential locations. Correlations are computed between Google change timeseries, for time spent at home and visits to workplaces, and the same quantities as inferred from the Cuebiq data. Comparisons for the change in number of visits to venue categories are performed by using Foursquare change visit data and the inferred visits to POI categories from Cuebiq and OpenStreetMap data.}
	\label{fig:sup3}
\end{figure}

The reliability of the Cuebiq mobility data is shown in Fig.~\ref{fig:sup3}, where we can see that Cuebiq mobility data show high agreement in the changes of mobility patterns with both Google and Foursquare data. Google data are used to compare the change in time spent at Residential locations and the change in the number of visits to user workplaces. Foursquare data, in contrast, are used as a comparison to check the reliability of the change in the number of visits to Foursquare venue categories.

\section{Behaviour change metrics}\label{sec:behaviour_and_change_metrics}
As the threat of the COVID-19 pandemic unfolds in different states, there is a consistent reduction in the population mobility. This effect is reflected at various levels in different key quantities capturing several aspects of human mobility.
Thus, we extract some additional mobility metrics to complement the analysis on the changes in individual mobility patterns presented in the main paper.

\begin{figure}[!h]
	\centering
	\includegraphics[width=\textwidth]{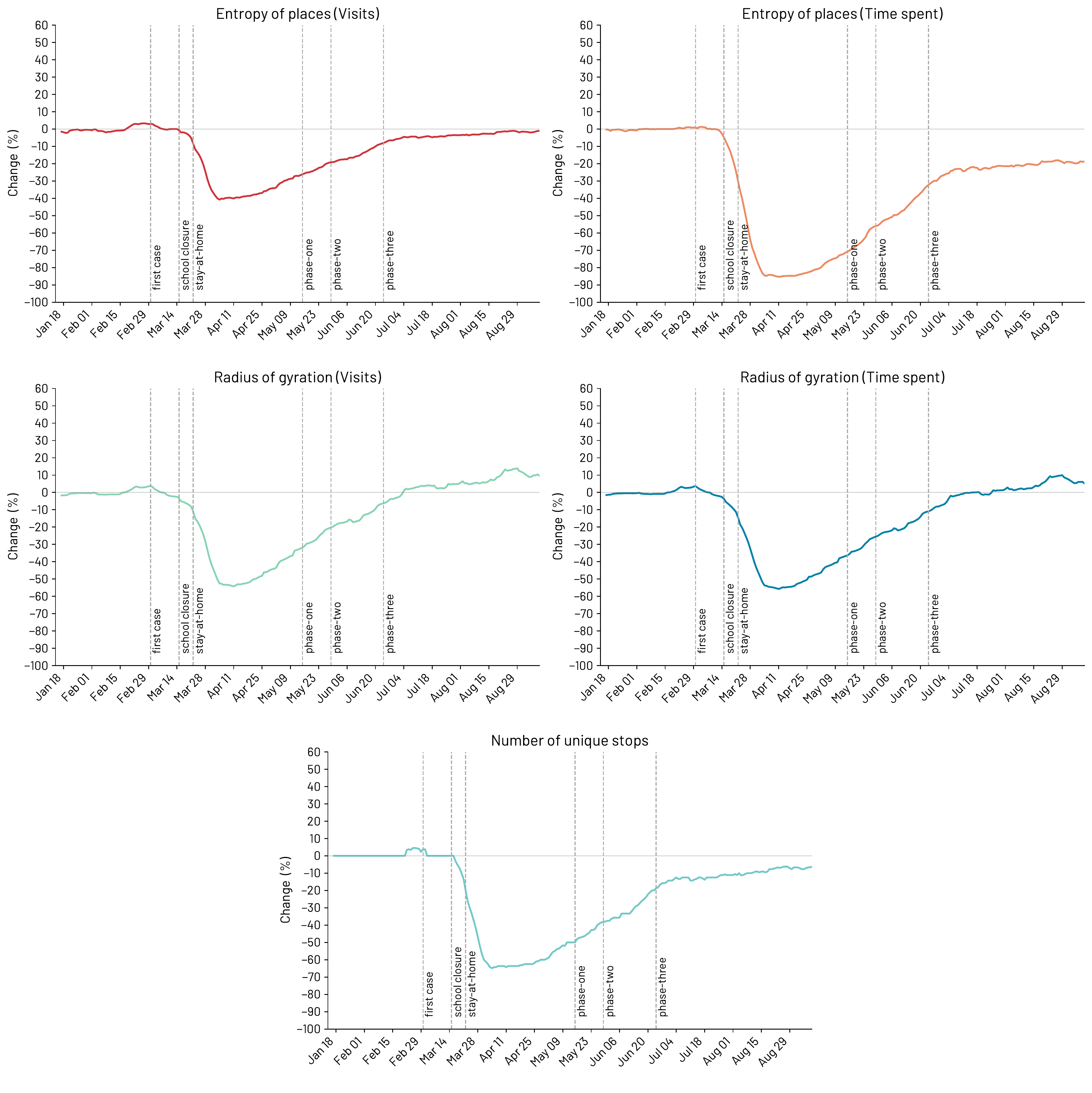}
	\caption{Mobility behaviour change for individuals in the New York state. Median change in entropy of places (visits), entropy of places (time spent), radius of gyration (visits), radius of gyration (time spent), and number of unique stops computed on a 2-weeks window with 1-day shift}
	\label{fig:beh_change_NY}
\end{figure}

As shown in Fig~\ref{fig:beh_change_NY}, for each individual, we computed the following quantities:
\begin{itemize}
	\item Number of unique stop locations;
	\item Diversity of visited locations weighted by the number of visits to an individual's locations; 
	\item Diversity of visited locations weighted by the time spent in an individual's locations; 
	\item Radius of gyration weighted by the number of visits to an individual's locations; 
	\item Radius of gyration weighted by the time spent in an individual's locations. 
\end{itemize}

We then aggregate these metrics, averaging over all individuals. Finally, we compute the percentage change for each metric $m$  as: 
$$m_{change} = \frac{m_{w(t)} - m_{b}}{m_{b}} * 100$$ where $m_b$ is the median of the aggregated metric $m$ during the baseline period before the pandemic, and $m_{w(t)}$ is the median of the aggregated metric $m$ for a window $w(t)$ for all individuals.

To compute the percentage of change for all the metrics we described, we used a time window of 2 weeks that we shift by 1 day, setting a baseline period that starts on 3 January 2020 and ends on 29 February 2020. 
We centre the last window at 29 February 2020, so that the last window included in the baseline period ends on 7 March 2020 (still before the stay-at-home measures).

\subsection{Entropy of visited locations}\label{sec:entropy}
We introduce the entropy of visited locations as a measure of the diversity of an individual's patterns of visit:
$$ D_{locations}(i)= -\frac{\sum_{j=1}^k p_{ij} \log(p_{ij})}{\log k}.$$
Here, $k$ is the total number of unique visited locations of an individual $i$, $p_{ij} = \frac{V_{ij}}{\sum_{j=1}^k V_{ij}}$ and $V_{ij}$ is either (i) the number of times individual $i$ visits location $j$ (weighted by number of visits), or (ii) the total time individual $i$ spend in location $j$ (weighted by time spent).

\subsection{Radius of gyration}\label{sec:radius_gyration}
The radius of gyration \cite{gonzalez2008understanding, pappalardo2015returners} is defined as:

$$r_g = \sqrt{\frac{1}{N} \sum_{l \in L} n_l|\mathbf{r_l} - \mathbf{r_{cm}}|^2}$$

Depending on the type of weighting we apply, $N$ is the total number of visits (or total time spent) a particular individual made to all their visited locations; $n_l$ are the visits (time spent) to location $l$; $L$ is the set of stop locations within a time window,; $\mathbf{r_l}$ is a two-dimensional vector representing the location’s GPS position recorded as latitude and longitude; and $\mathbf{r_{cm}}$ is the center of mass of the trajectories, defined as $\mathbf{r_{cm}} = \frac{1}{N} \sum_{l=1}^{N} \mathbf{r_l}$.

\section{Mobility routines}\label{SI:sec:sequitur_supplementary}

\subsection{Comments on significant routines}
\label{SI:significant_routines}
The Sequitur compression algorithm is used in this work as a tool for detecting recurrent patterns in the sequence of location type visits \cite{nevill1997identifying}. However, since we are interested in understanding how individual routines changed as the pandemic unfolded, it is important to ensure that the recurrent patterns identified are not spurious and represent meaningful patterns in terms of individual routines' description power. Given the nature of mobile phone data, random sequences can be erroneously identified as relevant routines. Following the approach of Di Clemente \textit{et al.}~\cite{di2018sequences}, we remove routines whose number of occurrences in the original sequence, if compared to a set of randomized versions of the original sequence, does not significantly differ. This process consists of generating $1\,000$ copies of each individual's original sequence of visits (including their homes, workplaces, uncategorized locations, and all venue categories following the temporal order of visits) and independently shuffling each of them. We then compare the occurrences of recurrent patterns identified from the original routine with the average occurrence they show in the randomized sequences. Operationally, we compute the z-score, using each recurrent pattern standard deviation and their average number of occurrences, and keep as "significant routines" only those with a score greater than 2.

While this procedure ensures that the routines we are analyzing are representative of an individual's recurrent patterns and thus help in understanding how deeply rooted behaviours were affected by the pandemic, it is also of interest to understand the changes that occurred to the overall sequences of visits. This is performed in the next section with a thorough analysis of the direct compression of the original sequence for the two time windows under study. 

\subsection{Sequitur compression ratio}
\label{SI:sequitur_compression}
\begin{figure}[!ht]
	\centering
	\includegraphics[width=0.7\textwidth]{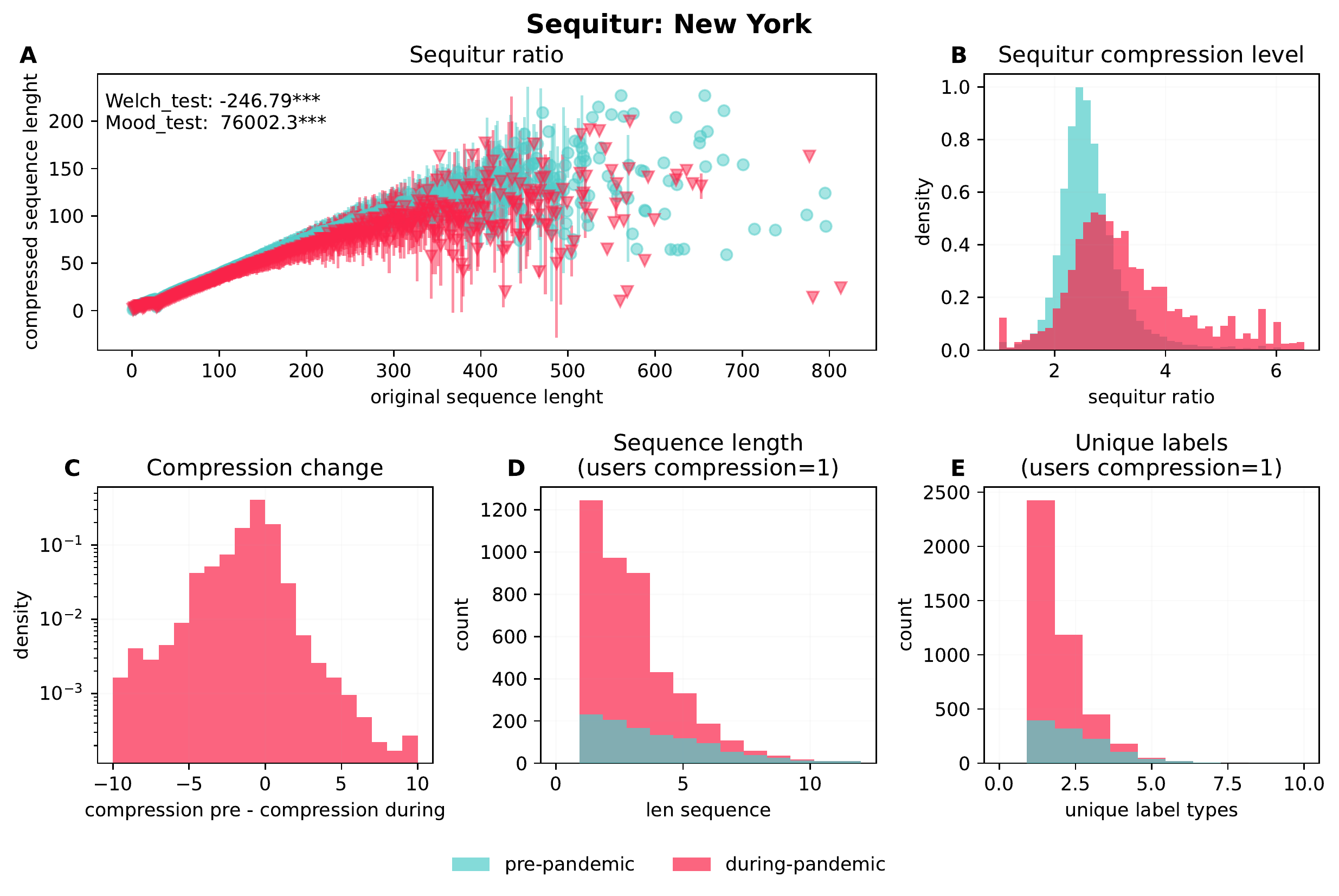}
	\caption{Sequitur compression for individuals living in the New York state between two 28-days periods namely pre- and during-pandemic. 
		A) scatters the median (and 95\% C.I.) of compressed sequences' length as a function of their original sequence length.
		B) shows the distribution of the sequitur ratio while C) their element-wise difference (pre-pandemic compression ratio minus its value during-pandemic). D) and E) show the number of users with $compression\_ratio = 1$ as a function of sequence length and unique stop locations respectively.}
	\label{SI:fig:sequitur_ratios_NY}
\end{figure}

\begin{figure}[!ht]
	\centering
	\includegraphics[width=0.7\textwidth]{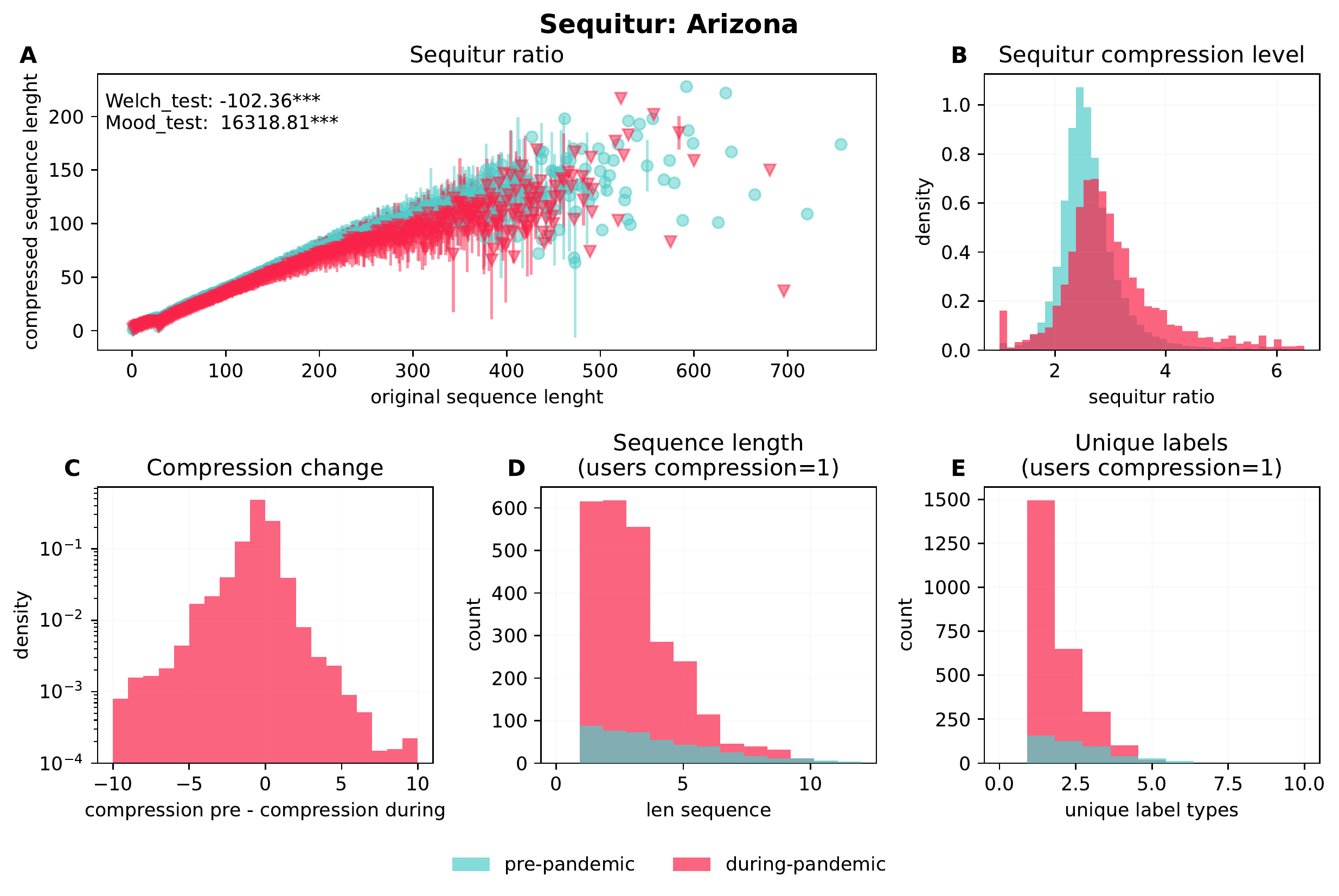}
	\caption{Sequitur compression for individuals living in Arizona between two 28-days periods namely pre- and during-pandemic. 
		A) scatters the median (and 95\% C.I.) of compressed sequences' length as a function of their original sequence length.
		B) shows the distribution of the sequitur ratio while C) their element-wise difference (pre-pandemic compression ratio minus its value during-pandemic). D) and E) show the number of users with $compression\_ratio = 1$ as a function of sequence length and unique stop locations respectively.}
	\label{SI:fig:sequitur_ratios_AZ}
\end{figure}

\begin{figure}[!ht]
	\centering
	\includegraphics[width=0.7\textwidth]{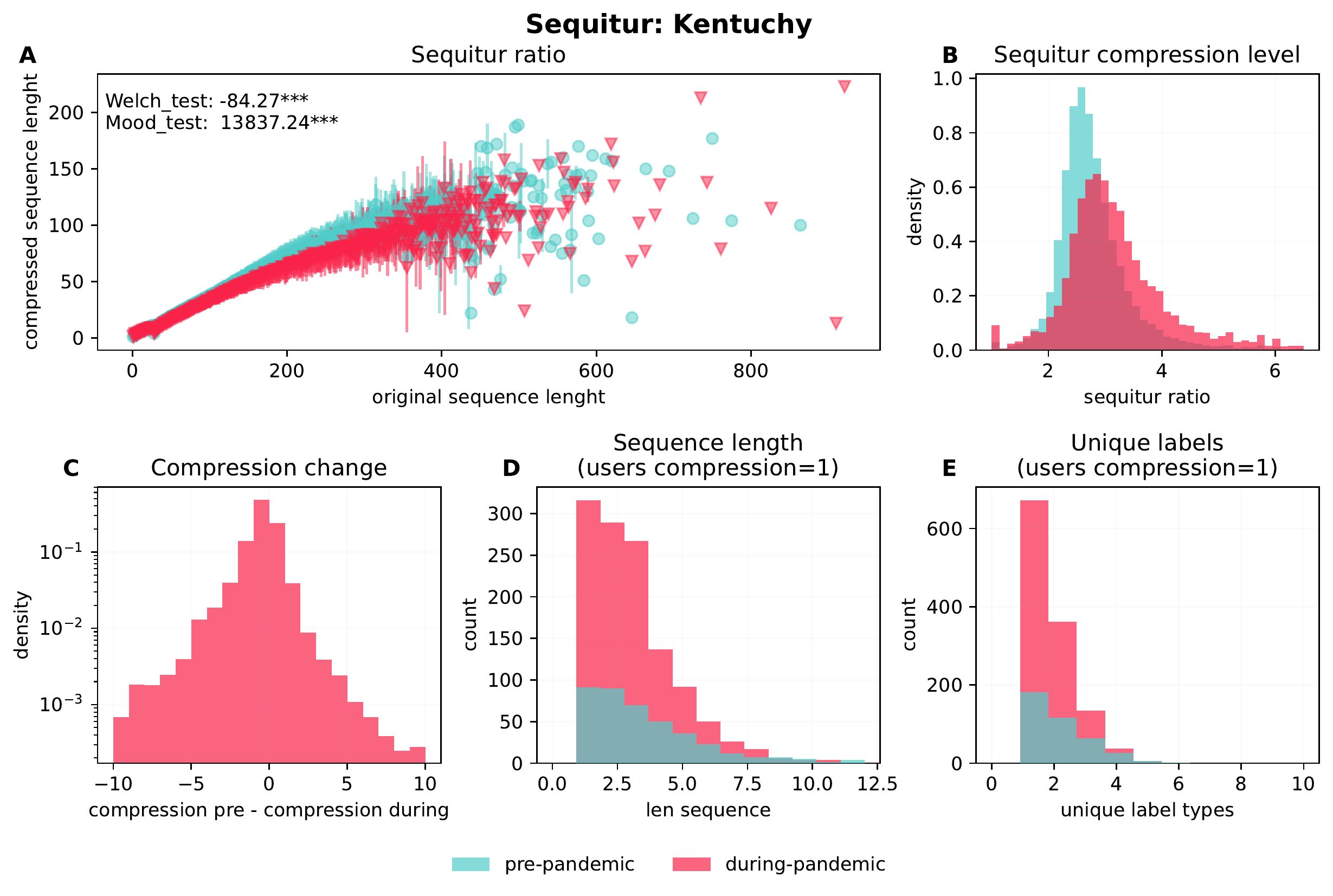}
	\caption{Sequitur compression for individuals living in Kentucky between two 28-days periods namely pre- and during-pandemic. 
		A) scatters the median (and 95\% C.I.) of compressed sequences' length as a function of their original sequence length.
		B) shows the distribution of the sequitur ratio while C) their element-wise difference (pre-pandemic compression ratio minus its value during-pandemic). D) and E) show the number of users with $compression\_ratio = 1$ as a function of sequence length and unique stop locations respectively.}
	\label{SI:fig:sequitur_ratios_KY}
\end{figure}

\begin{figure}[!ht]
	\centering
	\includegraphics[width=0.7\textwidth]{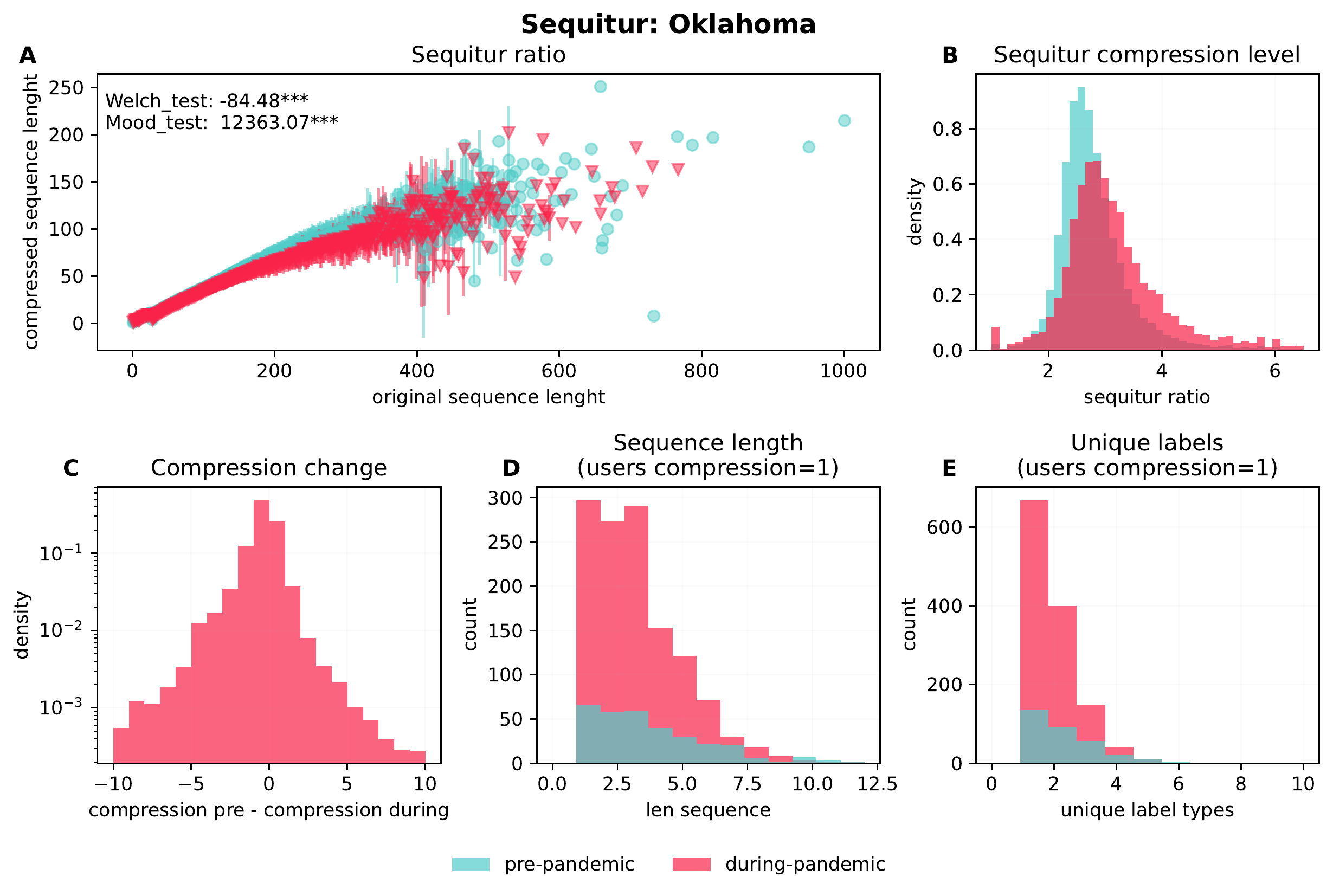}
	\caption{Sequitur compression for individuals living in Oklahoma between two 28-days periods namely pre- and during-pandemic. 
		A) scatters the median (and 95\% C.I.) of compressed sequences' length as a function of their original sequence length.
		B) shows the distribution of the sequitur ratio while C) their element-wise difference (pre-pandemic compression ratio minus its value during-pandemic). D) and E) show the number of users with $compression\_ratio = 1$ as a function of sequence length and unique stop locations respectively.}
	\label{SI:fig:sequitur_ratios_OK}
\end{figure}

As reported in the main text, encoding individual visit patterns in a sequence of visited categories help us to better understand the extent to which human behaviour adapted to the pandemic threat under different environmental circumstances. \Cref{SI:fig:sequitur_ratios_NY}, \Cref{SI:fig:sequitur_ratios_AZ}, \Cref{SI:fig:sequitur_ratios_KY}, and \Cref{SI:fig:sequitur_ratios_OK} show the results of the compression process, for the four states included in our analysis, highlighting the difference between sequences in the pre-pandemic period (from February 1, 2020, to February 28, 2020) and in the during-pandemic period (from March 25, 2020, to April 21, 2020). We find a significant reduction in sequence length from the pre-pandemic period to the during-pandemic one for all four states. This result is mirrored in the length of the compressed sequences (see panel A in \Cref{SI:fig:sequitur_ratios_NY}, \Cref{SI:fig:sequitur_ratios_AZ}, \Cref{SI:fig:sequitur_ratios_KY}, and \Cref{SI:fig:sequitur_ratios_OK}). We summarize this result by computing the compression ratio, which is defined as the original sequence length, $l_o$, divided by the compressed sequence length, $l_c$, for each individual in both periods. We provide evidence of a significant increment of sequence compressibility of during-pandemic visit patterns when compared with the pre-pandemic period (under both a Welch test and Mood test at a significance level of $\alpha=0.001$)  (panels C and D).
We also note that most of the non-compressible sequences (i.e., with compression ratios equal to 0) are short: during-pandemic, in $71\%$ of the cases the length is smaller than four. Importantly, these short sequences are mostly composed by a single stop category: during-pandemic, in $71\%$ of the cases this category corresponds to an individual's "Residential location".

\subsection{Routine reduction}
\begin{figure}[!ht]
	\centering
	\includegraphics[width=0.8\textwidth]{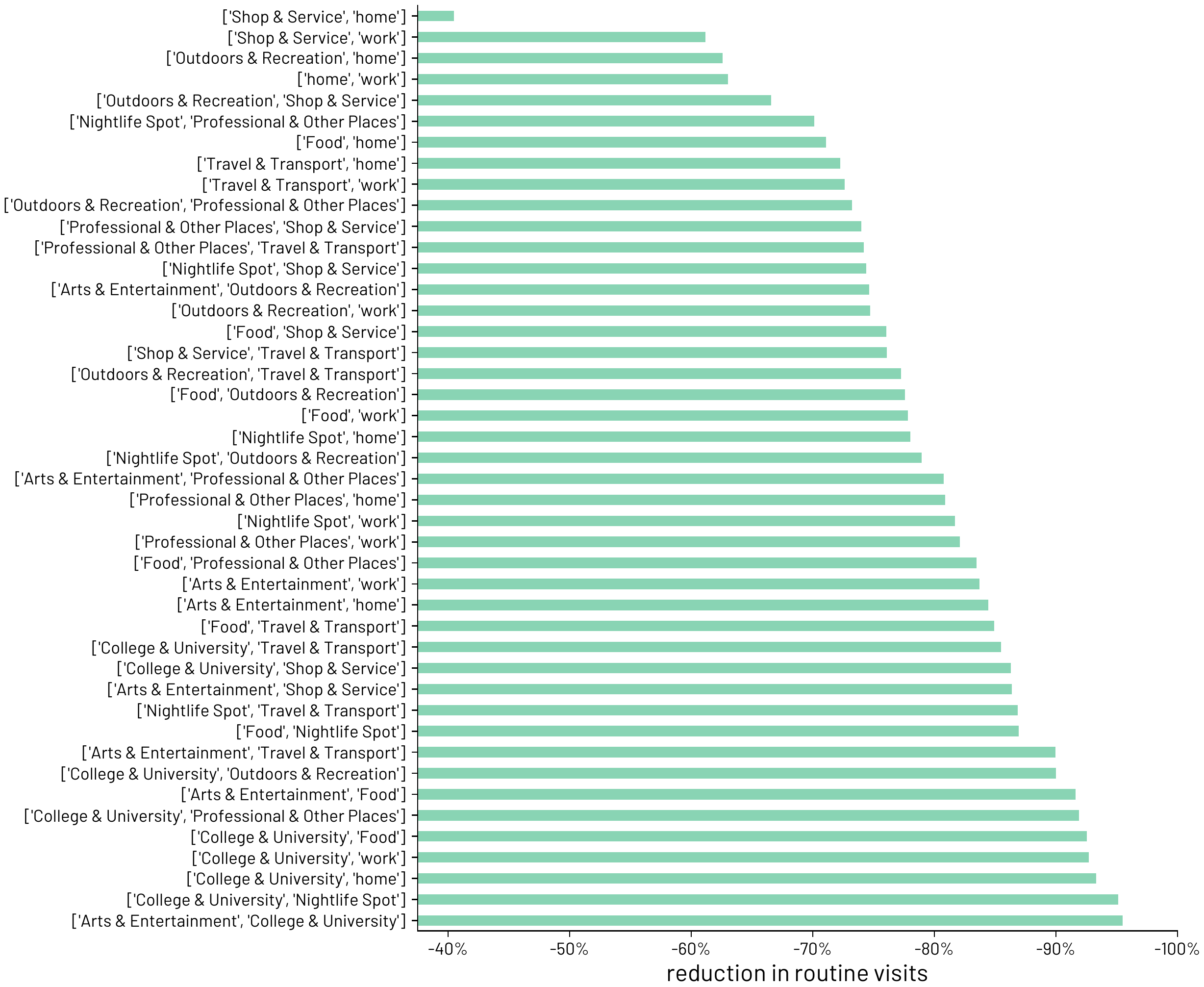}
	\caption{Undirected routine reductions.}
	\label{fig:SI_routine_reduction_analysis}
\end{figure}

From the pre-pandemic to the during-pandemic period, we find a global decrease in all the mobility routines. 
In \Cref{fig:SI_routine_reduction_analysis} we report the reductions faced by each link. \emph{College \& Universities} closure exposed all the activities linked to that category to the largest reduction. A similar behaviour hit the \emph{Arts and Entertainment} category due to the restrictions and closures of museums and other cultural places.

\subsection{Jaccard similarity and clustering}\label{SI:clustering_distance_silhouette}
\begin{figure}[!ht]
	\centering
	\includegraphics[width=0.8\textwidth]{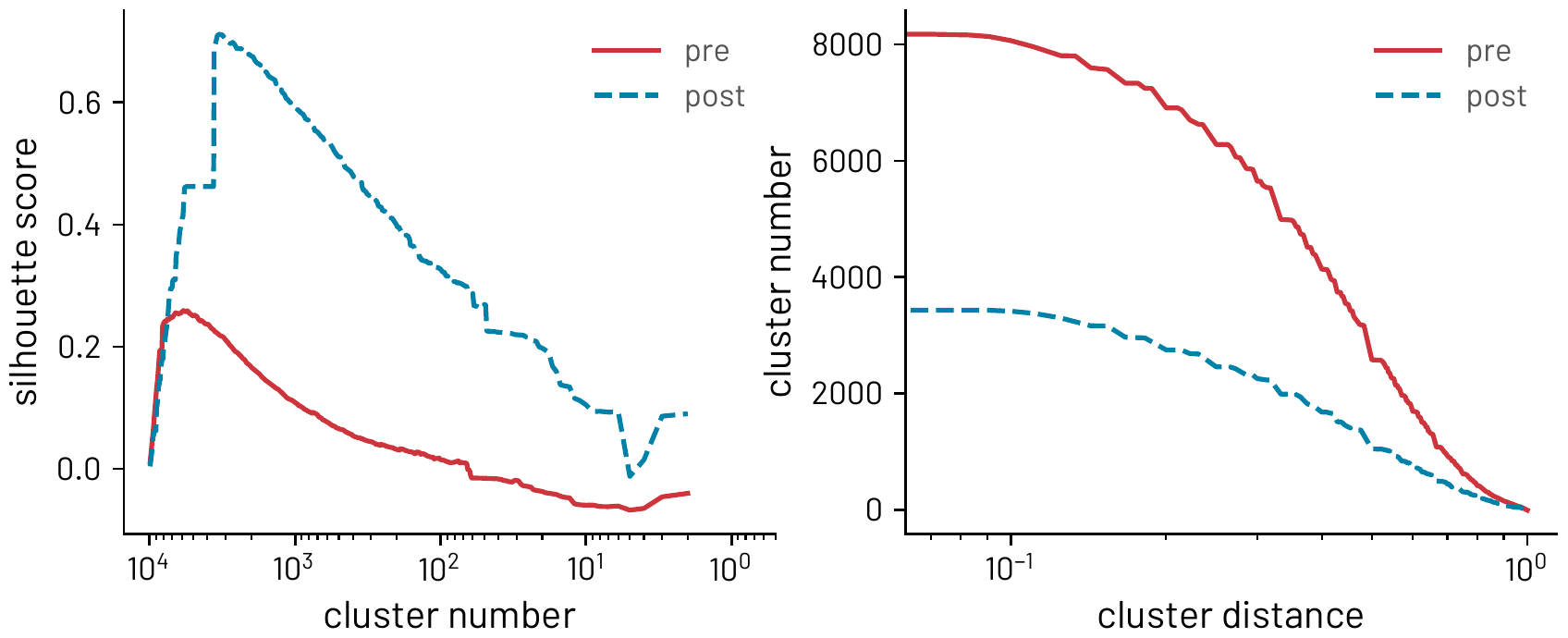}
	\caption{Clustering pre and during the pandemic. A) shows the silhouette score as a function of the number of clusters hierarchically assembled. The red line represents the silhouette score in the pre-pandemic period, while the blue one during the pandemic. B) shows the number of clusters as a function of the largest distance between aggregated components.}
	\label{SI:fig:clustering_silhouette}
\end{figure}

The silhouette score is a metric that compares an object cohesion to its own cluster with its distance from other clusters. Silhouette score ranges from -1 to +1. A high value indicates a good match and in general a well matched set of cluster components. For each unit in a system, silhouette is computed as the average distance of one unit with all other units outside its own cluster $\bar{d}_{out}(i)$ minus the average distance of the same unit with the other components from its own cluster $\bar{d}_{in}(i)$: $$s(i)=\frac{\bar{d}_{out}(i)-\bar{d}_{in}(i)}{\max{\bar{d}_{out}(i),\bar{d}_{in}(i)}}.$$
Figure~\ref{SI:fig:clustering_silhouette} shows in panel A the average silhouette score, for the state of New York, as a function of the number of clusters as the hierarchical process progresses. As expected the clustering process is able to better group different individuals' routines during the pandemic period at all steps of the agglomerative process. Panel B shows the number of clusters as a function of the relative distance between the two elements involved in the latest clustering step. At the same distance values, we show that clustering of routines during the pandemic period have been aggregate twice as fast as during the pre-pandemic period.

\subsection{Top clusters}
\label{SI:sec:topclusters}
\begin{figure}[!ht]
	\centering
	\subfloat[Cluster 1, 24\% users]{\includegraphics[width=0.31\textwidth]{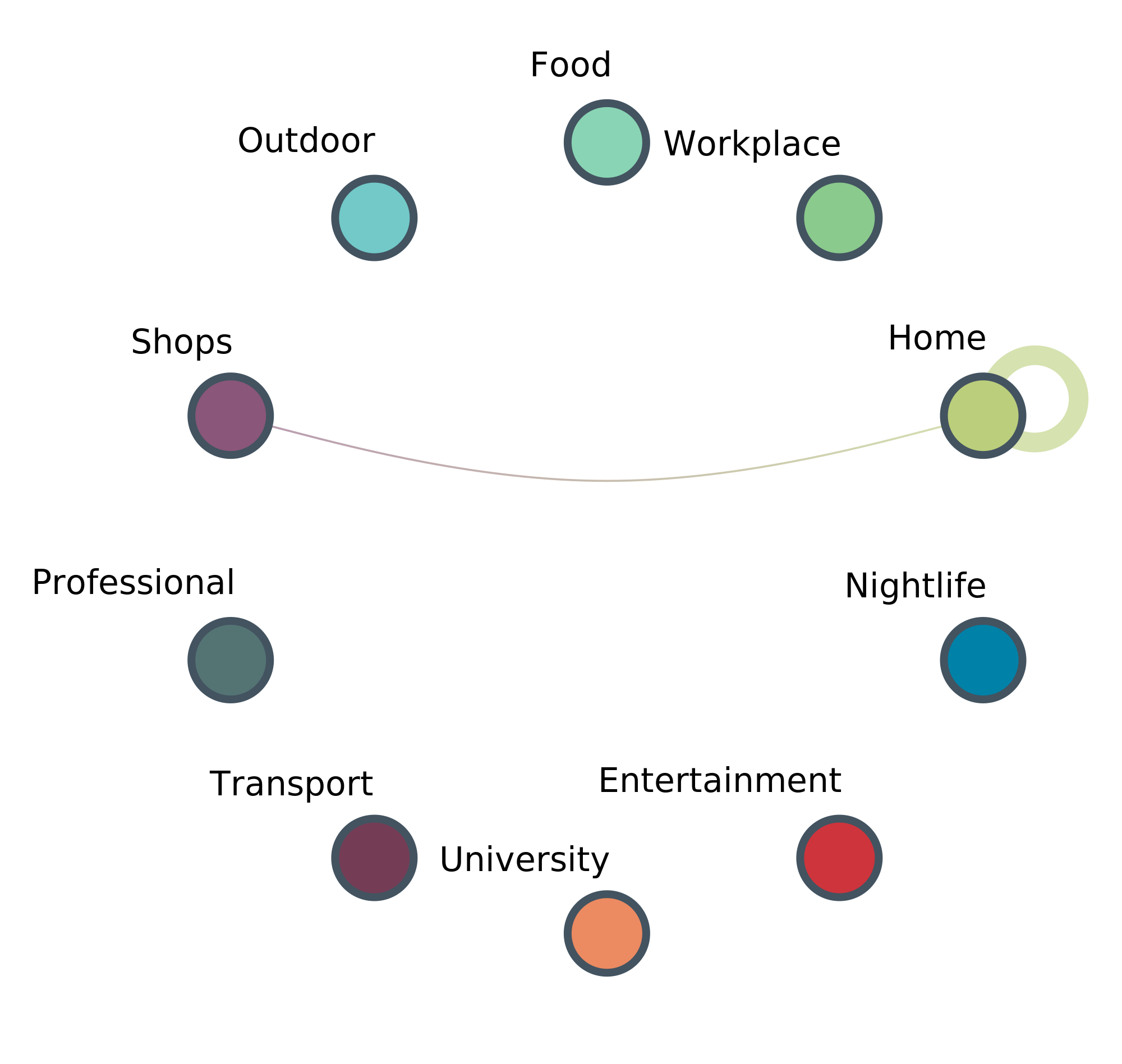}}
	\subfloat[Cluster 2, 9.6\% users]{\includegraphics[width=0.31\textwidth]{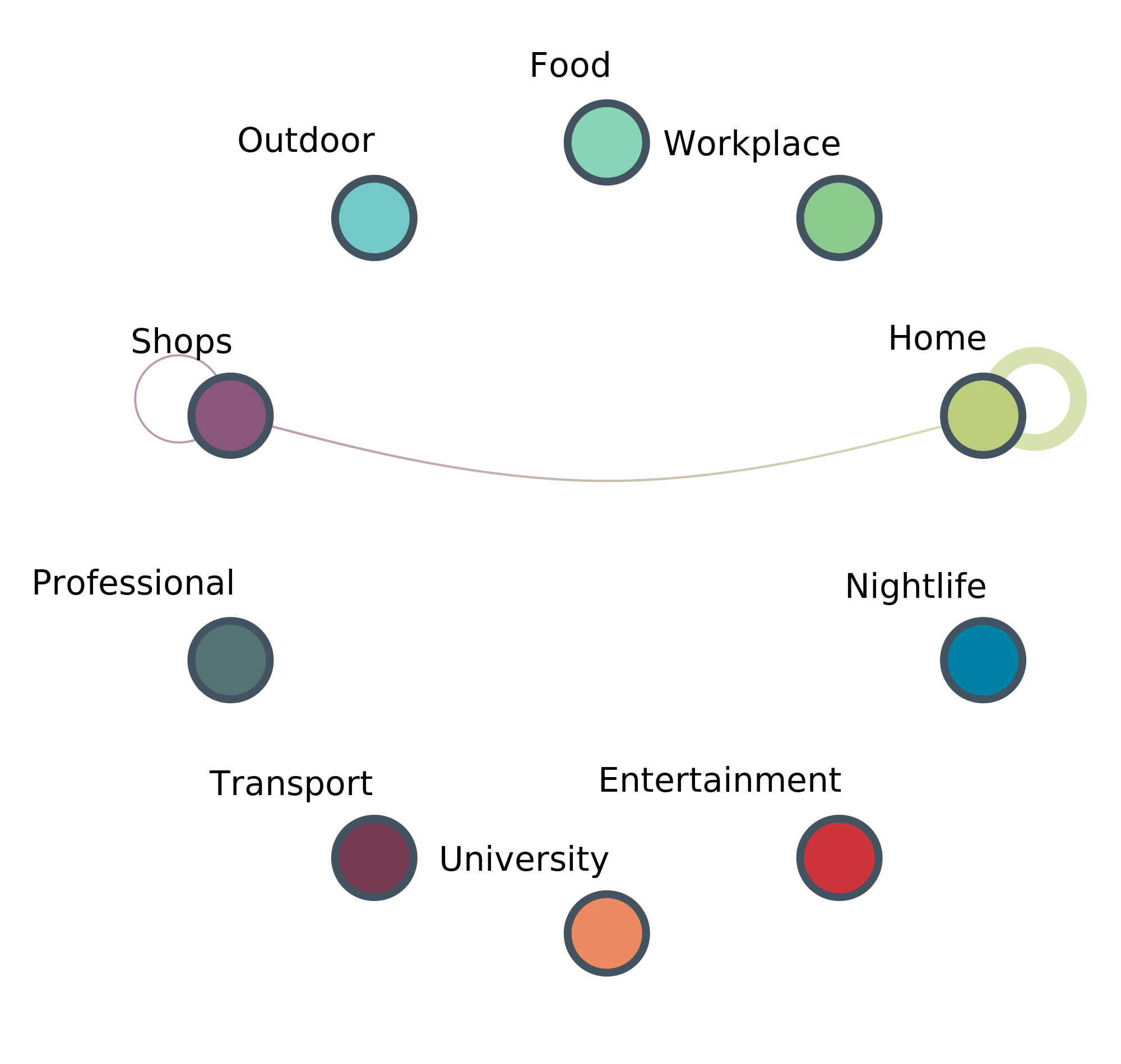}}
	\subfloat[Cluster 3, 6.6\% users]{\includegraphics[width=0.31\textwidth]{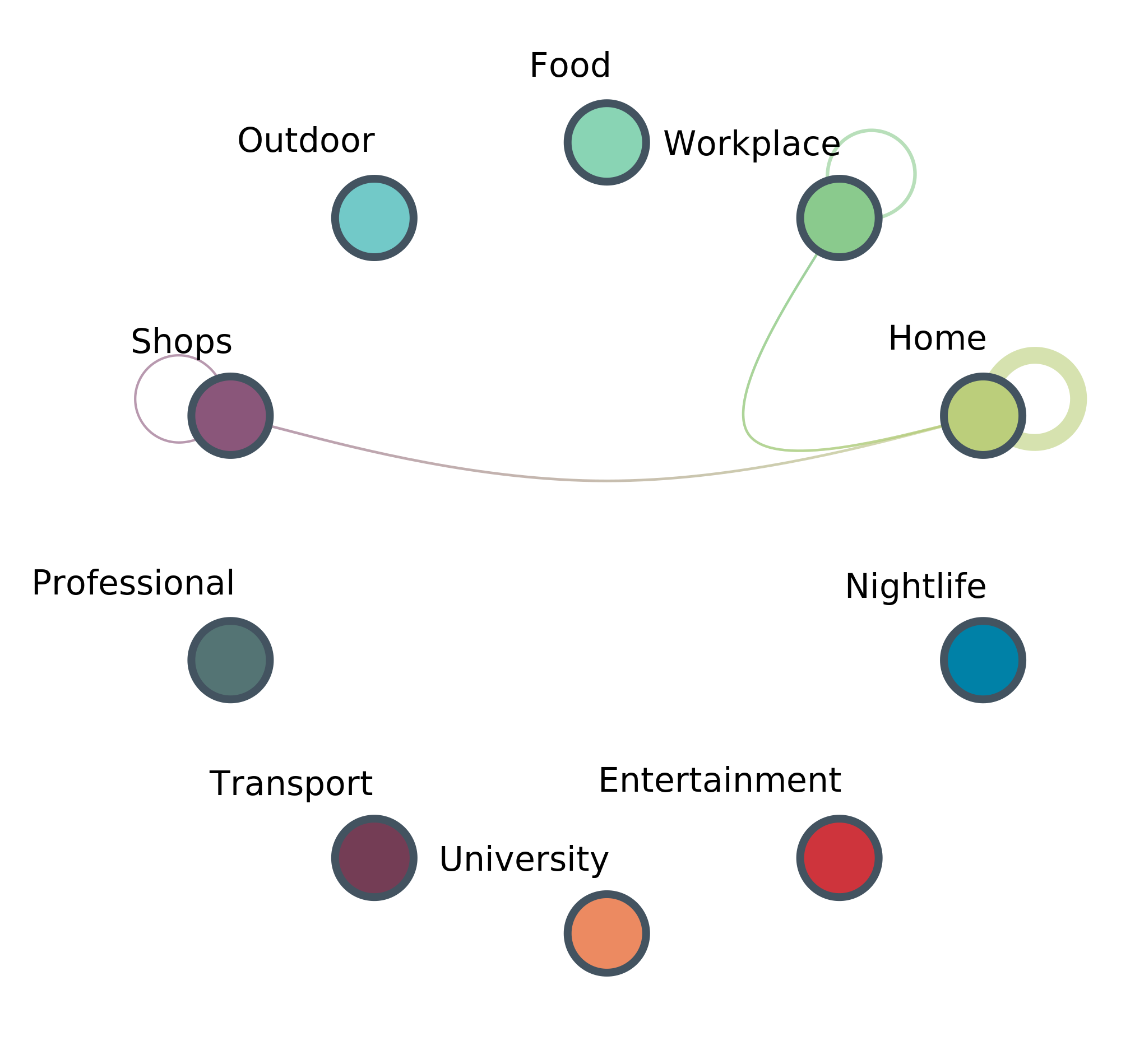}}\\
	\subfloat[Cluster 4, 6.3\% users]{\includegraphics[width=0.31\textwidth]{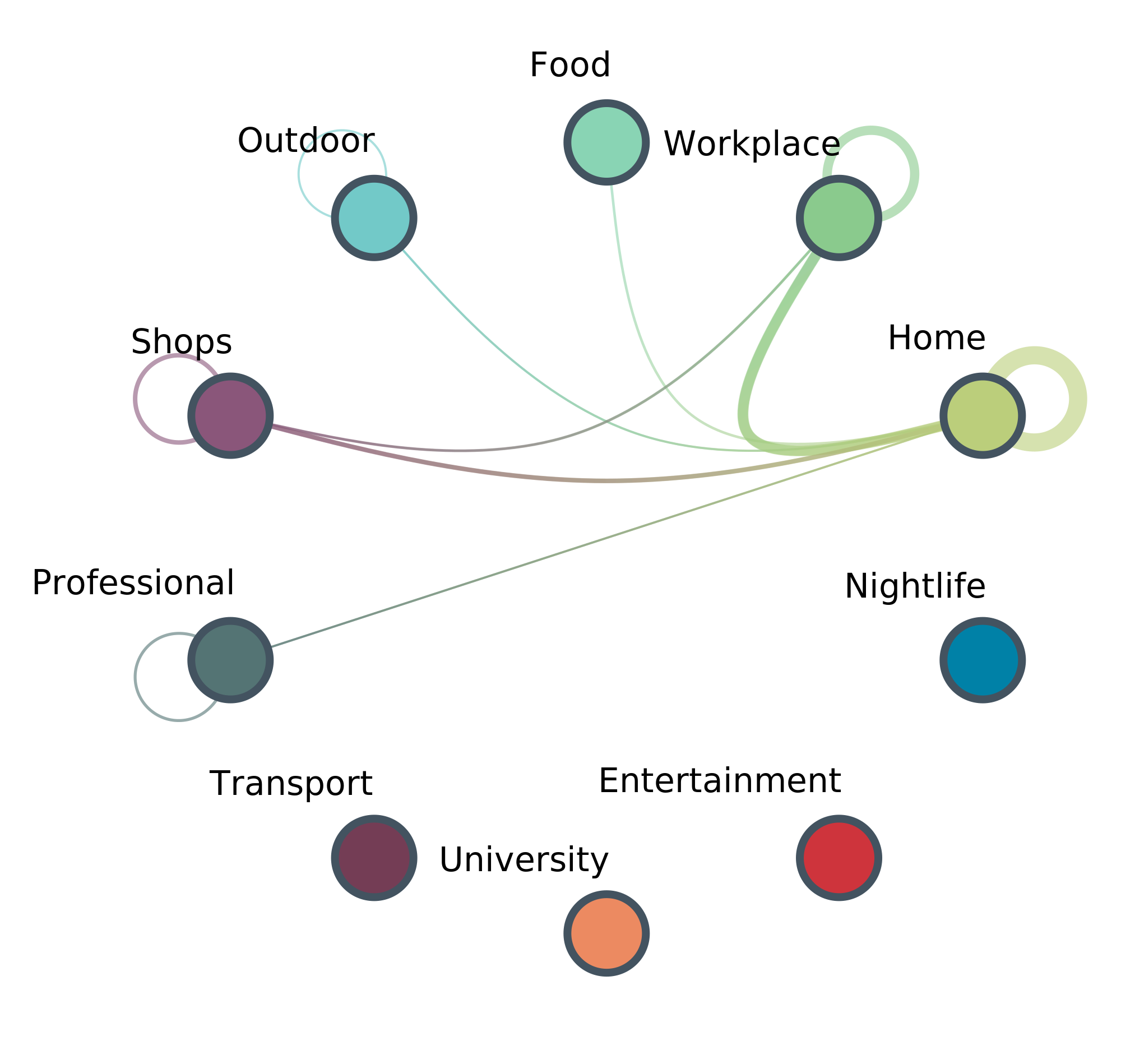}}
	\subfloat[Cluster 5, 5.0\% users]{\includegraphics[width=0.31\textwidth]{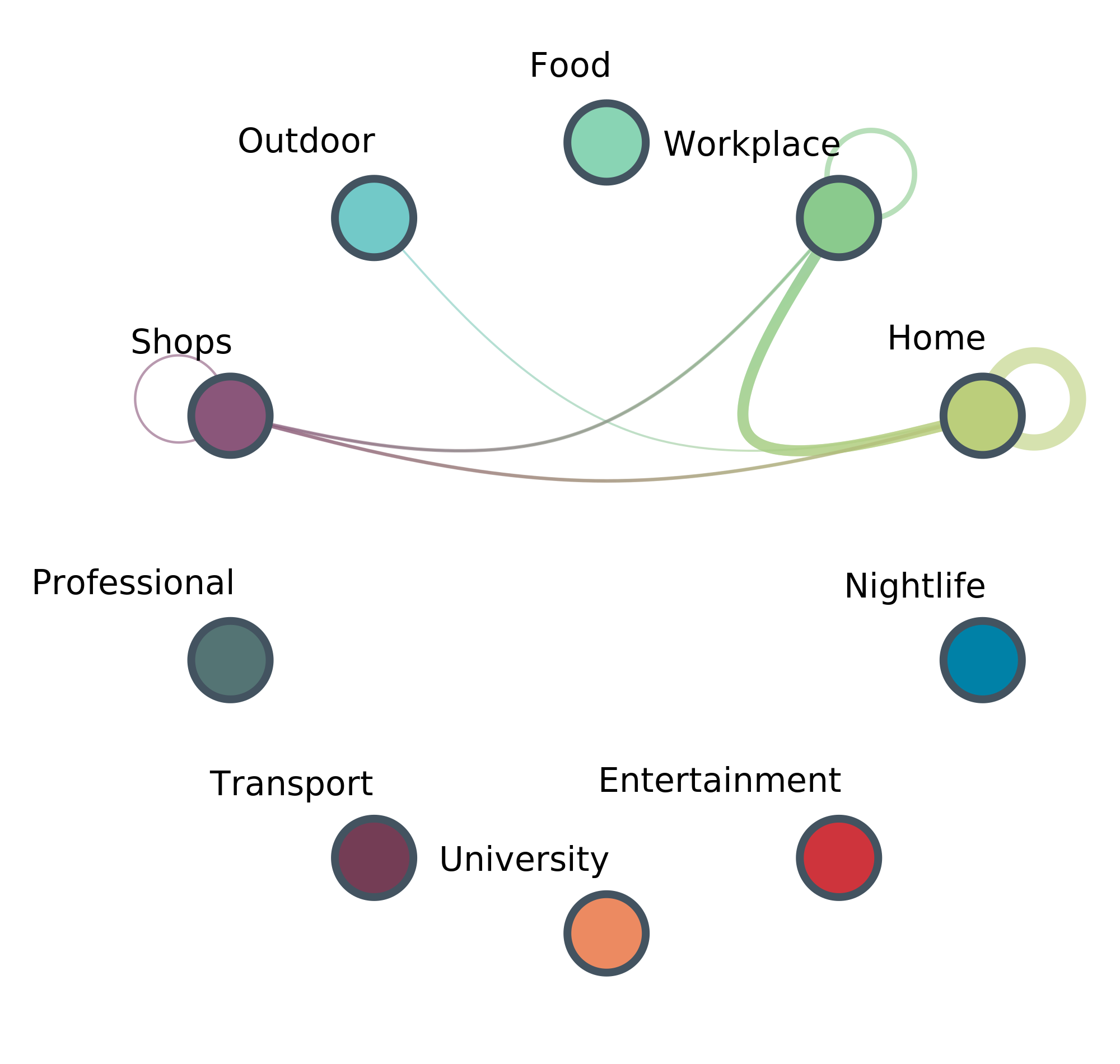}}
	\subfloat[Cluster 6, 4.9\% users]{\includegraphics[width=0.31\textwidth]{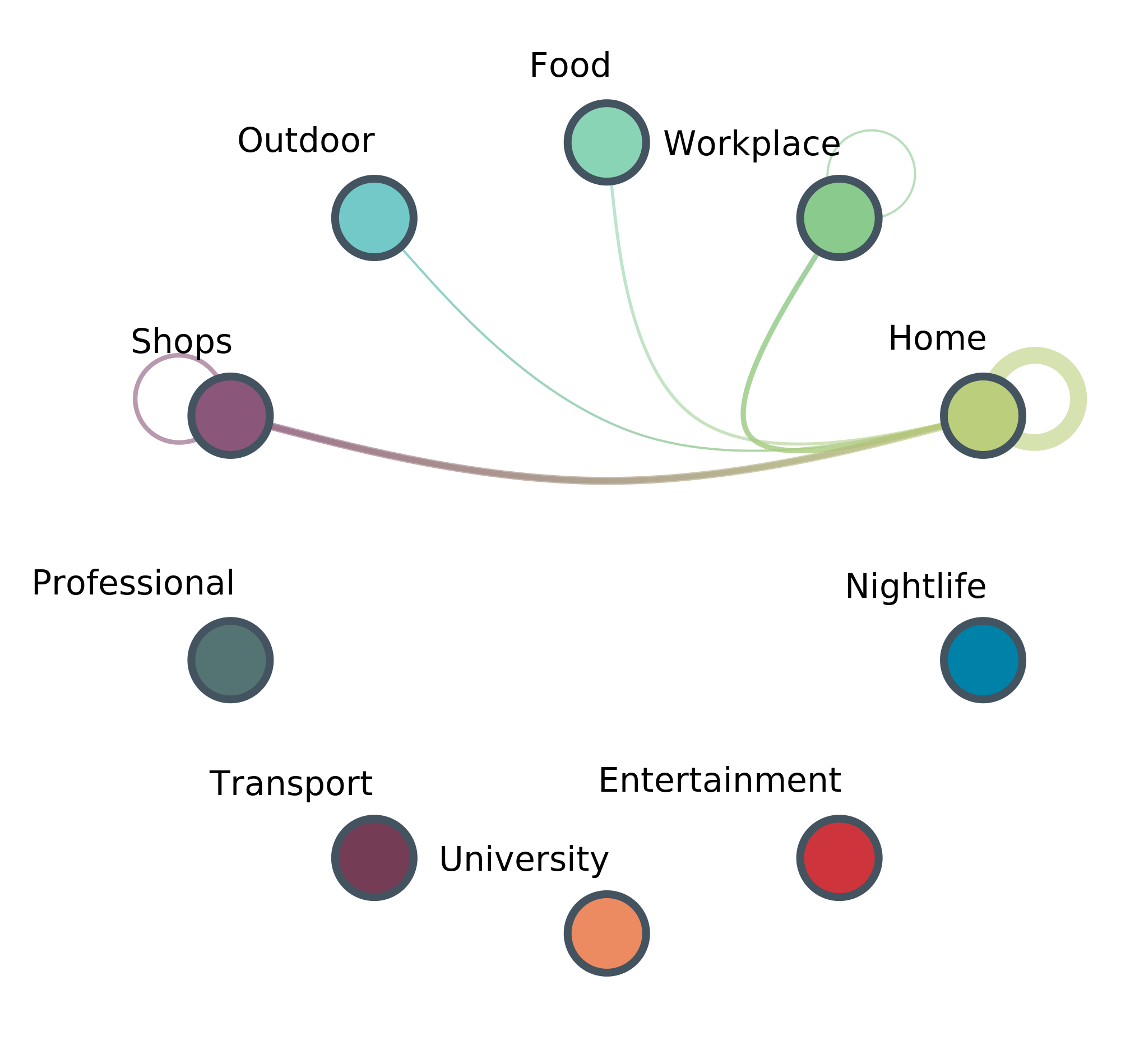}}
	\caption{The top six clusters with most individuals in the post-pandemic period.}
	\label{fig:top_clusters}
\end{figure}

To better understand the routine characteristics of people during the pandemic, we display in \Cref{fig:top_clusters} the network of subsequent visits, extracted from the top six clusters defined from the routine activity of Sequitur.

We select the clusters to visualize by selecting the top six clusters at the 95\% height of the dendogram formed by the complete linkage agglomerative clustering described in the manuscript. For clarity reasons, we filter out the links with less than 5\% of intensity.

\Cref{fig:top_clusters} shows that most of the users, which belong to Cluster 1 (24\%) and Cluster 2 (9.6\%), have their significant locations between \emph{Residential} and \emph{Shops \& Services}. Other clusters include more diversity in mobility, including also the work places.

\section{Co-location events}\label{SI:sec:co-locations}

Two individuals are said to be "co-located" if they both have a stop location that is at most
50 meters far from the other one, and these stop locations are overlapping on date and time for at least 15 minutes. 
A "co-location event" is thus defined as one event in time when two or more individuals are co-located. 

These \emph{co-location events} are grouped
in four different categories depending on the place where the possible social contact took place: 
\begin{itemize}
	\item \textit{Residential}, a \emph{co-location event} where one and only one of the two individuals have the venue marked as Residential location;
	\item \textit{Workplace}, a \emph{co-location event} that happened in a venue labeled as a workplace for both the individuals;
	\item \textit{POI}, a \emph{co-location event} where both the individuals are in the same POI (same ID of the POI);
	\item \textit{Other}, a \emph{co-location event} in which the two individuals meet in a place that it is neither a \textit{Residential} nor a \textit{Workplace} nor a \textit{POI}.
\end{itemize}

\subsection{Percent change of the number of Co-location events}  
\label{SI:sec:co-location-number}
\begin{figure}[!ht]
	\centering
	\includegraphics[width=\textwidth]{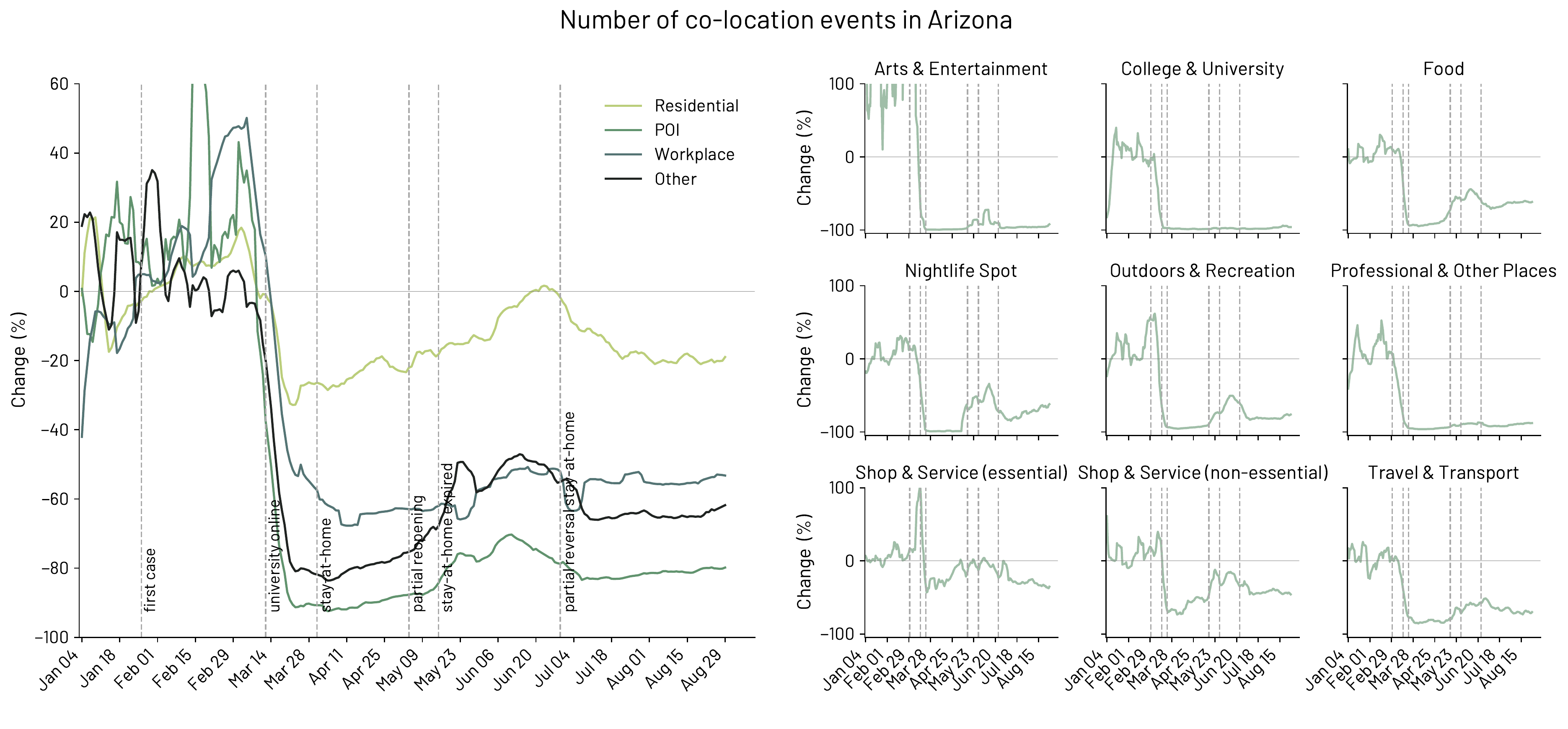}
	\caption{Percent change in the number of co-location events from the baseline period (until 29 February 2020) in Arizona. (left) Percent change for Residential areas (one of the two individuals in proximity is at their own residential area), POI (both individuals are at the same POI), Business (both of having the same work location) and Other co-location events. (right) Percent change for all the categories of POIs.}
	\label{SI:fig:contacts-AZ}
\end{figure}

\begin{figure}[!ht]
	\centering
	\includegraphics[width=\textwidth]{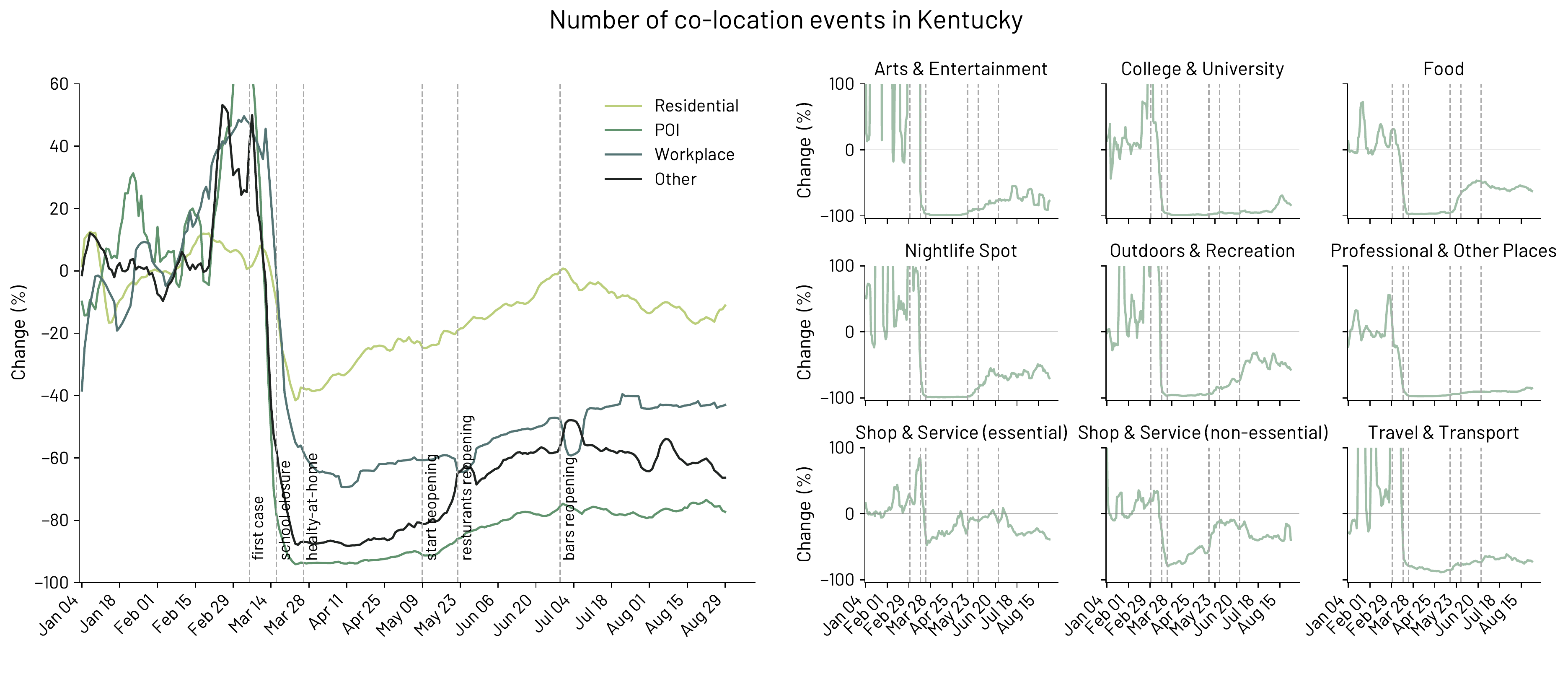}
	\caption{Percent change in the number of co-location events from the baseline period (until 29 February 2020) in the Kentucky. (left) Percent change for Residential areas (one of the two individuals in proximity is at their own residential area), POI (both individuals are at the same POI), Business (both of having the same work location) and Other co-location events. (right) Percent change for all the categories of POIs.}
	\label{SI:fig:contacts-KY}
\end{figure}

\begin{figure}[!ht]
	\centering
	\includegraphics[width=\textwidth]{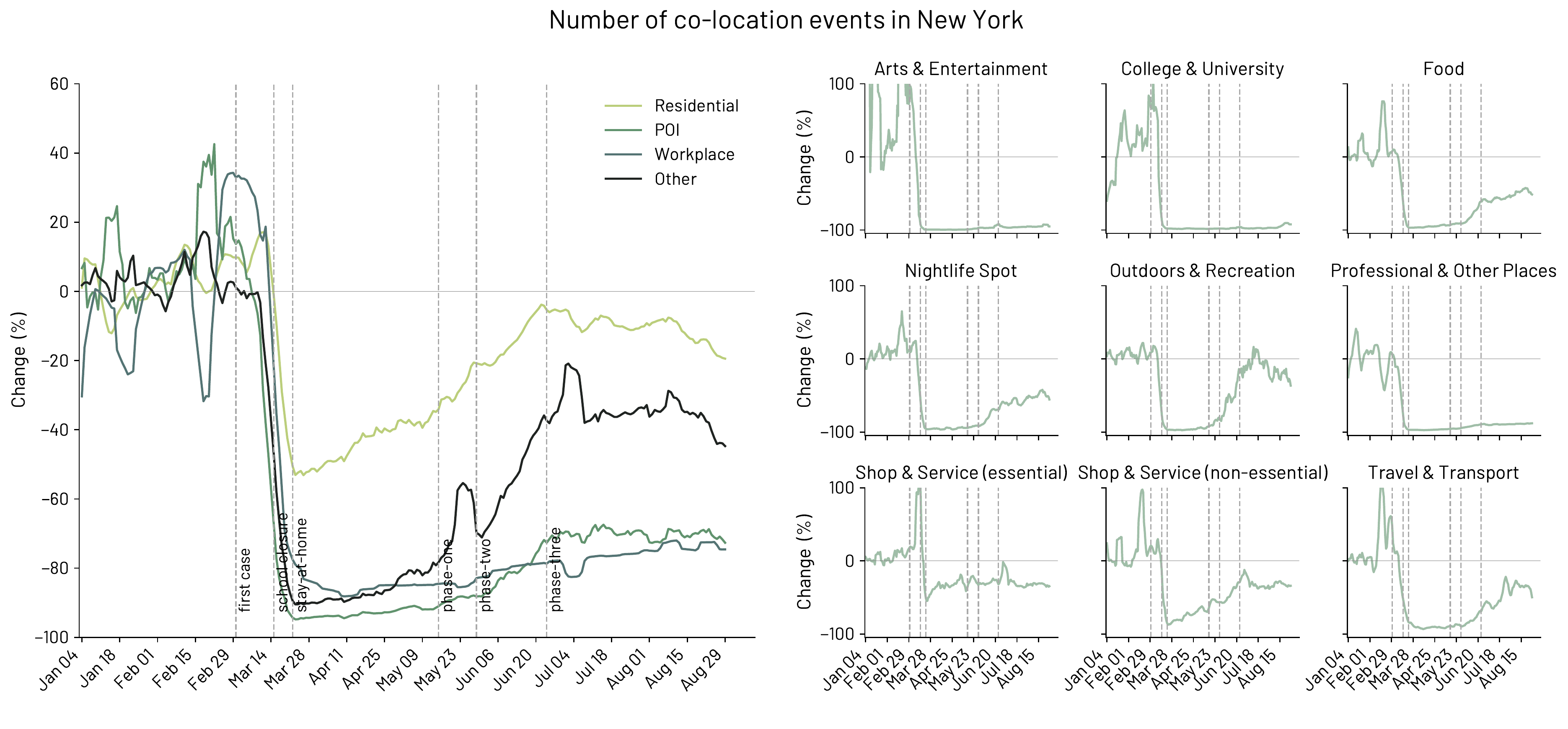}
	\caption{Percent change in the number of co-location events from the baseline period (until 29 February 2020) in the New York. (left) Percent change for Residential areas (one of the two individuals in proximity is at their own residential area), POI (both individuals are at the same POI), Business (both of having the same work location) and Other co-location events. (right) Percent change for all the categories of POIs.}
	\label{SI:fig:contacts-NY}
\end{figure}

\begin{figure}[!ht]
	\centering
	\includegraphics[width=\textwidth]{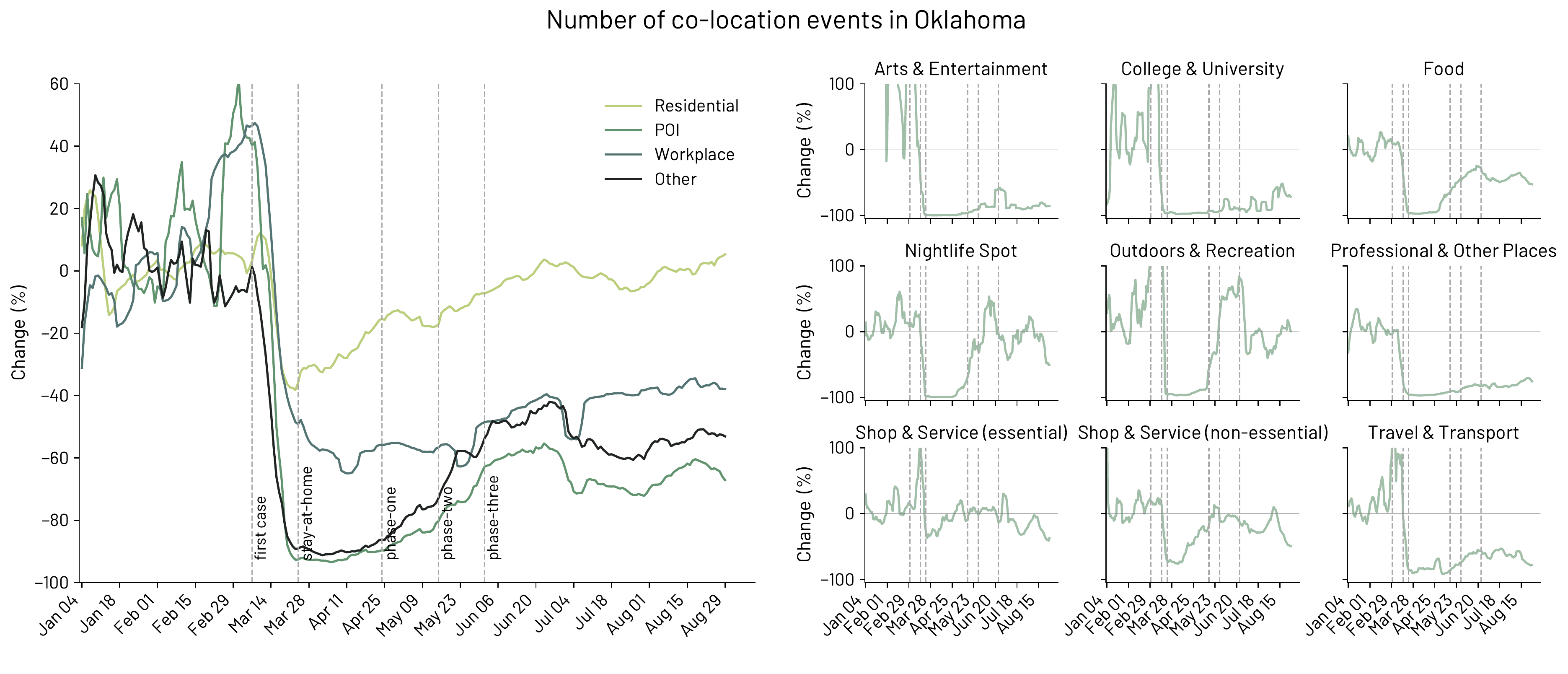}
	\caption{Percent change in the number of co-location events from the baseline period (until 29 February 2020) in the Oklahoma. (left) Percent change for Residential areas (one of the two individuals in proximity is at their own residential area), POI (both individuals are at the same POI), Business (both of having the same work location) and Other co-location events. (right) Percent change for all the categories of POIs.}
	\label{SI:fig:contacts-OK}
\end{figure}

\Cref{SI:fig:contacts-AZ}, \Cref{SI:fig:contacts-KY}, \Cref{SI:fig:contacts-NY} and \Cref{SI:fig:contacts-OK} show the percent change in the number of co-location events in the states of New York, Kentucky, Oklahoma and Arizona respectively.

\subsection{Percent change of the duration of Co-location events} 
\label{SI:sec:co-location-duration}
\begin{figure}[!ht]
	\centering
	\includegraphics[width=\textwidth]{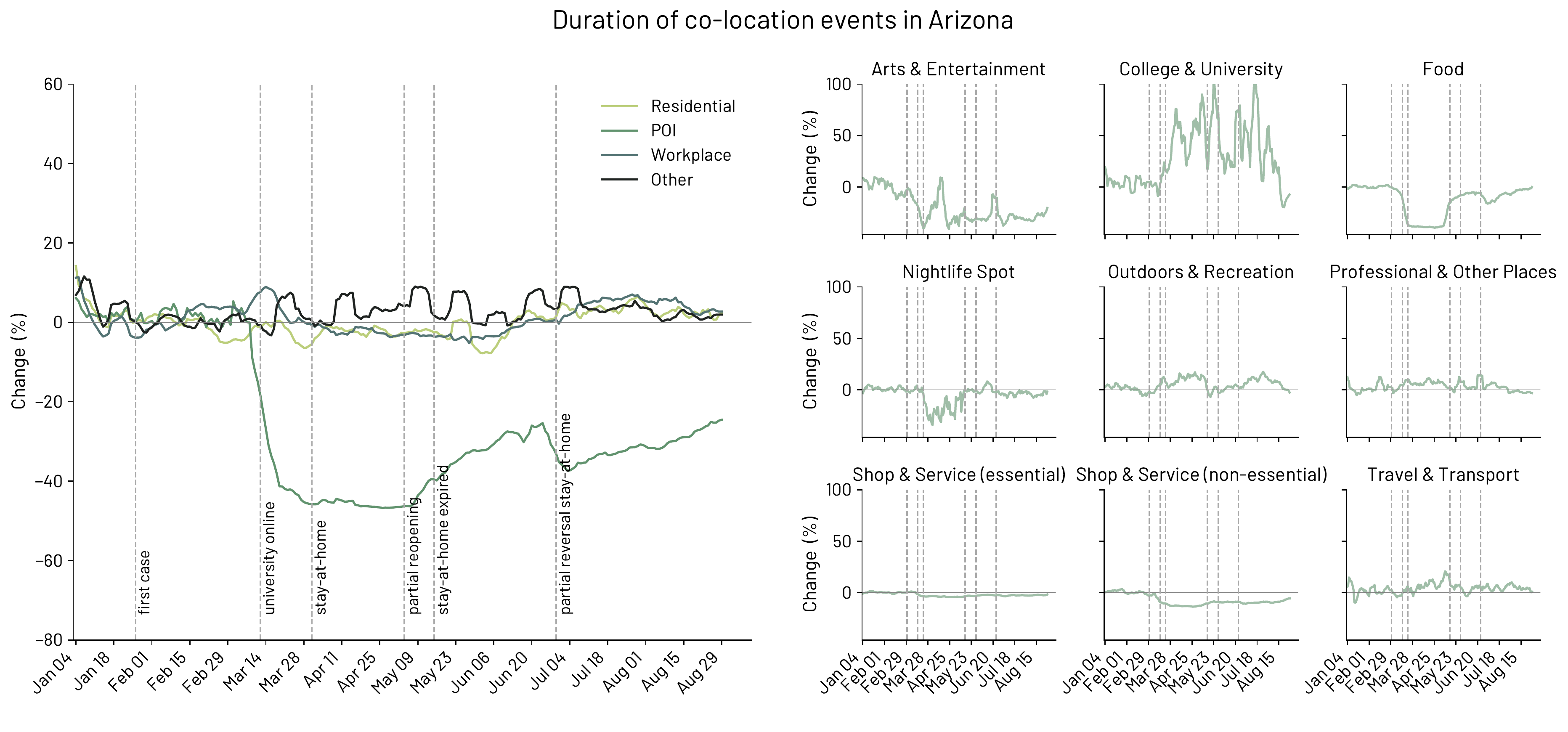}
	\caption{Percent change of the duration of co-location events from the baseline period (until 29 February 2020) in the Arizona state. (left) Percent change for Residential areas (one of the two individuals in proximity is at their own residential area), POI (both individuals are at the same POI), Business (both of having the same work location) and Other co-location events. (right) Percent change for all the categories of POIs.}
	\label{SI:fig:contacts-duration-AZ}
\end{figure}

\begin{figure}[!ht]
	\centering
	\includegraphics[width=\textwidth]{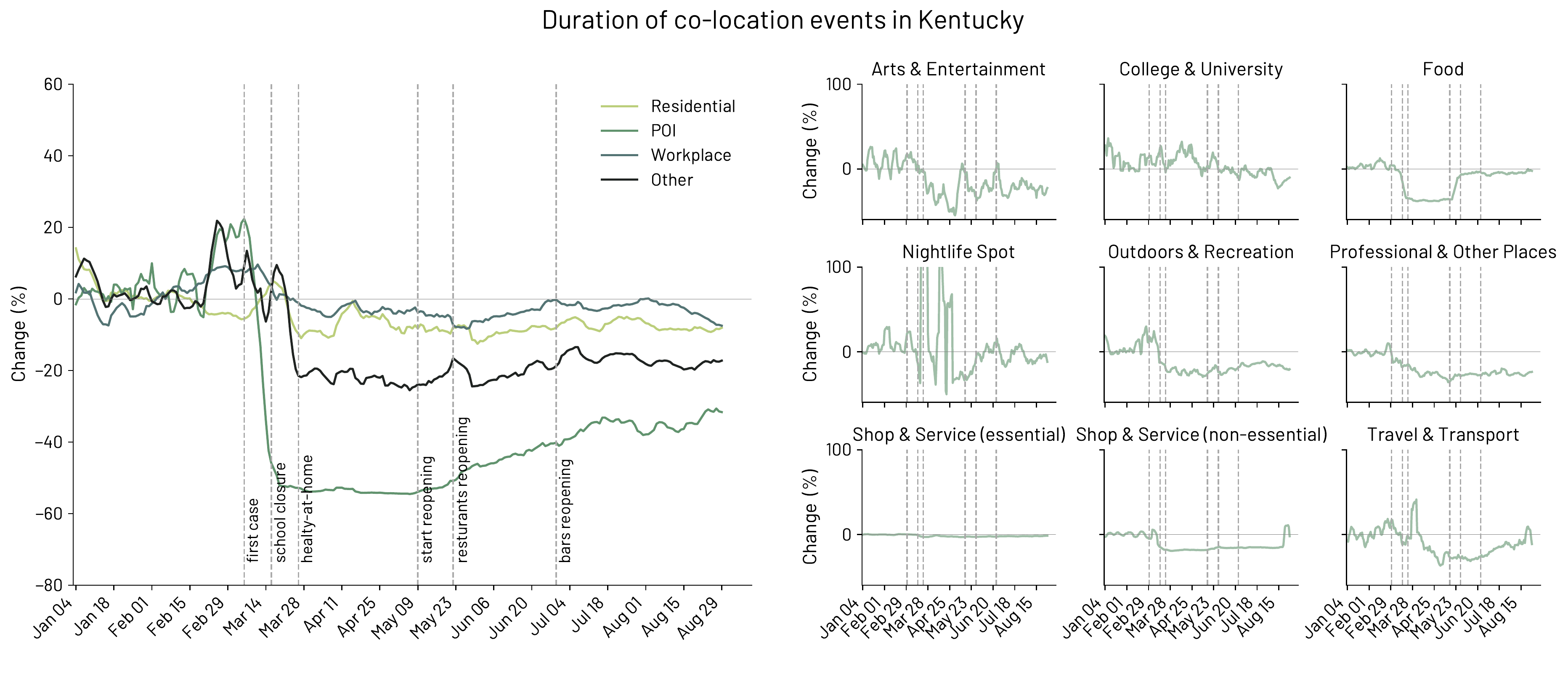}
	\caption{Percent change of the duration of co-location events from the baseline period (until 29 February 2020) in the Kentucky state. (left) Percent change for Residential areas (one of the two individuals in proximity is at their own residential area), POI (both individuals are at the same POI), Business (both of having the same work location) and Other co-location events. (right) Percent change for all the categories of POIs.}
	\label{SI:fig:contacts-duration-KY}
\end{figure}

\begin{figure}[!ht]
	\centering
	\includegraphics[width=\textwidth]{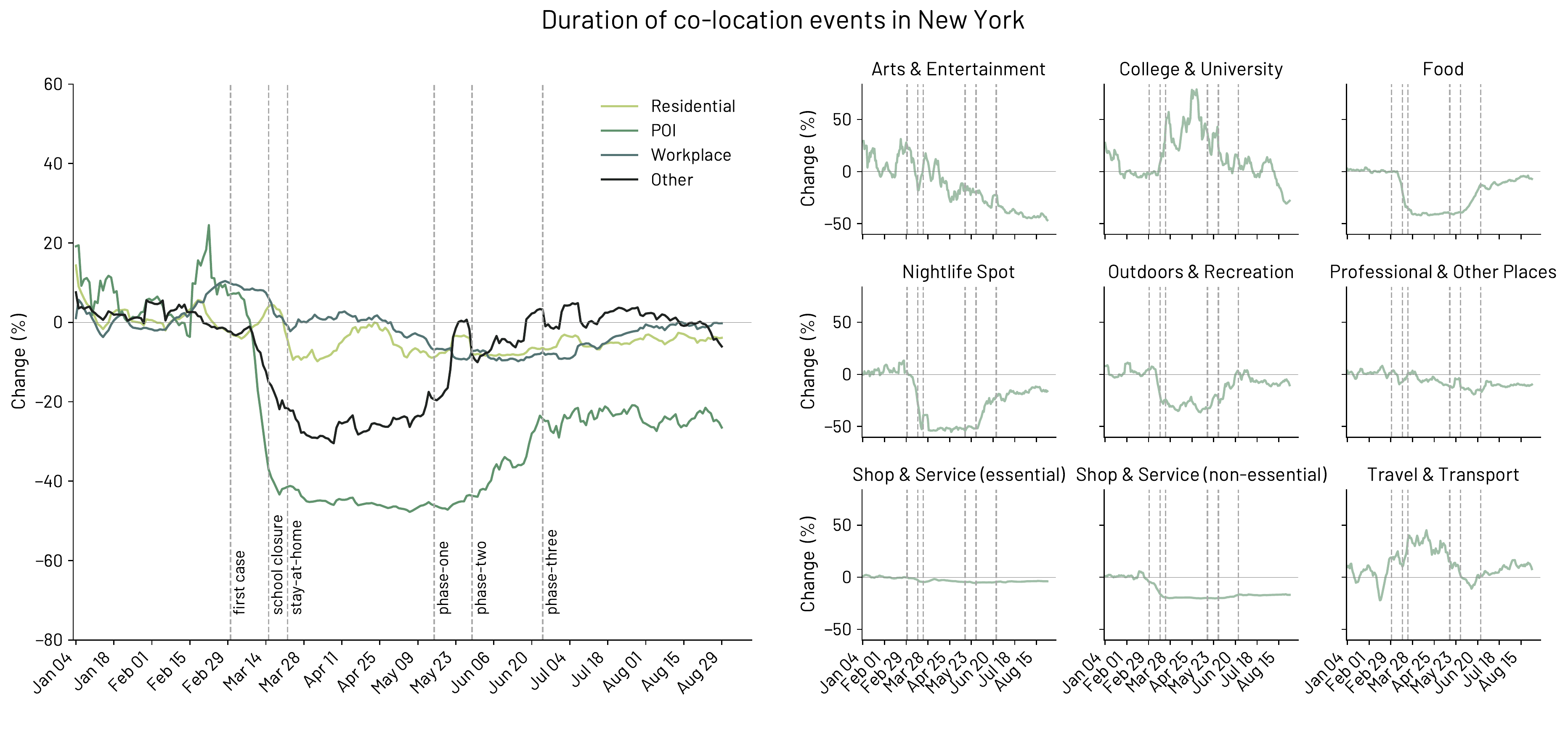}
	\caption{Percent change of the duration of co-location events from the baseline period (until 29 February 2020) in the New York state. (left) Percent change for Residential areas (one of the two individuals in proximity is at their own residential area), POI (both individuals are at the same POI), Business (both of having the same work location) and Other co-location events. (right) Percent change for all the categories of POIs.}
	\label{SI:fig:contacts-duration-NY}
\end{figure}

\begin{figure}[!ht]
	\centering
	\includegraphics[width=\textwidth]{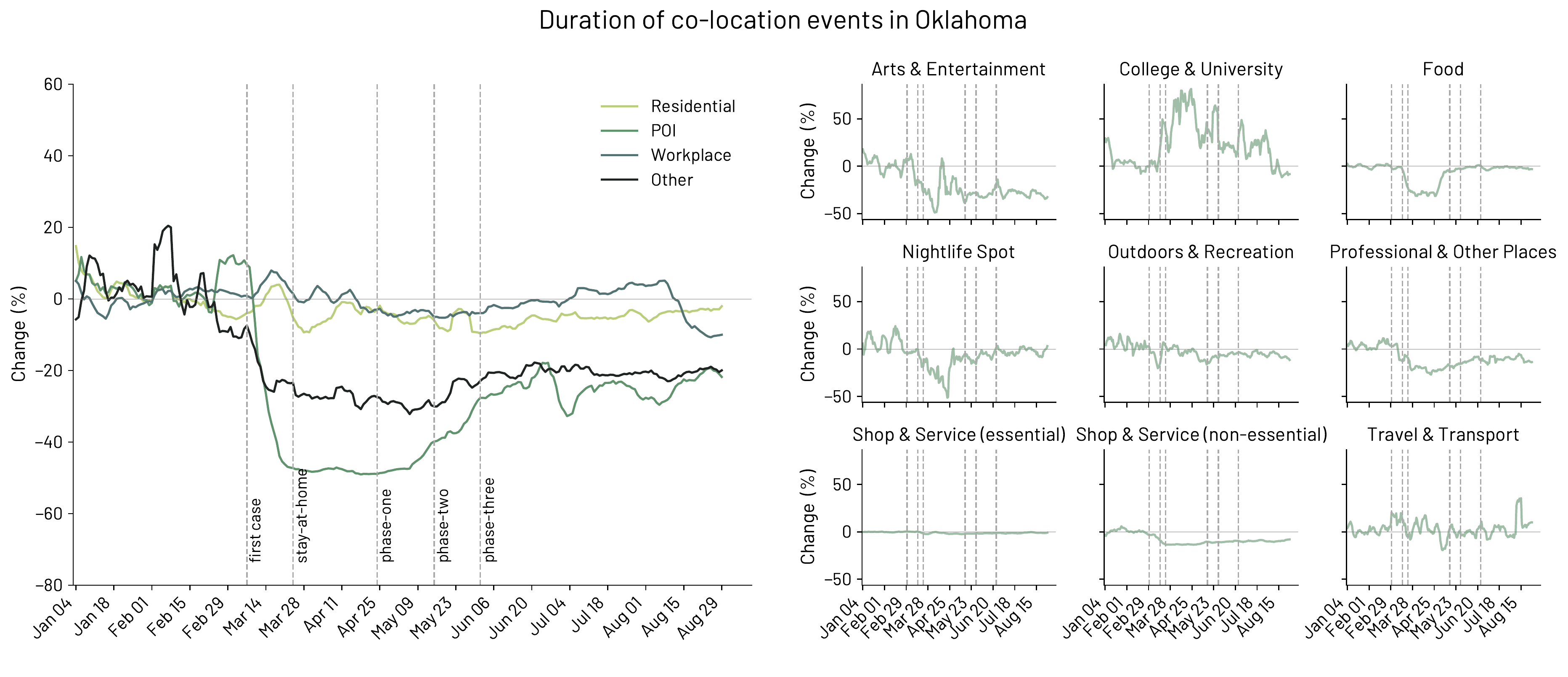}
	\caption{Percent change of the duration of co-location events from the baseline period (until 29 February 2020) in the Oklahoma state. (left) Percent change for Residential areas (one of the two individuals in proximity is at their own residential area), POI (both individuals are at the same POI), Business (both of having the same work location) and Other co-location events. (right) Percent change for all the categories of POIs.}
	\label{SI:fig:contacts-duration-OK}
\end{figure}

\Cref{SI:fig:contacts-duration-AZ}, \Cref{SI:fig:contacts-duration-KY}, \Cref{SI:fig:contacts-duration-NY} and \Cref{SI:fig:contacts-duration-OK} show the percent change in the duration of co-location events in the states of New York, Kentucky, Oklahoma and Arizona respectively.

\subsection{Null model}
\label{SI:sec:co-location-null}
As the pandemic unfolds in the US, its impact on the number of visits to POIs grows, reducing by about $80\%$ their number during the first phase period.
This reduction of the number and duration of visits are expected to impact the number of co-location events and their duration strongly. To this extent, we propose two null models that aim at capturing the expected number and duration of co-locations given the reduction in the number of visits to POIs, the time spent at POIs, and the duration of those stops during which a co-location event occurred.

\noindent\textbf{Number of co-location events null-model.} To provide an estimate for the daily number of co-location events, we focus on co-locations happening at a single POI $i$ on a specific day $d$. The number of co-location events occurring at that location on a specific day $e_{i,d}$ can be estimated from both the number of individuals visiting the POI $n_{i,d}$, and the median duration of their stops there $\bar{d}_{i,d}$.

We can then provide an estimate for $e_{i,d}$ using the following:
\begin{equation}
	e_{i,d} = \binom{n_{i,d}}{2} p_{i,d}
\end{equation}
where $p_{i,d}$ is the probability of having a co-location event given two individuals visiting POI $i$ on day $d$. $p_{i,d}$ is computed assuming a Uniform distribution for the time-interval of visit of two individuals potentially having a co-location. This means that, if two individuals visited the same POI $i$ on day $d$, they will have a non-null probability of visiting $i$ with partially (or completely) overlapping time-intervals of visit. The probability that the two intervals have an overlap of at least $\epsilon$ time is then computed as a fraction of the length of their stop duration and the duration of a day with open boundary conditions:
\begin{equation}
	p_{i,d} = 
	\begin{cases}
		0 & \bar{d}_{i,d} < \epsilon\\
		\frac{2(\bar{d}_{i,d}-\epsilon)}{\Tilde{d}} & \epsilon \le \bar{d}_{i,d} < \Tilde{d}/2\\
		1 & \Tilde{d}/2 \le \bar{d}_{i,d}\le \Tilde{d}/2
	\end{cases}
\end{equation}
where $\epsilon$ is a minimum number of minutes the two time-interval are required to overlap in order for a co-location to be counted as a co-location event, and $\Tilde{d}$ is the duration of a day in minutes.

\noindent\textbf{Co-location event duration null-model.}
The expected duration of a co-location event is estimated following a similar reasoning. Conditioning on the fact that two individuals are co-located at POI $i$ on day $d$, we use the average length of their stops there, namely $\hat{d}_{i,d}$, to estimate their expected time-overlap under a uniformly distributed probability for the displacement of their visit over the entire day. Thus, our null-model assumes that individuals visit POIs without the specific intent of meeting with someone. Under this assumption, the expected temporal overlap, $o_{i,d}$, which corresponds the expected duration of a co-location events, can be easily derived as the average overlap between two time-intervals moving over the same domain:
\begin{align}
	o_{i,d} & = \frac{1}{2(\hat{d}_{i,d}-\epsilon)}\int_{-\hat{d}_{i,d}+\epsilon}^{\hat{d}_{i,d}-\epsilon}(\hat{d}_{i,d}-|x|)\cdot dx\\
	& = \frac{\hat{d}_{i,d}+\epsilon}{2}
\end{align}
where $x$ is the distance between the center of the two intervals of equal length $\hat{d}_{i,d}$.

\begin{figure}[!ht]
	\centering
	\includegraphics[width=\textwidth]{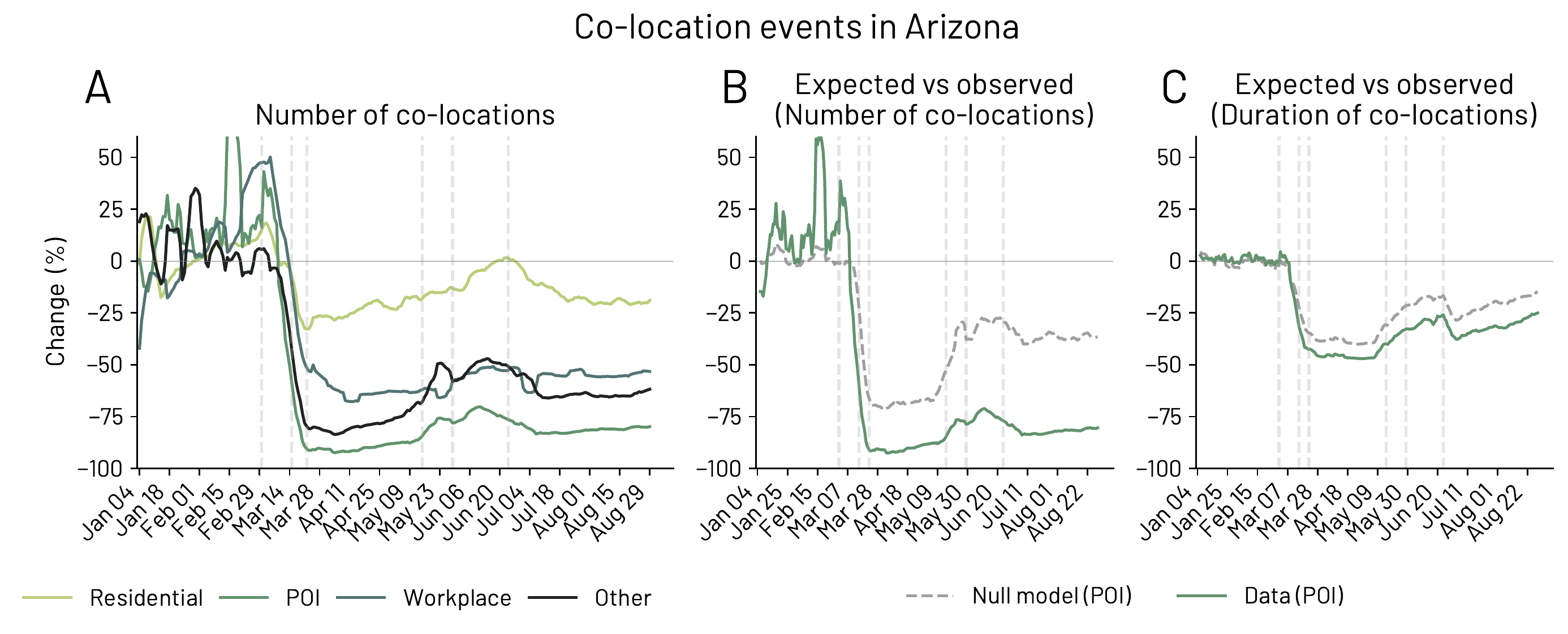}
	\caption{A) Percentage change of co-location events from the baseline period (until 29 February 2020) in the Arizona state. We measure the change for Residential areas (one of the two individuals in proximity is at their own residential area), POI (both individuals are at the same POI), Business (both of having the same work location) and Other co-location events. B-C) We compare the difference between the expected and observed number and duration of co-location events. The figure shows that during the pandemic, individuals tend to have less co-locations than expected. However, it seems that co-locations are longer than expected.}
	\label{SI:fig:null-contact-model-AZ}
\end{figure}

\begin{figure}[!ht]
	\centering
	\includegraphics[width=\textwidth]{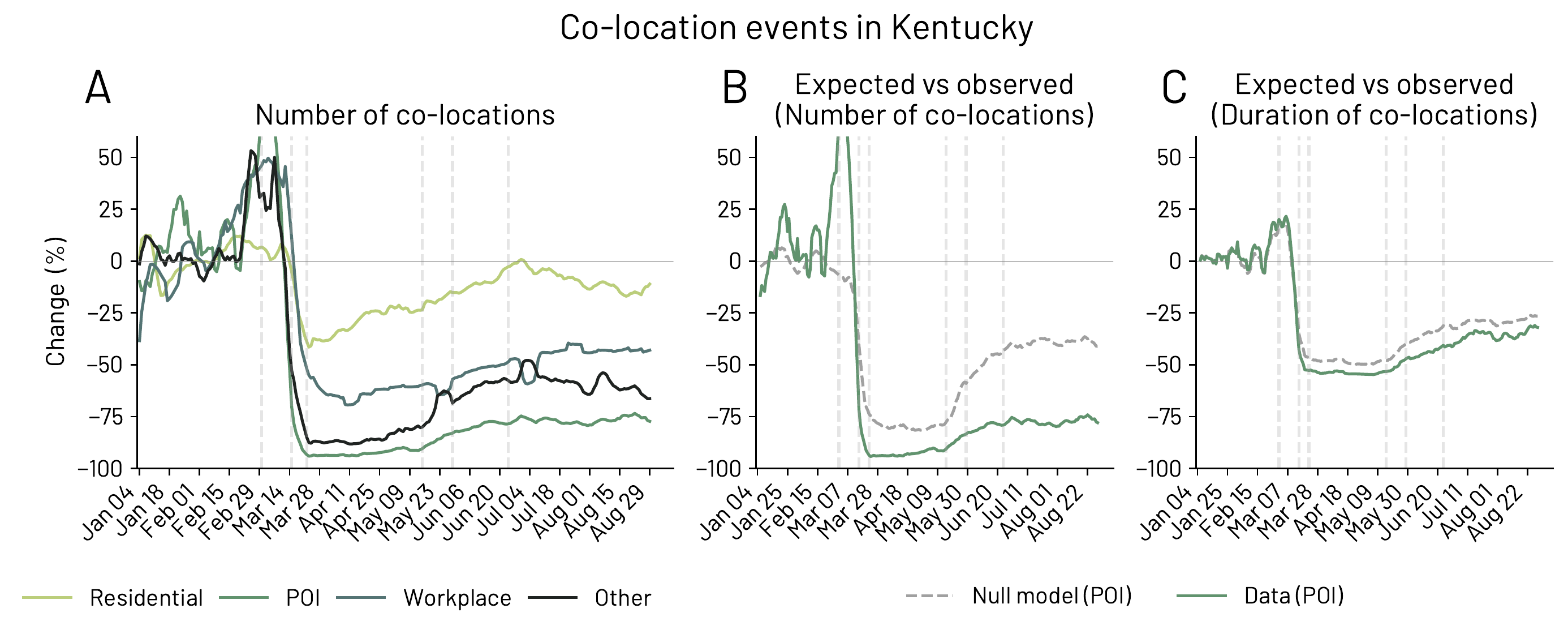}
	\caption{A) Percentage change of co-location events from the baseline period (until 29 February 2020) in the Kentucky state. We measure the change for Residential areas (one of the two individuals in proximity is at their own residential area), POI (both individuals are at the same POI), Business (both of having the same work location) and Other co-location events. B-C) We compare the difference between the expected and observed number and duration of co-location events. The figure shows that during the pandemic, individuals tend to have less co-locations than expected. However, it seems that co-locations are longer than expected.}
	\label{SI:fig:null-contact-model-KY}
\end{figure}

\begin{figure}[!ht]
	\centering
	\includegraphics[width=\textwidth]{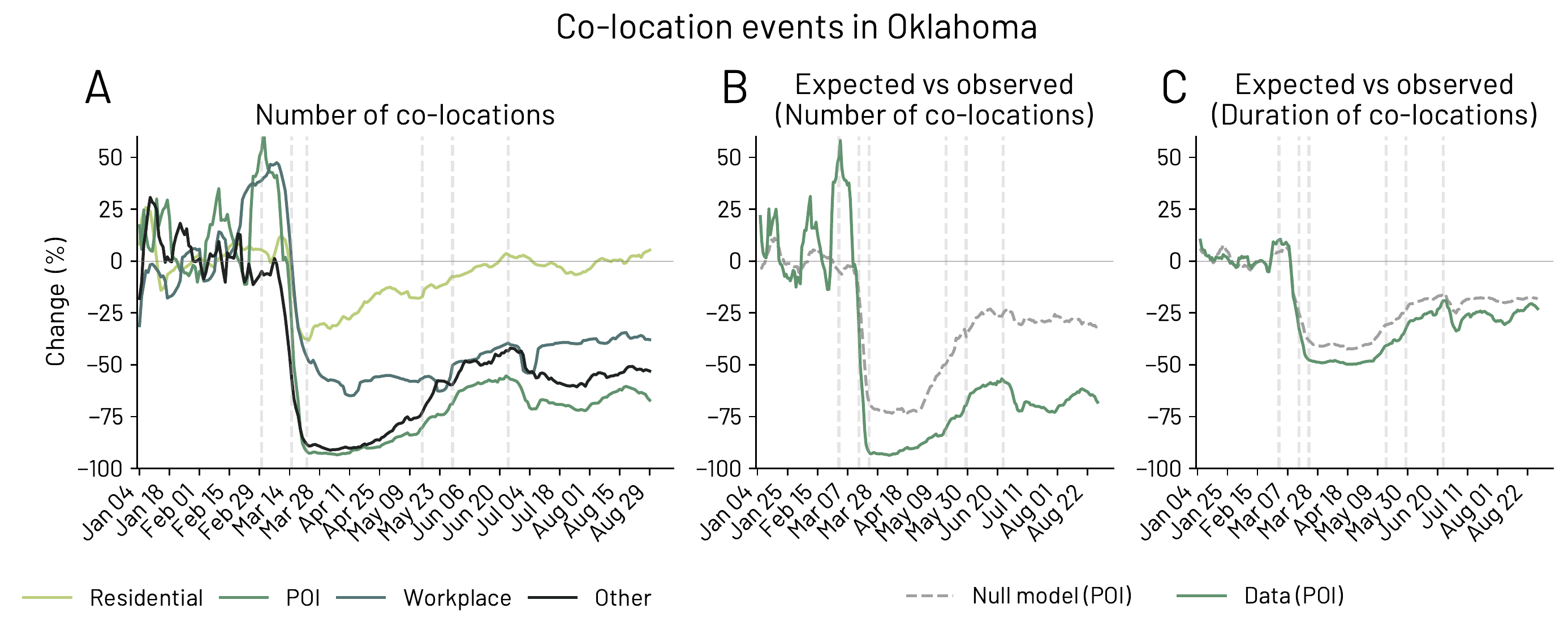}
	\caption{A) Percentage change of co-location events from the baseline period (until 29 February 2020) in Oklahoma state. We measure the change for Residential areas (one of the two individuals in proximity is at their residential area), POI (both individuals are at the same POI), Business (both of having the same work location) and Other co-location events. B-C) We compare the difference between the expected and observed number and duration of co-location events. The figure shows that during the pandemic, individuals tend to have fewer co-locations than expected. However, it seems that co-locations are longer than expected.}
	\label{SI:fig:null-contact-model-OK}
\end{figure}

\vspace{0.75cm}
\Cref{SI:fig:null-contact-model-AZ}, \Cref{SI:fig:null-contact-model-KY}, and \Cref{SI:fig:null-contact-model-OK} show the percent change in: A) the number of co-location events divided per co-location event type; B) the percent change in co-location events happening in POIs (full line) against the expected co-location events (dashed line); C) the percent change in the duration (i.e. the time overlap) of co-location events happening in POIs (full line) against the expected duration of co-location events (dashed line). Figures \Cref{SI:fig:null-contact-model-AZ}, \Cref{SI:fig:null-contact-model-KY}, and \Cref{SI:fig:null-contact-model-OK} show the results for the state of Kentucky, Oklahoma and Arizona respectively.

\section{Visits to POIs: model construction and model selection}\label{SI:sec:model_selection_all}
This section reports all the models tested and their construction in terms of prior distributions and sampling algorithms. 

\noindent\textbf{Prior selection:} Here, we introduce the prior distributions adopted to fit the full model. The baseline and weather models use the same prior distribution without the additional parameters introduced in the full model only. Therefore, for the sake of simplicity, we report here the complete list of priors without differentiating for the three main models presented and discussed in the main paper.
\begin{eqnarray*}
	\alpha_s &\sim& Normal(\mu=0,\sigma=1)\\
	\beta_{policy} &\sim& Normal(\mu=-1, \sigma=0.1)\\
	\beta_{deaths} &\sim& Normal(\mu=-1, \sigma=0.1)\\
	\rho_i &\sim& Normal(\mu=0, \sigma=0.1)\\
	\beta_{temp} &\sim& Normal(\mu=0, \sigma=0.1)\\
	\beta_{prec} &\sim& Normal(\mu=0, \sigma=0.1)\\
	\beta_{adapt} &\sim& HalfNormal(\sigma=0.1)\\
	\gamma_s &\sim& 0.01 + HalfNormal(\sigma=0.1)\\
	\phi_s &\sim& 90 + HalfNormal(\sigma=10)\\
	\epsilon &\sim& HalfNormal(\sigma=0.05)\\
	Y &\sim& Student(\mu = y_{model},\sigma=\epsilon,\nu=3).
	\label{si:eq:full_model}
\end{eqnarray*}

\noindent\textbf{Cumulated stringency model:} 
Our research question aims at understanding if a model accounting for a behavioural adaptation to risk after a sustained period of exposure can better describe the individual's tendency to visit POIs or not. The formulation of the full model takes this effect into account by adding a dependency on the time elapsed after the first pandemic restrictions. To confirm and test the robustness of the full model results, we also test for a different model, namely the "cumulated stringency model", which instead of adding a dependency on time, uses the value of the cumulated restrictions intensity as a proxy of the pandemic burden experienced so far. Table~\ref{tab:model_results_mean_visits} shows that also this model better captures the reduction and subsequent recovery of the number of visits to POIs.

\noindent\textbf{Modeling "time not at home" :} As an additional robustness check, we also tested the performances of these four models in describing the timeseries of the time not spent in residential areas. Also in this case, Tab.~\ref{tab:model_results_no_home_mean_minutes_spent} summarizes the better performance of both models accounting for a behavioural adaptation component. 

\begin{table}
	\centering
	\setlength{\tabcolsep}{6pt}
	\label{tab:model_results_mean_visits}
	\begin{tabular}{rrrrrrrr}
		\toprule
		\textbf{model} &       \textbf{metric} &     \textbf{loo} &     \textbf{se} &    \textbf{r2} &  \textbf{r2std} & \textbf{weight}\\
		\midrule
		\textbf{full} &  mean visits & 846.770 & 29.598 & 0.783 &  0.024 &  0.999 \\
		\textbf{cumulated strin.} & mean visits & 811.043 & 26.941 & 0.758 &  0.017 &  0.001 \\
		\textbf{weather} & mean visits & 747.450 & 28.780 & 0.728 &  0.019 &  0.000 \\
		\textbf{baseline} & mean visits & 610.603 & 31.943 & 0.674 &  0.022 &  0.000 \\
		\bottomrule
	\end{tabular}
	\caption{\textbf{Summary results for the models describing the average number of visits per individual per day.} Results show that the full model outperforms all the other in terms of the Pareto-smoothed importance sampling Leave-One-Out cross-validation information criterion (PSIS-LOO)~\cite{vehtari2017practical}. The \emph{cumulated Stringency} model following a similar idea as the full model shows better performances if compared with the weather and the baseline models.}
\end{table}

\begin{table}
	\centering
	\setlength{\tabcolsep}{6pt}
	\label{tab:model_results_no_home_mean_minutes_spent}
	\begin{tabular}{rrrrrrrr}
		\toprule
		\textbf{model} &       \textbf{metric} &     \textbf{loo} &     \textbf{se} &    \textbf{r2} &  \textbf{r2std} & \textbf{weight}\\
		\midrule
		\textbf{full} &  time not in residential areas & 328.691 & 42.865 & 0.858 &  0.013 &  1.000 \\
		\textbf{cumulated strin.} &  time not in residential areas & 234.250 & 37.602 & 0.820 &  0.018 &  0.000 \\
		\textbf{weather} &  time not in residential areas & 192.226 & 40.513 & 0.807 &  0.018 &  0.000 \\
		\textbf{baseline} &  time not in residential areas & 138.522 & 46.064 & 0.797 &  0.020 &  0.000 \\
		\bottomrule
	\end{tabular}
	\caption{\textbf{Summary results for the models describing the average time spent outside the residential area per individual per day.} Results show that the full model outperforms all the other in terms of the Pareto-smoothed importance sampling Leave-One-Out cross-validation information criterion (PSIS-LOO)~\cite{vehtari2017practical}. The \emph{cumulated Stringency} model, following a similar idea as the full model, shows better performances if compared with the weather and the baseline models.}
\end{table}





\end{document}